\documentclass[resetfootnote,trackchanges, twocolumn, tighten]{aastex701}
\usepackage{tikz}
\usepackage{ulem}
\usepackage{mhchem}

\newcommand\subs[1]{\textsubscript{#1}}

\newcommand\rh[1]{\textcolor{black}{{\textit{r}\subs{\textit{H}}}#1}}

\newcommand\um[1]{\textcolor{black}{{$\mu$m}#1}}

\definecolor{gold}{rgb}{0.64,0.54,0.29}

\usepackage{color}

\newcommand{\gtsimeq}{\raisebox{-0.6ex}{$\,\stackrel
        {\raisebox{-.2ex}{$\textstyle >$}}{\sim}\,$}}

\usepackage{ulem}
\usepackage{hyperref}
\usepackage{datetime2}
\usepackage{graphicx}
\graphicspath{{figures_paper2/}{./}}
\usepackage{caption} 

\usepackage{amsmath}

\received{Sept. 27, 2026}
\submitjournal{The Planetary Science Journal}

\makeatletter
\renewcommand{\frontmatter@title@above}{}
\makeatother
\shorttitle{A JWST Study of Stardust. II.}
\shortauthors{Roth et al.}

\begin{document}


\title{A JWST Study of Stardust. II. Polycyclic Aromatic Hydrocarbons in Comet 81P/Wild 2 and Implications for Primitive Solar System Carbon.}

\correspondingauthor{Nathan X. Roth}
\email{nathaniel.x.roth@nasa.gov}

\author[0000-0002-6006-9574]{Nathan X. Roth}
\affiliation{Solar System Exploration Division, Astrochemistry Laboratory Code 691, NASA Goddard Space Flight Center, 8800 Greenbelt Rd, Greenbelt, MD 20771, USA}
\affiliation{Department of Physics, American University, 4400 Massachusetts Ave NW, Washington, DC 20016, USA}
\email{nathaniel.x.roth@nasa.gov}

\author[0000-0001-7694-4129]{Stefanie N. Milam}
\affiliation{Solar System Exploration Division, Astrochemistry Laboratory Code 691, NASA Goddard Space Flight Center, 8800 Greenbelt Rd, Greenbelt, MD 20771, USA}
\email{stefanie.n.milam@nasa.gov}

\author[0000-0002-1888-7258]{Diane H. Wooden}
\affiliation{NASA Ames Research Center, Astrophysics Division, MS 245-6, Moffett Field, CA 94035-1000, USA}
\email{diane.h.wooden@nasa.gov}

\author[0000-0002-2541-1602]{Els Peeters}
\affiliation{Department of Physics and Astronomy, University of Western Ontario, London, Ontario N6G 2V4, Canada}
\affiliation{Institute for Earth and Space Exploration, University of Western Ontario, London, Ontario N6A 5B7, Canada}
\affiliation{SETI Institute, Mountain View, CA 94043, USA}
\email{epeeters@uwo.ca}

\author[0000-0001-6567-627X]{Charles E. Woodward}
\affiliation{Minnesota Institute for Astrophysics, School of Physics and Astronomy, 116 Church Street SE, University of Minnesota, Minneapolis, MN 55455, USA}
\email{chickw024@gmail.com}

\author[0000-0002-8130-0974]{Dominique Bockelée-Morvan}
\affiliation{LIRA, Observatoire de Paris, Université PSL, CNRS, Sorbonne Université, Université Paris Cité, 5 place Jules Janssen, 92195 Meudon, France}
\email{dominique.bockelee@obspm.fr}

\author[0000-0002-6702-7676]{Michael S. P. Kelley}
\affiliation{Department of Astronomy, University of Maryland, College Park, MD 20742-0001, USA}
\email{msk@astro.umd.edu}

\author[0000-0001-6752-5109]{Steven B. Charnley}
\affiliation{Solar System Exploration Division, Astrochemistry Laboratory Code 691, NASA Goddard Space Flight Center, 8800 Greenbelt Rd, Greenbelt, MD 20771, USA}
\email{steven.b.charnley@nasa.gov}

\author[0000-0002-5187-1653]{Simon J. Clemett}
\affiliation{NASA Johnson Space Center, 2224 Bay Area Blvd, Houston, TX, 77058, USA}
\email{simon.j.clemett@nasa.gov}

\author[0000-0001-8233-2436]{Martin A. Cordiner}
\affiliation{Solar System Exploration Division, Astrochemistry Laboratory Code 691, NASA Goddard Space Flight Center, 8800 Greenbelt Rd, Greenbelt, MD 20771, USA}
\affiliation{Department of Physics, The Catholic University of America, 620 Michigan Ave., N.E. Washington, DC 20064, USA}
\email{martin.cordiner@nasa.gov}


\author[0000-0002-9667-5904]{Perry A. Gerakines}
\affiliation{Solar System Exploration Division, Astrochemistry Laboratory Code 691, NASA Goddard Space Flight Center, 8800 Greenbelt Rd, Greenbelt, MD 20771, USA}
\email{perry.a.gerakines@nasa.gov}

\author[0000-0001-6397-9082]{David E. Harker}
\affiliation{Department of Astronomy and Astrophysics, University of California, San Diego, 9500 Gilman Drive, MC 0424, La Jolla, CA 92093-0424, USA}
\email{dharker@ucsd.edu}


\author[0000-0002-1883-552X]{Ella Sciamma-O’Brien}
\affiliation{NASA Ames Research Center, Astrophysics Division, MS 245-6, Moffett Field, CA 94035-1000, USA}
\email{ella.m.sciammaobrien@nasa.gov}

\author[0000-0001-6752-5109]{Geronimo L. Villanueva}
\affiliation{Solar System Exploration Division, Code 690, NASA Goddard Space Flight Center, 8800 Greenbelt Rd, Greenbelt, MD 20771, USA}
\email{geronimo.l.villanueva@nasa.gov}




\begin{abstract}

We report observations of comet 81P/Wild 2, target of the Stardust sample return mission, on UT 2023 March 20 and 24 at a heliocentric distance (\rh{}) of 1.85 au using the NIRSpec and MIRI integral field unit spectrographs on board the James Webb Space Telescope (JWST). This  study is the first compositional comparison between JWST remote-sensing spectroscopy of a solar system object against terrestrial analysis of its returned samples. Recent work determined contributions of molecular emission from coma volatiles and thermal emission from the nucleus and coma dust grains to these spectra. Subtracting the molecular and thermal emission models produces a residual spectrum with a clear resonance near 3.37 \um{} and multiple others spanning the $6 - 12$ \um{} region. Analysis of these residuals provides strong evidence for the presence of polycyclic aromatic hydrocarbons (PAHs) in the coma of 81P/Wild 2. The resulting PAH populations reflect material incorporated into the nucleus of 81P/Wild 2 that was inherited from the diffuse interstellar medium, the interstellar medium, and the protoplanetary disk. The PAHs sensed by JWST are compared to those detected in laboratory analyses of the Stardust mission returned samples.

\end{abstract}



\section{Introduction}
\label{sec:intro}
Comets may serve as ``living fossils'' of solar system formation, with the
volatile composition of their nuclei reflecting the chemistry and prevailing
conditions present where and when they formed
\citep{2004come.book..391B,2011ARA&A..49..471M}. The James Webb Space
Telescope \cite[JWST;][]{2023PASP..135d8001R,2023PASP..135f8001G} is opening
new frontiers in cometary science, providing sensitive measures of the
volatile and refractory components of the coma, including evidence for
polycyclic aromatic hydrocarbon (PAH) species \citep{2025PSJ.....6..139W}.
Here we report JWST MIRI and NIRSpec integral field unit (IFU) observations of
Jupiter-family comet 81P/Wild 2 (hereafter 81P). Comet 81P represented an
exceptional JWST target, being also the target of the Stardust sample return
mission
\citep[e.g.,][]{2006Sci...314.1720S,2010M&PS...45..701C,2017M&PS...52..471B}
and complemented by the availability of ground-based, high spectral resolution
near-infrared compositional studies during its previous perihelion passages
\citep{2014Icar..238..125D}. These combined measurements reported here
represent the first opportunity to integrate the results of comet sample
return analyses with a full near--mid infrared cometary spectrum measured with
the unparalleled sensitivity of JWST.

\citet[][hereafter Paper I]{2026arXiv260802190R} reported our radiative
transfer modeling of molecular emission from volatiles in the coma and thermal
modeling of continuum emission from the nucleus and coma dust grains. The
best-fit models for all three of these quantities were subtracted to form a
residual spectrum, which is analyzed in this work (Paper II) to test for the
presence of PAHs. Section~\ref{sec:obs} provides a brief overview of the
observations and of the derivation of the residual spectrum.
Section~\ref{sec:pah} describes the PAH features of the residual spectrum and
the modeling formalism for determining contributions from PAHs in the coma.
Section~\ref{sec:results} presents the PAH models explored, the best-fit
model, and the searches for very small N-bearing PAHs and for carbon
nanograins. Section~\ref{sec:pah:origins} compares the PAHs detected in the
coma of 81P with JWST to those found in protoplanetary disks and the
interstellar medium, and Section~\ref{sec:pah:stardust} makes a comparison
with those measured in the Stardust returned samples.
Section~\ref{sec:pah:summary} summarizes the findings.
\section{Observations and the Residual Spectrum}
\label{sec:obs}

Comet 81P was observed post-perihelion with the JWST MIRI Medium
Resolution Spectroscopy (MRS) IFU on UT 2023 March 20 (4.9--28.1~\micron{},
$\lambda/\Delta\lambda \sim 3000$) and with the NIRSpec IFU in the
G395H/F290LP setting on UT 2023 March 24 (2.87--5.27~\micron{},
$\lambda/\Delta\lambda \sim 2700$). The heliocentric distances were 
$r_{\rm H} = 1.87$ and 1.85~au and the JWST--comet distance was 
$\Delta = 1.43$~au (Paper~I, \citealt{2026arXiv260802190R}, Table~1). 
The data were reduced with version\
1.14.0 of the JWST pipeline and the \texttt{jwstComet} package, and
spectra were extracted in a 1\farcs41 diameter aperture centered on the
photocenter. In this aperture the nucleus supplies about 0.75 of the MRS flux.
The coma flux was obtained by subtracting a NEATM model of the nucleus,
aperture-corrected for the MRS. At NIRSpec wavelengths the nucleus model was
scaled by 0.96 and the resulting coma flux by 1.042 so that the NIRSpec and
MRS coma fluxes agree in their region of overlap. This scaling is consistent
with a small change in the projected area of the rotating nucleus between the
two epochs (Paper~I, Section~6.2).

The residual spectrum analyzed here is the coma flux minus three modeled
components: the molecular line emission, the dust thermal emission, and the
scattered sunlight. Molecular lines were modeled with the Planetary Spectrum
Generator \citep[PSG;][]{Villanueva2018} together with a piecewise continuous baseline (Paper~I,
Sections~4 and 5). Residual H$_{2}$O line cores in MRS Channel~1 were removed
by masking and median filtering. The \ce{CH3OH} $\nu_9$ band is intrinsically challenging to model, yet the application of its fluorescence model plays a significant role in determining the shape of the residuals near 3.4 \um{}. We provide additional details for the PSG \ce{CH3OH} model and its application to 81P in Appendix~\ref{sec:methanol}.

The dust thermal model, five materials with a Hanner grain size distribution
\citep[Paper~I, Section~6.3;][]{2023PSJ.....4..242H}, and the scattered light
model, a linear model in normalized reflectance anchored at 3.6~\micron{}, were fitted iteratively. Neither component can be fitted from one instrument alone, and
both must be determined iteratively. Thermal emission extends into the
long-wavelength end of NIRSpec, scattered light reaches into the
short-wavelength end of MRS, and near the region of overlap the two are
roughly equal. Each iteration fitted the scattered light to the flux minima
between the molecular bands in NIRSpec, extrapolated it to MRS and subtracted
it, then refitted the thermal model to the scattered-light-subtracted flux
down to 4.1~\micron{}. The iteration was repeated until the two models and the
NIRSpec-to-MRS scaling were mutually consistent, ending with a final scattered
light fit. Of three NIRSpec scaling factors tested, 1.042 gave the smallest
total residual in the overlap region.

The thermal model was not fitted to the
measured flux directly where PAH emission is expected. Over 5.60--8.65~\micron{} two estimates of the local
continuum beneath the bands were obtained by sliding the amorphous carbon and
the amorphous olivine components of an initial thermal model under the
emission. An uncertainty in each estimate was set at
$2.5\times10^{-21}$~W~cm$^{-2}$~\micron$^{-1}$, representing the
pixel-to-pixel undulations of the Channel~1 data. The midpoint of the two
estimates and the extrema of their uncertainty envelopes were used to
constrain the thermal model. In addition, the uncertainties of the
short-wavelength shoulder of the silicate feature (8.65--10~\micron{}) and of
the 12.5--16.5~\micron{} region were enlarged by 40$\times$, because the
shoulder could hold emission from organics and because the thermal model's
Akaike Information Criterion (AIC) value improved when the
12.5--16.5~\micron{} region was underweighted.

The thermal model was then adopted as the baseline under the 5.6--8.65,
8.65--10, and 12.5--16.5~\micron{} regions. The coma flux minus the thermal
and scattered light models is the PAH residual spectrum of
Figure~\ref{fig:figure-1}. The contrast of the residual over 5.5--27~\micron{}
depends on the placement of the continuum set by the thermal model, and over
5.6--8.65~\micron{} on the local continuum estimated beneath the bands. The
adjustments made to the residual before fitting the PAH model are described in
Section~\ref{sec:pah:method}.

The uncertainties of the thermal and scattered light models are not propagated
formally, but the differences among the residuals for the three NIRSpec
scaling factors are small compared with the MRS instrumental uncertainties
(Paper~I, Section~6.3). The uncertainties of the residual are those of the
coma flux density (Paper~I, Section~6.1). For the MRS these are the
instrumental uncertainties added in quadrature with the uncertainty of the
nucleus subtraction. That uncertainty was assessed at each wavelength as the
absolute value of the difference between the coma flux in the 1\farcs41
aperture and the coma flux in the 1\farcs22 aperture scaled by 1.19, the
factor that best matches the larger aperture. Each coma spectrum had its own
aperture correction factor for the nucleus. The 1\farcs00 diameter aperture
was judged less consistent in spectral shape with the other two and was
dropped from the error estimate.

\section{PAHdb Modeling and Analyses}
\label{sec:pah}

\subsection{PAH Spectral Features of Comet 81P}
\label{sec:pah:intro}

\begin{figure*}[ht!]
\figurenum{1}
\begin{center}
\includegraphics[trim=0.07cm 0.05cm 0.07cm 0.07cm, clip, width=0.800\textwidth]{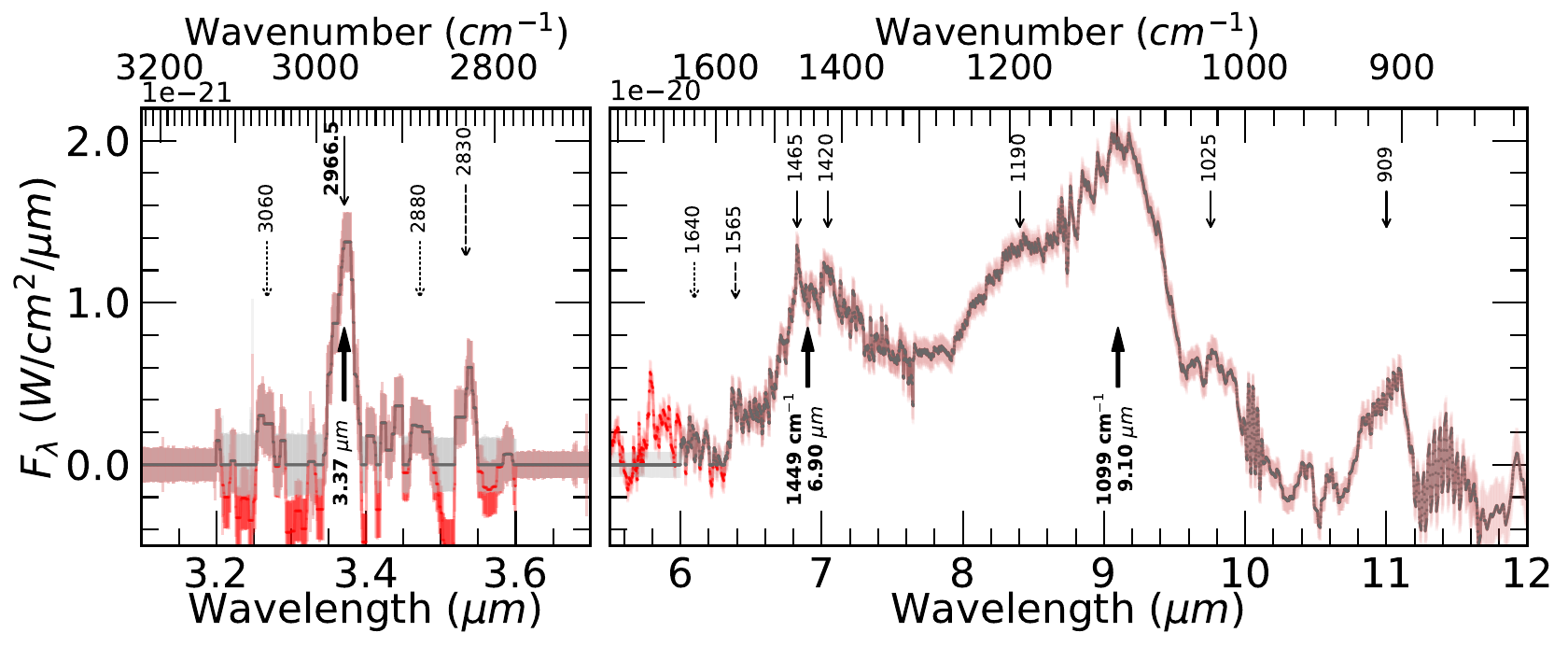}
\caption{Comet 81P's PAH residual spectrum, shown (left) over the NIRSpec
range, 3.2--3.6~\micron, and (right) over the 5.5--12~\micron{} portion of the
MRS range (4.9--27~\micron) that contains the most determinate features for
fitting the PAH model. Red shows the coma flux density with 1$\sigma$
uncertainties; gray shows the modified data fitted by the PAH model, with flux
density set to zero across 5--6~\micron{} and negative points across
3.2--3.6~\micron{} clipped to zero. A spectral feature and two spectral
complexes (filled arrows) are centered near 3.37~\micron{} (2966.5~cm$^{-1}$),
6.90~\micron{} (1449~cm$^{-1}$), and 9.10~\micron{} (1099~cm$^{-1}$). Solid,
dashed, and dotted arrows mark strong features, weak features, and upper
limits, respectively. The NIRSpec data and 1$\sigma$ instrumental
uncertainties are scaled by a factor of 2 (\ref{sec:pah:method}).}
\label{fig:figure-1}
\end{center}
\end{figure*}

The emission spectrum of organics in comet 81P is dominated by broad, strongly
asymmetric features. These are presumed to arise from polycyclic aromatic
hydrocarbons (PAHs) or carbonaceous nanograins because only these species ---
as measured in space, in the laboratory, or theoretically computed --- have
resonances corresponding to the observed features in comets. In the
interstellar medium (ISM), such as in the Orion Bar, PAHs and hydrogenated
amorphous carbon (a-C:H) grains produce distinct bands at wavelengths commonly
cited as 3.3~\micron, 6.2~\micron, 7.7~\micron, 8.6~\micron, and
11.2~\micron{} \citep[][and references therein]{2022MNRAS.509.3523K}. In
contrast, comet 81P's spectrum shows broad, strongly asymmetric, and
ostensibly overlapping features (Figure~\ref{fig:figure-1}). The sole prominent NIRSpec feature in 81P at 3.37~\micron{} is weak in flux relative to the
flux in the 6--10~\micron{} region: a factor of $\sim$10 times weaker in flux
density, \textit{F}$_{\lambda}$ (W~cm$^{-2}$ $\mu$m$^{-1}$), or $\sim$85 times
weaker in \textit{F}$_{\nu}$ units (Jy). The 3.37~\micron{} feature has its
peak at 3.375~\micron{} (2963~cm$^{-1}$, 7.33$\sigma$ at the peak) with a
gaussian-fit centroid at 2966.5 cm$^{-1}$ and FWHM $\simeq 51.8$~cm$^{-1}$.
There is an isolated weak peak at 3.53~\micron{} (2830 cm$^{-1}$,
2.54$\sigma$), an upper limit at 3.27~\micron{} (3060 cm$^{-1}$, 1.37$\sigma$
at the peak), and an upper limit at 3.47~\micron{} (2880 cm$^{-1}$,
1.43$\sigma$ at peak).

In the mid-IR the prominent features are referred to as the 6.9~\micron{}
complex and the 9.1~\micron{} complex. The 6.9~\micron{} complex peaks from
6.83 -- 7.04 \micron{} (1460-1420~cm$^{-1}$). Throughout, we refer to the
emission between 6.7 and 7.6~\micron{} as the 6.9~\micron{} complex. We define
the complex by these limits, rather than by a profile fit, because the
emission in this range is blended and does not separate into individual
features at the resolution of these data. In the vicinity of the 9.1~\micron{}
complex that peaks at 9.10~\micron{} (1099~cm$^{-1}$), there is a
8.4~\micron{} (1190~cm$^{-1}$) feature, and a 9.76~\micron{} feature
(1025~cm$^{-1}$) on the long wavelength shoulder of the 9.1~\micron{} complex.
There also is an isolated feature at 11.0~\micron{} (909 cm$^{-1}$). The main
objectives of our analysis here are to ascertain the population of aromatic
molecules that can explain the absence of the 3.3~\micron{} feature, a weak
3.37~\micron, and the strong emissions observed between 6.9 and 10~\micron.

Three PAH populations are central to our analysis: large PAHs, which produce
little near-infrared emission \citep{2024ApJ...968..128R}; dehydrogenated PAHs
(DPAHs), which produce neither 3.3~\micron{} nor 3.4~\micron{} emission
because there are no peripheral hydrogens \citep{2015ApJ...799..131M}; and
heavily hydrogenated PAHs (H$_{n}$-PAHs), which produce 3.4~\micron{} emission
and little or no 3.3~\micron{} emission \citep{2013ApJS..205....8S}. Modeling
comet 81P's spectrum with the PAHdb
PythonSuite\footnote{\url{https://github.com/PAHdb/AmesPAHdbPythonSuite}}
\citep{2018ApJS..234...32B, 2014ApJS..211....8B, 2020ApJS..251...22M} and the
PAHdb V4.00 database\footnote{\url{http://www.astrochemistry.org/PAHdb/}}
\citep{2026ApJS..282....7R}, we demonstrate that comet 81P has a combination
of these three PAH subpopulations: H$_{n}$-PAHs, DPAHs, and large PAHs. The
H$_{n}$-PAHs in the best-fit model are very small ($\rm < 20$ carbon atoms),
although small ones ($20 <\rm{C} \leq 50$) were included in the search. The
DPAHs are typically small ($20 <$ C $\leq 50$) yet include C$_{70}$
(Fullerene-C70), and large PAHs have more than 70 carbon atoms (C $> 70$).

In what follows, we describe the modeling approach (Section~\ref{sec:pah:method}) and present a simplified
model with these three subpopulations of PAHs (Section~\ref{sec:results}). We demonstrate that a
three-subpopulation PAH model that includes PAHs with --CH$_{2}$ side groups
(H$_{n}$-PAHs) rather than --CH$_{3}$ side groups (methyl-PAHs or Me-PAHs)
reproduces better the observed emission in the SED of comet 81P. We present
the best-fit model and describe the relative fractions of PAHs as
characterized by the fraction of integrated flux density and by the number
fraction. Compared with the three- and four-subpopulation models, the best-fit
model adds unsubstituted PAHs, i.e., regular PAHs with a single hydrogen at
each peripheral site, and medium PAHs ($50 < \rm{C} \leq 70$). A search for
very small and small N-bearing PAHs is performed in the best-fit model run.
The 14.5~\micron{} complex in 81P's spectrum is compared to candidate
carbonaceous nanograin spectra. The potential astronomical regions from which
these PAHs originated are then discussed (Section~\ref{sec:pah:origins}): large PAHs from the interstellar
medium (ISM), DPAHs from the diffuse interstellar medium (DISM), and
H$_{n}$-PAHs and very small regular PAHs, which are not expected to survive in
either the ISM or the DISM, from the prenatal cloud or our protoplanetary
disk. PAH measurements of Stardust samples by 2-step laser ionization mass
spectrometry and their relation to the JWST remote sensing observations of
PAHs in the coma of comet 81P are summarized.

\subsection{Fitting Procedure and Data Adjustments}
\label{sec:pah:method}

The PAHdb PythonSuite coupled with the PAHdb V4.00 database forward models PAH
emission in response to excitation by sunlight (5770 K). A fluorescent cascade
is calculated for each PAH, and the resultant transitions are convolved with a
15~cm$^{-1}$ FWHM Gaussian. 
The forward model is fitted to the comet 81P infrared spectrum
in flux (Jy) units where the weights per PAH are determined by the PAHdb
PythonSuite's least-squares minimization.\footnote{\url{https://github.com/PAHdb/AmesPAHdbPythonSuite}, Ames PAHdbPythonSuite version 0.5.0.post87.dev0+g25879cb} Neutral and cationic PAHs are
selected from the PAHdb V4.00 database using the search command in the PAHdb
PythonSuite. For PAH models that yield repetition of formulae with multiple
structures the unique identifier or uid are cited for clarity.

One thousand Monte Carlo trials were run for a given model implementation, and
the subset of PAHs contributing 99\% of the flux (Jy) defines the population
of PAHs in the coma. Per PAH and per trial, we first compute the fraction of
integrated flux over NIRSpec, MRS, and combined NIRSpec and MRS wavelengths, and
then evaluate the mean fraction of integrated flux over 1000 trials. The
supposition adopted here is that the same set of PAH subpopulations produces
both the NIR and MIR flux from the comet. 
Because the NIR/MIR flux ratio is small, we first sort the PAHs in descending
order of contribution to the NIRSpec region and keep the subset summing to
99\% of the NIRSpec flux. The remaining PAHs are then sorted by descending
fraction of integrated flux at MRS wavelengths while keeping the subset sum to
95\% of the MRS flux. In succession, these PAHs sum to 99\% of the total
integrated flux across NIRSpec and MRS, defining ``the PAH model'' for the PAH
emission spectrum of the coma of comet 81P. For this subset, the number
fraction per PAH --- the model weight (number of molecules of that type
required to account for the flux) divided by the summed weights per trial ---
is computed, and the mean is evaluated across 1000 trials. Because the number
fractions are computed after the low-contribution PAHs are omitted, they sum
to unity. Typically, $\sim$50--70\% of the PAHs considered across all 1000
trials populate the final model selection. About thirty PAHs appear in fewer
than 100 of the 1000 trials, and none of those is retained. Amongst several
repetitions of the best-fit run, about 100 PAHs were selected across the 1000
trials, of which 57 were retained because their cumulative fractions of
integrated flux summed to 99\% of the total flux. Of the 57 species retained
in the best-fit model, 51 are selected in at least 900 of the 1000 trials and
together contribute 98\% of the modeled NIRSpec flux and 98\% of the MRS flux.
Three are in 500--899 trials and contribute 1.0\% of the NIRSpec flux and
0.6\% of the MRS flux, and three are marginal, selected in 185--379 trials and
contributing 0.3\% and 0.1\%, respectively.

We call a species \textit{marginal} if it is retained in fewer than 400 of the
1000 Monte Carlo trials. Repeating the Monte Carlo best-fit run, each with
different random number draws, reproduces all species that are not marginal and
changes only the marginal ones. The threshold follows from the trial counts
themselves, which fall into two groups separated by an empty interval between
379 and 821. The retained set has changed by a single species, an exchange
between 1-C$_{10}$H$_{9}$N$^{+}$ (uid=476) and C$_{31}$ (uid=2748), which have
the same integrated flux to two digits,
$1.1\times10^{-5}$ of the model, and compete for the same place at the 99\%
cumulative boundary. The marginal species are included in the sums over
spectral decomposition groups but are not distinguished separately in the
figures.

The best-fit model is the one with the lowest Akaike Information Criterion
\citep[AIC,][]{doi:10.1177/0049124104268644, 2023PSJ.....4..242H} value. A
variety of models were run by modifying the search criterion. AIC permits
comparison of the same model fitted to different renditions of the data, as
well as different models fitted to the same data. All data points with
reliable flux calibration are used, 3.2--27~\micron{} (explicitly,
3.2--3.6~\micron{} and 5.5--27~\micron{}), with the adjustments described
below. Using all data points carries more information than a subset, tested by
fitting the 3.2---8.6~\micron, 3.2---12.5~\micron, and 3.2---27~\micron{}
regions. Inclusion of the 12.5---16.5~\micron{} region always worsened the AIC
metric, so the instrumental uncertainties were multiplied by 40$\times$ to
underweight those points --- in essence, the PAHdb V4.00 database lacks the
ingredients to fit that portion of the spectrum.

Flux values were adjusted to be commensurate with the limitations of the best
available computational models of PAH species emission
\citep{2026ApJS..282....7R}. Negative flux values shortward of 6.312~\micron{}
were set to zero, including the negative NIRSpec fluxes set by the NASA
PSG-determination of the baseline under the molecular lines. We were advised
that the PAHdb fitting algorithms behave better in the regions of low flux
when negative fluxes are zeroed (C. Boersma 2025, private communication).
Negative values longward of 10~\micron{} were retained. Both treatments are
applied identically to all models compared here and cancel in the $\Delta$AIC
shown (Appendix~\ref{sec:appendix1-aic}). The NIRSpec fluxes and instrumental
uncertainties (modeled over 3.2--3.6~\micron) were multiplied by a factor of
two because current harmonic theoretical calculations in the PAHdb V4.00 database 
of PAH flux in the NIR are overpredicted ($\sim$34\%) \citep{2023FaDi..245..380L},
The PAHdb Anharmonic V1.00
database\footnote{\url{https://www.astrochemistry.org/pahdb/anharmonic/}} does
not include enough PAHs to fit the comet's spectrum. An attempt to fit only
the NIRSpec data yielded only one PAH and it did not account for the NIRSpec
features. The current PAH harmonic computational results \citep{2026ApJS..282....7R}
do not produce combination bands relevant to the 5--6~\micron{} region,
specifically those potentially causing the 5.25~\micron{} and 5.75~\micron{}
features from PAHs in the ISM \citep{2025A&A...698A..86C}, hence the
5--6~\micron{} flux was set to zero. Comet 81P appears to lack a
5.25~\micron{} feature but shows some 5.85~\micron{} emission. Ultimately we
find that the best-fit model predicts 5.85~\micron{} emission consistent with
the observed flux, even though this flux is zeroed in the data fitted by the
model. The final data modifications necessary for good model convergence are:
2$\times$flux and 2$\times\sigma$ 3.2--3.6~\micron, 0$\times$flux
5--6~\micron, 40$\times\sigma$ 12.5--16.5~\micron.

A range of PAHs was explored in our modeling. PAH sizes are delineated by C
(or N$_{\rm{C}}$), the number of carbon atoms \citep[cf. Table
2,][]{2024ApJ...968..128R}: very small (C$<$20), small ($20 < \rm{C} \leq
50$), medium ($50 < \rm{C} \leq 70$), and large (C $> 70$). Model parameters
and PAH geometries from the model runs were translated from output tables into
astropy tables, and predicted fluxes (Jy) were translated to flux densities
(W~cm$^{-2}$~$\micron ^{-1}$), enabling analysis of different PAH parameters. These models are presented in
Figure~\ref{fig:figure-2}(a)-(f) and Figure~\ref{fig:figure-4}(a)-(g) and
described in Section~\ref{sec:results}.



\section{Results}
\label{sec:results}

\subsection{Three- and Four-Subpopulation PAH Models}
\label{sec:pah:3pop}

A model using the three PAH subpopulations -- moderately- to
heavily-hydrogenated H$_{n}$-PAHs, defined below as those with the fraction of
peripheral H bonding sites carrying CH$_{2}$ side groups $\geq 0.35$, DPAHs,
large PAHs -- is fit to comet 81P's observed spectrum and is shown in
Figure~\ref{fig:figure-2}(a). Each of the three subpopulations dominates a
distinct wavelength region. The 3.37~\micron{} feature is fitted by very small
neutrals, very small cations, and small cations, all with CH$_{2}$ side
groups, the so-called H$_{n}$-PAHs. The 6.9~\micron{} region is dominated by
dehydrogenated PAHs (H=0, DPAHs), mostly small in size. The 8.5--10~\micron{}
region, including the 9.1~\micron{} complex, is dominated by large neutrals
and cations. Overall, the height of the 3.37~\micron{} feature is
underpredicted, the isolated weak 3.53~\micron{} feature is well-reproduced,
and the predicted flux at 3.27~\micron{} lies within the 1$\sigma$
uncertainties of the observations. The model also passes through the observed
flux at 3.47~\micron{} (2880 cm$^{-1}$), where the signal-to-noise is too low,
1.43$\sigma$ at the peak, for the emission to be seen by eye as a feature.
This model comprises 59 PAHs: 10 H$_{n}$-PAHs with mean N$_{\rm{C}}$=12.9, 23
DPAHs with mean N$_{\rm{C}}$= 34.7, and 26 large PAHs with mean
N$_{\rm{C}}$=114.0.

We examine the contribution of each subpopulation by fitting all three to
NIRSpec and MRS. We also fit just the particular subpopulation of very small
and small molecules to the NIRSpec data by increasing the MRS uncertainties by
15$\times$ to illuminate their potential contributions to the NIR spectrum of
comet 81P. The model with only H$_{n}$-PAHs is shown in
Figure~\ref{fig:figure-2}(b). In addition, a modified three-subpopulation
model -- very small Me-PAHs, DPAHs, large PAHs -- is fitted and presented in
Figure~\ref{fig:figure-2}(d) and its companion model of only Me-PAHs in
Figure~\ref{fig:figure-2}(e). In Figure~\ref{fig:figure-2}(f), the
four-subpopulation model -- H$_{n}$-PAHs (peripheral CH$_{2}$ fraction $\geq
0.35$, N$_{\rm{C}}$ $\leq$ 50), Me-PAHs, DPAHs, large PAHs -- is presented.
The models show the interplay between the PAH subpopulations and set the
framework for the best-fit model.

For H$_{n}$-PAHs, increasing the degree of hydrogenation shifts the emission
from near 3.3~\micron{} to near 3.4~\micron{} and reduces the
near-3.4~\micron{} emission strength because the energy is distributed across
a greater number of peripheral bonds \citep{2013ApJS..205....8S}. The PAHdb
V4.00 database contains 9487 PAHs in total, with 4476 neutrals and 5011
cations. Of these, 259 have CH$_{2}$ side groups (95 neutrals and 164
cations), of which 170 are very small and small PAHs with N$_{\rm{C}}$ $\leq$
50. To focus on the moderately- to heavily-hydrogenated H$_{n}$-PAHs, we
define the fraction of peripheral bonding sites carrying a CH$_{2}$ group as
(N$_{\rm{CH2}}$/2)/(N$_{\rm{H}}$ $-$ N$_{\rm{CH2}}$/2), where N$_{\rm{H}}$ $-$
N$_{\rm{CH2}}$/2 is the total number of peripheral bonding sites and each
CH$_{2}$ side group's bonding site has N$_{\rm{CH2}}$=2 (H atoms). Applying a
threshold of 0.35 yields 31 PAHs spanning 10 to 96 C atoms; of these, 26 have
N$_{\rm{C}}$ $\leq$ 50, including 10 neutrals and 16 cations. These 26 PAHs
form the H$_{n}$-PAH subpopulation used in this demonstration of the
three-subpopulation model. Figure~\ref{fig:figure-2}(c) shows the distribution
of this fraction as a function of carbon atom number for the full database and
for this subset.

Similarly, the subset with H=0 contains 810 DPAHs, of which 799 have
N$_{\rm{C}}$ $\leq$ 50 and 759 have 25 $\leq$ N$_{\rm{C}} \leq$ 35. The
database also contains 3004 large PAHs (N$_{\rm{C}} > 70$), comprising 1011
neutrals and 1903 cations.

Even though the database contains far fewer H$_{n}$-PAHs than DPAHs or large
PAHs, their role in fitting the 3.37~\micron{} feature is clearly
demonstrated. The very small and small H$_{n}$-PAHs span a broad range in the
fraction of peripheral CH$_{2}$ bonds {\it versus} N$_{\rm{C}}$ and densely
sample N$_{\rm{C}}$ (Figure~\ref{fig:figure-2}(c)).
In contrast, medium and large H$_{n}$-PAHs sparsely sample both the fraction
of peripheral CH$_{2}$ bonds and the range of N$_{\rm{C}}$.

\subsubsection{H$_{n}$-PAHs and the 3.37~\micron{} feature}
H$_{n}$-PAH cations dominate because the NIR/MIR flux ratio for cations is
lower than for neutrals. This allows them to contribute to both wavelength
regions simultaneously. For neutrals, the NIR/MIR ratio is high enough that
their contribution to the 3.37~\micron{} feature is accompanied by a
negligible contribution to the MIR. In Figure~\ref{fig:figure-2}(a), very
small neutrals appear in the 3.37~\micron{} feature with no visible
contribution in the MIR. Very small cations, by contrast, contribute to the
3.37~\micron{} feature while contributing minimally to the 6.9~\micron{}
feature.

\subsubsection{H$_{n}$-PAHs fitted only to the NIRSpec spectrum}
A model fit using only H$_{n}$-PAHs (peripheral CH$_{2}$ fraction $\geq 0.35$,
N$_{\rm{C}}$ $\leq$ 50) and constrained primarily by NIRSpec data (employing
15$\times \sigma$ for MRS outside of the 12.5--16.5~\micron{} region) is shown
in Figure~\ref{fig:figure-2}(b) to show the potential role of H$_{n}$-PAHs in
the best-fit. Relaxing the MRS constraint selects a different set of nine
H$_{n}$-PAHs (Figure~\ref{fig:figure-2}(b)). Two are small cations that the
three-subpopulation model does not retain, C$_{24}$H$_{23}^{+}$ (midnight
blue) and the marginal C$_{24}$H$_{24}^{+2}$. Two are the very small neutrals
C$_{10}$H$_{18}$ and C$_{19}$H$_{18}$, retained in both models. Five are very
small cations, of which C$_{10}$H$_{18}^{+2}$, C$_{16}$H$_{16}^{+}$ and
C$_{16}$H$_{16}^{+2}$ are shared with the three-subpopulation model and
C$_{19}$H$_{16}^{+2}$ and C$_{14}$H$_{16}^{+2}$ are new to this fit. The very
small cation C$_{10}$H$_{12}^{+}$ of the three-subpopulation model is not
retained here. The small cation H$_{n}$-PAHs supply 0.376 of the NIRSpec flux
and most of the MRS flux (0.618). This one-subpopulation model provides the
best match to the 3.37~\micron{} feature and to the NIRSpec spectrum of the
models presented here (over the 602 NIRSpec points, $\chi^{2}$ is 771, against
1058 for the three-subpopulation model and 1258 for the best-fit model), at
the cost of the MIR, which is significantly underweighted in the fit. The
major contributor, C$_{24}$H$_{23}^{+}$ (uid=584), produces half the flux of
the 3.37~\micron{} feature and peaks at its center wavelength. It also
slightly overpredicts at 3.47~\micron{} at about the 1.5$\sigma$ level and
places its MIR feature at 6.66~\micron{}, where it reaches the lower 1$\sigma$
envelope. The very small neutral H$_{n}$-PAHs supply 0.223 of the NIRSpec
flux, down from 0.427 in the three-subpopulation model. This fit shows the
potential contribution H$_{n}$-PAHs can make to the NIRSpec spectrum when they
are not constrained by the MIR. The DPAHs and large PAHs of
Figure~\ref{fig:figure-2}(a) account for the MIR well and preclude a
6.66~\micron{} feature. As shown in the three-subpopulation model, the large
neutral PAHs that fit the MIR add emission at 3.27~\micron{}, where only an
upper limit is measured (Section~\ref{sec:pah:intro}). The same set of PAHs
must account for both regions, which is the supposition adopted in
Section~\ref{sec:pah:method}. This one-subpopulation fit produces no
significant emission at 3.2--3.3~\micron{}, which supports the selection of
H$_{n}$-PAHs with peripheral CH$_{2}$ fractions above 0.35
(Figure~\ref{fig:figure-2}(c)). The N$_{\rm{C}}$ $\leq$ 50 limit is supported
by the five large H$_{n}$-PAHs with peripheral CH$_{2}$ fraction greater than
0.35 (Figure~\ref{fig:figure-2}({c})), which, although not plotted here, also
do not contribute to the 3.37~\micron{} feature. Their average spectrum peaks
near 7.2~\micron{}, with only a very weak, broad feature near 3.52~\micron{}
and a narrow feature near 3.38~\micron{}, neither of which aligns with the
observed spectral features.

\subsubsection{DPAHs and the 6.9~\micron{} complex}
DPAHs with H=0 are in the three subpopulation model, the modified three
subpopulation model, and the four-subpopulation models shown in
Figure~\ref{fig:figure-2}(a, {d}, {f}). DPAHs produce no flux in the
3.4~\micron{} region because there are no peripheral hydrogens. DPAHs
reproduce the height and width of the 6.9~\micron{} feature but cannot
simultaneously account for the 9.1~\micron{} feature or the 11.0~\micron{}
feature that is commonly attributed to single peripheral hydrogens. DPAHs also
contribute flux near 5.85~\micron{} that is consistent with the observed flux,
although we exclude this region from the fitting process due to theoretical
limitations (i.e., specifically the absence of combination bands as described
in Section~\ref{sec:pah:method}). The DPAHs in this model comprise only a
small subset of the nearly 800 small dehydrogenated PAHs in the database.
DPAHs have been noted for their potential contributions to the 6--10~\micron{}
wavelength region \citep{2015ApJ...799..131M}.

A model in which all DPAHs are excluded from the search is presented in
Appendix~\ref{sec:appendix1-nodpah} and its model spectrum in
Figure~\ref{fig:appendix-1}. Its $\Delta $AIC demonstrates that the no-DPAHs
model fits significantly worse than the other models presented here.

\subsubsection{Large PAHs and the 7.5--10~\micron{} emission}
Large PAHs are in the three subpopulation model, the modified three
subpopulation model, and the four-subpopulation models shown in
Figure~\ref{fig:figure-2}(a, {d}, {f}). Large PAHs contribute to the
6.9~\micron{} feature and account for the 7.5--10~\micron{} emission through a
combination of neutrals and cations. The large neutrals are centered near
9.0~\micron, with cations contributing on either side. Large cations account
for the long-wavelength shoulder of the 9.1~\micron{} complex and the centroid
of the 11.0~\micron{} feature, while large neutrals contribute to the
9.76~\micron{} peak on the shoulder of the 9.1~\micron{} complex and to the
8.4~\micron{} shoulder. Both large neutrals and cations contribute to the
6.9~\micron{} complex but fall short of the observed peak at
6.83--7.04~\micron{} while overpredicting at 6.2--6.5~\micron{}. In both the
three-subpopulation and modified three-subpopulation models, large neutral
PAHs generate a 3.27~\micron{} feature at the level of the observed flux, a
1.37$\sigma$ upper limit. Minor contributions from the other subpopulations
bring the model slightly above the measured flux there but within its
1$\sigma$ uncertainties. These large neutrals do not contribute to the
3.37~\micron{} feature.

\subsubsection{Interplay between the subpopulations}
\begin{figure*} [t!]
\figurenum{2}
\begin{center}
\includegraphics[width=0.980\textwidth]{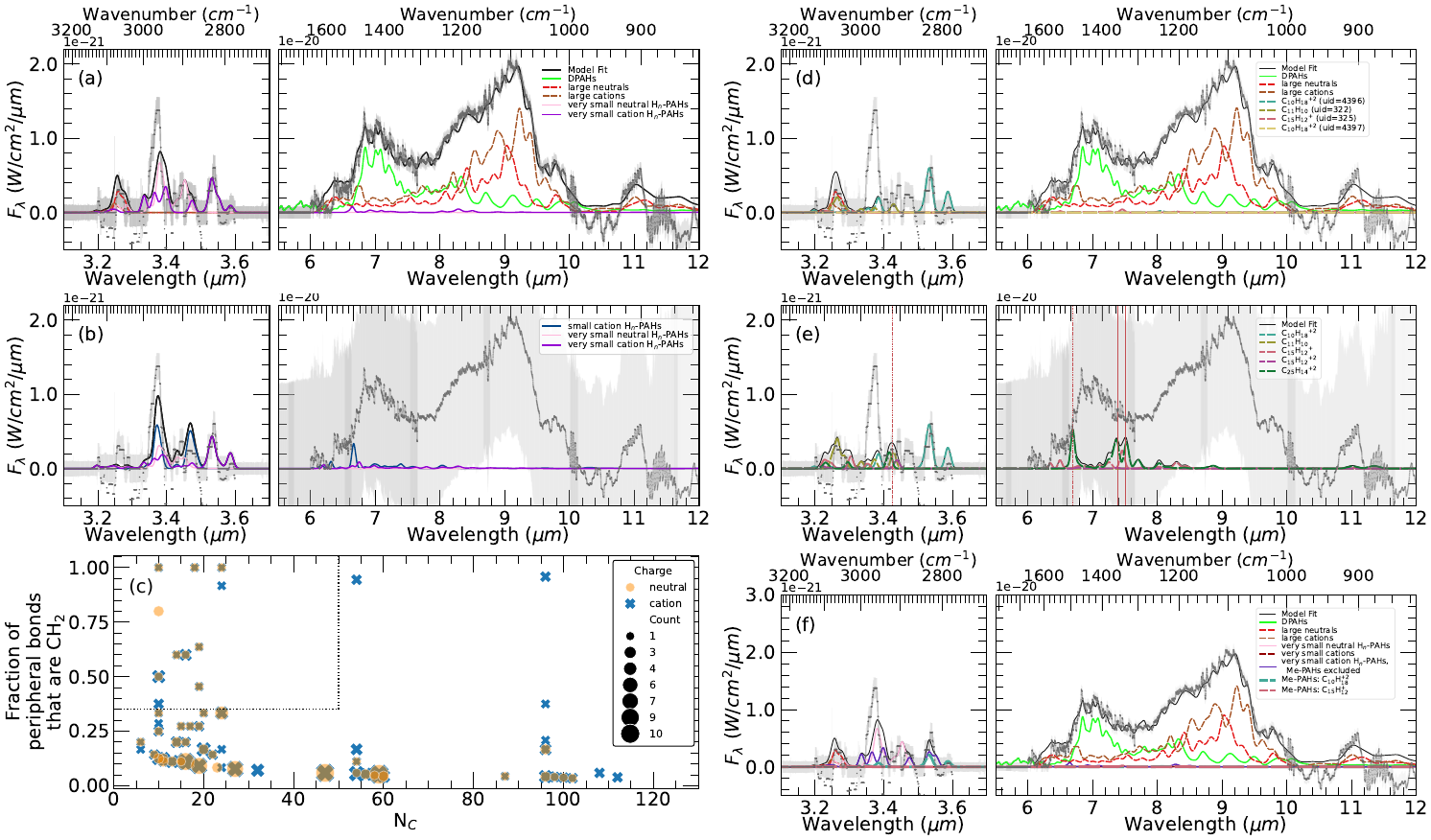}
\caption{Two three-subpopulation PAH models and a four-subpopulation PAH model
for comet 81P as well as their individual very small and small subpopulations.
(a)~A three-subpopulation PAH model comprising DPAHs, large PAHs (N$_{\rm{C}}$
$> 70$), and very small and small (N$_{\rm{C}}$ $\leq 50$) H$_{n}$-PAHs with
the fraction of peripheral H bonding sites carrying CH$_{2}$ side groups $\geq
0.35$. Species are DPAHs (lime), very small neutral and cationic H$_{n}$-PAHs
(pink, dark-violet), and neutral and cationic large PAHs (dashed sienna,
dashed crimson). (b)~A model fit using only very small and small H$_{n}$-PAHs,
primarily constrained by NIRSpec data. (c)~Fraction of peripheral bonds
occupied by CH$_{2}$ side groups versus number of carbon atoms (N$_{\rm{C}}$)
for the neutral and cationic H$_{n}$-PAHs in the PAHdb V4.00 database; symbol
size represents the count (e.g., counting isomers), and the dotted box in the
upper left corner encloses the H$_{n}$-PAH subset used in the subpopulation
models (a)-(b).
(d)~A modified three-subpopulation PAH model comprising DPAHs, large PAHs
(N$_{\rm{C}}$ $> 70$), and Me-PAHs. The four Me-PAHs retained in this model
fit are C$_{10}$H$_{18}^{+2}$ (uid=4396, medium turquoise), a second
C$_{10}$H$_{18}^{+2}$ isomer (uid=4397, khaki), C$_{11}$H$_{10}$ (olive), and
C$_{15}$H$_{12}^{+}$ (salmon). Uids are given because uid=4396 and uid=4397
share both formula and charge. (e)~The subpopulation model using only Me-PAHs
and primarily constrained by NIRSpec data. The five Me-PAHs are, in order of
N$_{\rm{C}}$ and then charge, C$_{10}$H$_{18}^{+2}$ (uid=4396, medium
turquoise), C$_{11}$H$_{10}$ (olive), C$_{15}$H$_{12}^{+}$ (salmon),
C$_{15}$H$_{12}^{+2}$ (redviolet), and C$_{25}$H$_{14}^{+2}$ (green),
illustrating their potential roles. (f)~A four-population model that combines
the three-subpopulation and modified-three-subpopulation models. This model
has DPAHs, large PAHs, very small H$_{n}$-PAHs, and two very small Me-PAH
cations. C$_{10}$H$_{18}^{+2}$ (uid=4396, medium turquoise) contributes only
very weakly to the 3.37~\micron{} feature and contributes to the
3.53~\micron{} feature. C$_{15}$H$_{12}^{+}$ (uid=325, salmon) is retained in
most of the MC trials and contributes 0.6\% of the modeled NIRSpec flux. It
produces no distinguishable feature at this scale. All very small cations
retained here are H$_{n}$-PAHs or Me-PAHs. No unsubstituted very small cation
appears in this model. }
\label{fig:figure-2}
\end{center}
\end{figure*}

The subpopulation of H$_{n}$-PAHs contributes less to the MIR because of their
small size but their role in the three-subpopulation model still depends on
the DPAHs and large PAHs, as shown by the comparison of
Figure~\ref{fig:figure-2}({a})-({b}). The two subpopulations of DPAHs and
large PAHs act co-dependently in the MIR and yet the two together cannot
produce the 3.37~\micron{} feature. DPAHs reproduce the peak flux of the
6.9~\micron{} complex, but they do not account for the full observed width and
contrast of that feature. The Me-PAH subpopulation, the modified
three-subpopulation model, and the four-subpopulation model that combines them
are examined next in Section~\ref{sec:pah:mepah}, where the three- and
four-subpopulation models are also compared quantitatively.



\subsection{Methyl-PAHs}
\label{sec:pah:mepah}

\subsubsection{Models using very small and small Methyl-PAHs}
\label{sec:pah:mepah:models}
Methyl-PAHs (Me-PAHs) may be relevant to comet 81P's infrared spectrum because
PAHs with $-$CH$_3$ groups have been identified in the mass spectra of
Stardust samples
\citep[Section~\ref{sec:pah:stardust},][]{2010M&PS...45..701C}. Following the
approach of Section~\ref{sec:pah:3pop}, we apply a modified
three-subpopulation PAH model using very small and small Me-PAHs (C $\leq $50)
bearing --CH$_{3}$ groups to test whether they can account for some of the NIR
emission in comet 81P. The PAHdb V4.00 database contains only singly
methylated Me-PAHs. The Me-PAHs in the relevant size range are
C$_{11}$H$_{10}$ (methylnaphthalene), C$_{15}$H$_{12}$ (methylanthracene, MA),
and C$_{25}$H$_{14}$ (methylcoronene), with 2, 3, and 1 peripheral --CH$_{3}$
positions, respectively. Combined with three charge states (neutral, +1, +2),
this yields 18 Me-PAH variants.

The database also includes C$_{10}$H$_{18}^{+2}$ (uid=4396, uid=4397) that
arise in the search of the PAHdb V4.00 database for CH$_{2} >0$ (H$_{n}$-PAHs)
and for CH$_{3} >0$ (Me-PAHs). C$_{10}$H$_{18}$ contains several CH$_2$ groups
and one CH$_3$ group, in a cyclohexane ring with four peripheral CH$_2$ groups
and an attached four-carbon chain, --CH$_2$--CH=CH--CH$_3$, with an internal
double bond. The side chain on C$_{10}$H$_{18}$ puts it among the alkylated
saturated rings rather than just among the Me-PAHs. This class of molecules
belongs to the cycloalkanes and alkyl-substituted aromatics that are important
in astrochemistry because such structures can be precursors to prebiotic
molecules \citep{2024ESC.....8.2380F}.

\subsubsection{The modified three-subpopulation model}
Figure~\ref{fig:figure-2}(d) shows the modified 3-population model with
Me-PAHs substituted for H$_{n}$-PAHs (i.e., compare to
Figure~\ref{fig:figure-2}(a)). As with the H$_{n}$-PAH version, the DPAHs and
large PAHs reproduce the MIR features reasonably well. However, the Me-PAHs do
not adequately fill in the 3.37~\micron{} feature. Four Me-PAHs are retained
in the fit, all very small, one neutral and three cations, together carrying
76\% of the modeled NIRSpec flux. Large PAH neutrals and cations contribute
the remaining 24\% of the modeled NIRSpec flux. In order of N$_{\rm{C}}$ and
then charge, the four Me-PAHs are C$_{10}$H$_{18}^{+2}$ (uid=4396), a second
C$_{10}$H$_{18}^{+2}$ isomer (uid=4397), C$_{11}$H$_{10}$, and
C$_{15}$H$_{12}^{+}$. Uids are given because uid=4396 and uid=4397 share both
formula and charge. C$_{10}$H$_{18}^{+2}$ (uid=4396,
Section~\ref{sec:pah:mepah:models}) supplies 53\% of the NIRSpec flux. This
cation generates the 3.53~\micron{} feature at the observed flux level and a
second feature at 3.59~\micron{} that rises through the upper 1$\sigma$
envelope, and contributes only weakly to the 3.37~\micron{} feature. Since
C$_{10}$H$_{18}^{+2}$ carries --CH$_{2}$ groups as well, this species is also
categorized as an H$_{n}$-PAH. The neutral C$_{11}$H$_{10}$ contributes 19\%
and is the only retained Me-PAH with an aromatic C--H stretch. It contributes
at 3.27~\micron{} together with the large neutral (dashed sienna) and large
cationic (dashed crimson) PAHs, the only wavelength at which any non-Me-PAH
contributes in the NIRSpec range, and again near 3.425~\micron{} at the lower
1$\sigma$ envelope. C$_{15}$H$_{12}^{+}$ contributes 2.5\% of the NIRSpec
flux. The second isomer, uid=4397, contributes 0.8\% and produces no
distinguishable feature at this scale. The model fluxes at these four
wavelengths 3.27, 3.425, 3.53, and 3.59~\micron{} lie at or near the 1$\sigma$
envelopes, while the strongest NIR feature at 3.37~\micron{} remains
unaccounted for.

\subsubsection{The one-subpopulation model with Me-PAHs}
Figure~\ref{fig:figure-2}(e) shows a one-subpopulation model fit using only
Me-PAHs, with C$_{10}$H$_{18}^{+2}$ (uid=4396) (medium turquoise,
(Section~\ref{sec:pah:mepah:models})), C$_{11}$H$_{10}$ (olive),
C$_{15}$H$_{12}^{+}$ (salmon), C$_{15}$H$_{12}^{+2}$ (redviolet), and
C$_{25}$H$_{14}^{+2}$ (green). Primarily cations are fitted rather than
neutrals because cations have a lower NIR/MIR flux ratio. Generally for
neutrals, the NIR features are too strong relative to the MIR. The two charge
states of C$_{15}$H$_{12}$ differ by a factor of five in NIR/MIR flux ratio,
0.38 for C$_{15}$H$_{12}^{+}$ and 1.91 for C$_{15}$H$_{12}^{+2}$. The singly
ionized state contributes mainly to the MIR and the doubly ionized state
mainly to the NIR.

The Me-PAH model significantly underpredicts the flux at 3.37~\micron{}:
together C$_{10}$H$_{18}^{+2}$, C$_{11}$H$_{10}$, C$_{15}$H$_{12}^{+}$,
C$_{15}$H$_{12}^{+2}$, and C$_{25}$H$_{14}^{+2}$ are insufficient to account
for the 3.37~\micron{} feature. The neutral C$_{11}$H$_{10}$ contributes the
most at 3.27~\micron{}, where side features from C$_{15}$H$_{12}^{+}$ and
C$_{25}$H$_{14}^{+2}$ make the model pass just above the 1$\sigma$ envelope on
either side. C$_{15}$H$_{12}^{+}$ also passes through the data points near
3.425~\micron{} and contributes a double peak at 7.38 and 7.50~\micron{}.
These MIR features fall where the comet has flux but fail to fill it in, apart
from the distinct C$_{25}$H$_{14}^{+2}$ band at 6.69~\micron{}, whose peak
reaches the observed level. The 3.53~\micron{} feature is contributed solely
by C$_{10}$H$_{18}^{+2}$ (uid=4396).

\subsubsection{The four-subpopulation model}
\label{sec:pah:mepah:4pop}
Figure~\ref{fig:figure-2}(f) shows a four-subpopulation model combining very
small and small H$_{n}$-PAHs (peripheral CH$_{2}$ fraction $\geq 0.35$,
N$_{\rm{C}}$ $\leq$ 50) with Me-PAHs, DPAHs, and large PAHs. Very small
H$_{n}$-PAHs, neutral and cationic, contribute to the 3.37~\micron{} feature.
Two Me-PAHs are retained in the four-subpopulation model.
C$_{10}$H$_{18}^{+2}$ (uid=4396, Section~\ref{sec:pah:mepah:models}) is the
same species retained in the modified three-subpopulation model and in the
best-fit model. It contributes very weakly to the 3.37~\micron{} feature but
about half the model flux to the 3.53~\micron{} (2830~cm$^{-1}$) feature. The
H$_{n}$-PAHs (Me-PAHs excluded) contribute the other half. The second,
C$_{15}$H$_{12}^{+}$ (uid=325), is retained in most of the MC trials and
produces no distinguishable feature. Neutral super-hydrogenated
1,2,3,4,5,6,7,8-octahydronaphthalene (C$_{10}$H$_{16}$, uid=333) has a feature
at 2838.6 cm$^{-1}$ that is at a slightly shorter wavelength (3.523~\micron{})
\citep{2013ApJS..205....8S}. However, very small neutral PAHs have too strong
NIR/MIR flux ratios to be major players in the PAH model for comet 81P
(Section~\ref{sec:pah:3pop}). The very small cations in this model are
predominantly H$_{n}$-PAH cations, with C$_{16}$H$_{16}^{+}$ (uid=362) the
largest contributor, together with the Me-PAH cations C$_{10}$H$_{18}^{+2}$
(uid=4396) and C$_{15}$H$_{12}^{+}$ (uid=325).

\subsubsection{Model limitations of Methyl-PAHs}
\begin{figure}[h]
\figurenum{3}
\begin{center}
\includegraphics[trim=0.07cm 0.0cm 0.07cm 0.0cm, clip, width=0.45\textwidth]{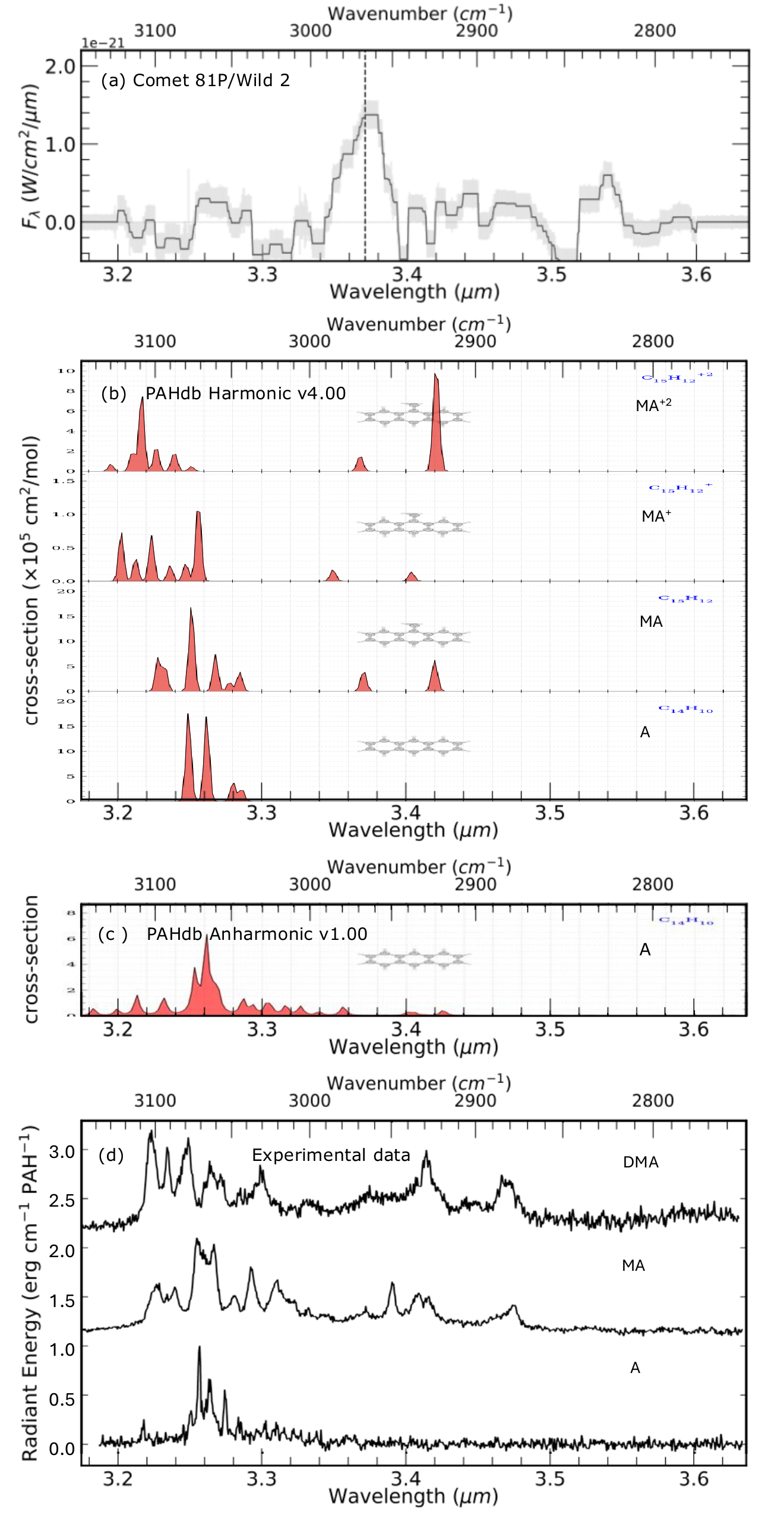}
\caption{Comparison of comet 81P data with PAHdb V4.00 theoretical spectra and
laboratory spectra of anthracene (A), 9-methylanthracene (MA), and
9,10-dimethylanthracene (DMA). (a)~Comet 81P data over 3.2--3.6~\micron{} (cf.
Fig.~\ref{fig:figure-1}). (b)~PAHdb V4.00 theoretical spectra (harmonic
calculations) (bottom to top): anthracene C$_{14}$H$_{10}$, then the three
charge states of 9-methylanthracene C$_{15}$H$_{12}$, C$_{15}$H$_{12}^{+}$,
and C$_{15}$H$_{12}^{+2}$. (c)~PAHdb Anharmonic V1.00 theoretical spectra of
anthracene C$_{14}$H$_{10}$. (d)~Laboratory spectra of A (C$_{14}$H$_{10}$),
MA (C$_{15}$H$_{12}$), and DMA (C$_{16}$H$_{14}$), adapted from Figure~6 of
\citet{2018A&A...610A..65M}.}
\label{fig:figure-3}
\end{center}
\end{figure}

However, two factors may limit the Me-PAH modeling: the sparse coverage of
Me-PAHs in the PAHdb V4.00 database, and the scarcity of anharmonic
theoretical calculations. Regarding the PAHdb V4.00 database coverage, only
the straight 3-ring PAH methylanthracene is included. The bent isomer
methylphenanthrene is absent, as are 4-ring Me-PAHs such as methylpyrene. The
recently released PAHdb Anharmonic V1.00 database has anthracene
(Figure~\ref{fig:figure-3}(c)) and C$_9$H$_8$.

Anharmonic calculations are especially important for --CH$_{3}$ groups, which
can twist as well as bend and stretch. \citet{2018A&A...610A..65M} studied
these effects experimentally and theoretically for anthracene (A),
9-methylanthracene (MA, C$_{15}$H$_{12}$ with a --CH$_{3}$ group at the center
of the straight 3-ring structure), and 9,10-dimethylanthracene (DMA,
C$_{16}$H$_{14}$ with --CH$_{3}$ groups on both central positions). For MA and
DMA, a broad absorption plateau spanning 3.215--3.484~\micron{} (3110--2870
cm$^{-1}$) arises from the high density of accessible vibrational states and
the ``free rotor'' character of the --CH$_{3}$ groups
\citep{2018A&A...610A..65M}; additional bands appear through coupling of the
aromatic C--H stretches with overtone and combination bands.
Figure~\ref{fig:figure-3}(d) shows the experimental spectra of A, MA, and DMA
\citep[bottom three spectra reproduced from Figure~6
of][]{2018A&A...610A..65M}.

Figure~\ref{fig:figure-3}(c) shows the PAHdb Anharmonic V1.00 spectrum of A,
and Figure~\ref{fig:figure-3}(b) the PAHdb V4.00 (harmonic) theoretical
spectra for A, MA, MA$^{+}$ and MA$^{+2}$, each convolved with a 4
cm$^{-1}$-FWHM Gaussian. For the strongest features for A, the anharmonic
calculation is more similar to the experimental data than the harmonic
calculation, being a single broader feature. The harmonic theoretical spectrum
of MA lacks the broad plateau seen in the experimental data, and the peaks
near 3.4~\micron{} are shifted relative to the measured peaks and the C--H
stretches are similarly displaced. For DMA, the experimental data show two
distinct peaks: one near 3.42~\micron{} and a longer-wavelength peak.
\footnote{The asymmetric alkyl CH-stretch vibrations of methylated PAHs (e.g.,
9-methylanthracene and 9,10-dimethylanthracene) are observed in the range of
3.409--3.414~\micron{} (2929--2933 cm$^{-1}$). However, these bands are
relatively insensitive to structural details, making them consistent
contributors to the 3.4~\micron{} band. -- \citet{2018A&A...610A..65M}.}

Figure~\ref{fig:figure-3}(b) shows the harmonic theoretical spectra of A, MA,
MA$^{+}$ and MA$^{+2}$. The MA$^{+2}$ 3.4~\micron{} feature is stronger than
in the other two charge states, which emphasizes the cation contribution to
this region of the spectrum. For comparison, Figure~\ref{fig:figure-3}(a)
shows the 3.2--3.6~\micron{} 81P spectrum (Figure~\ref{fig:figure-1}).

Anharmonic theoretical calculations for A, MA, and DMA
\citep{2018A&A...610A..65M}, covering the limited wavelength range shown in
Figure~\ref{fig:figure-3}(d), give significantly better agreement with their
experimental data. However, these calculations do not yet extend to the
wavelength range of interest (minimally, 3--12~\micron), nor are they
available within the PAHdb databases, where a full temperature cascade
emission model under sunlight (5770 K or 3.5 eV) could be computed to better
assess the contribution of Me-PAHs to the IR spectrum of comet 81P's coma.
Despite these limitations, the available experimental data suggest what would
be needed for Me-PAHs to account for the 3.37~\micron{} feature in comet 81P
(Figure~\ref{fig:figure-3}(a)). This would require the presence of either (i)
multiple methylated PAHs since they have resonances near 3.4~\micron{} or (ii)
heavily hydrogenated Me-PAHs, in which case the aromatic C--H features at
3.2--3.35~\micron{} from peripheral hydrogens would be suppressed. The
experimental data of A, MA, and DMA show prominent features at
3.2--3.35~\micron{} that are not present at comparable contrast in comet 81P.
Arguments for considering the presence of more heavily methylated species in
comets come from mass spectrometry of Stardust samples, which identified
Me-PAHs bearing one to three --CH$_{3}$ groups, designated for example as
C$_{3}$-Benzene, C$_{3}$-Naphthalene, and C$_{3}$-Pyrene \citep[cf. Table
2,][]{2010M&PS...45..701C}.

In summary, H$_{n}$-PAHs remain the most likely carriers of the 3.37~\micron{}
feature in comet 81P given the currently available theory and analysis
tool-sets. However, the presence of Me-PAHs cannot be ruled out. A more
definitive assessment will require anharmonic theoretical spectra covering a
broader range of Me-PAH sizes and --CH$_{3}$ peripheral bonding
configurations, including bent isomers like methylphenanthrene.

Adding Me-PAHs to the three-subpopulation model slightly improves the quality
of the fit, lowering the AIC by an amount only a few times greater than the
fluctuation in AIC among repeated best-fit runs
(Table~\ref{tab:model-comparison}, Appendix~\ref{sec:appendix1-aic}). Both the
three-subpopulation model (Figure 2(a)) and the four-subpopulation model
(Figure 2(f)) fit NIRSpec better and MRS worse than the best-fit model
(Section~\ref{sec:pah:bestfit}).

\subsection{Best-fit PAH Model}
\label{sec:pah:bestfit}

\begin{figure*}[h]
\figurenum{4}
\begin{center}
\includegraphics[trim=0.07cm 0.07cm 0.07cm 0.15cm, clip, width=0.775\textwidth]{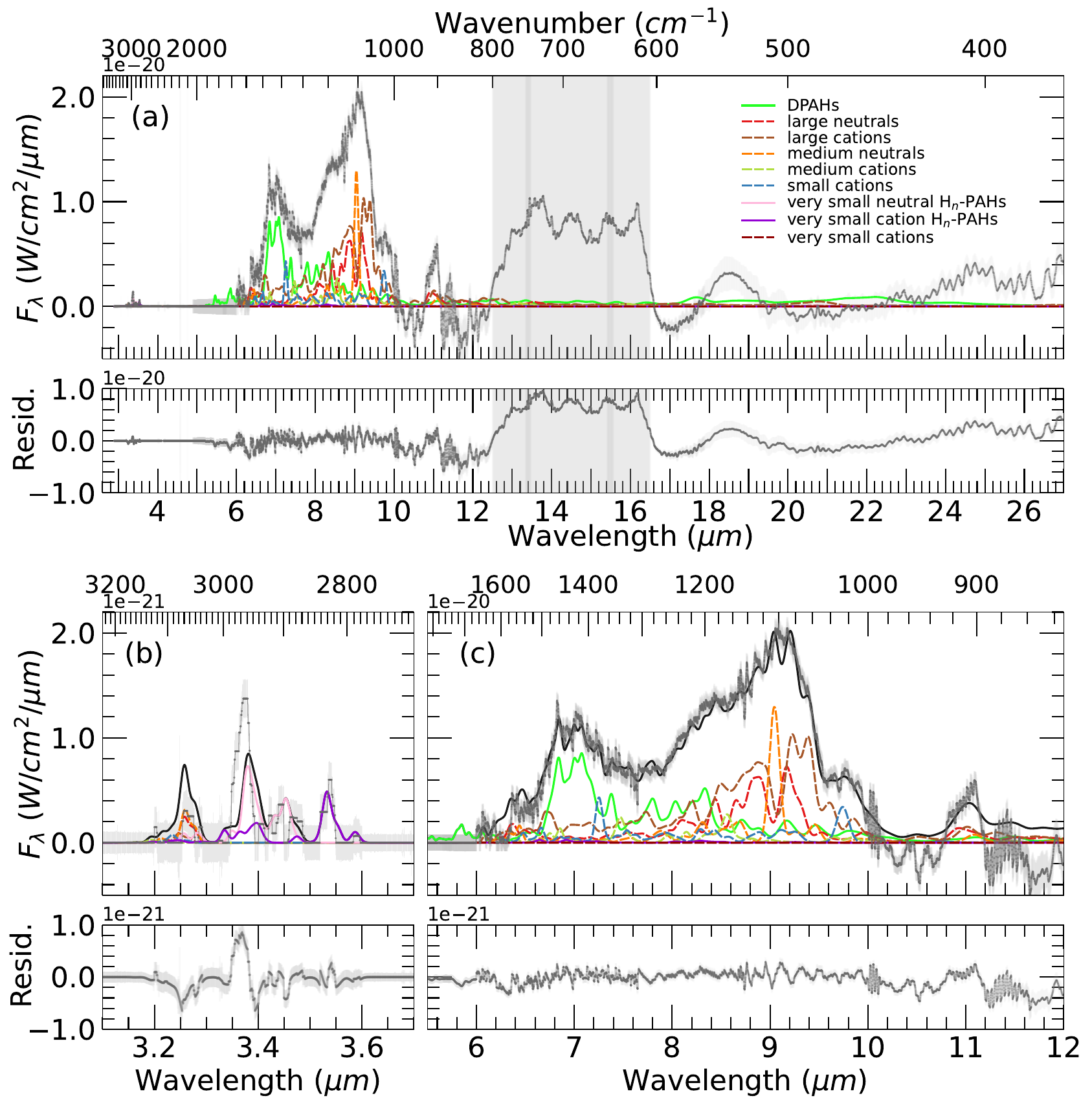}
\caption{Best-fit PAH model for comet 81P. (a)~PAH model forward prediction
over the full NIRSpec and MRS range (3.2--27~\micron{}; explicitly,
3.2--3.6~\micron{} and 5.5--27~\micron), with the residual below. The NIRSpec
model is evaluated over 3.2--3.6~\micron. Components are large neutrals and
cations (crimson and sienna dashed lines), medium neutrals and cations
(darkorange and yellowgreen dashed lines), small cations (steelblue dashed),
very small cations (darkred dashed), very small neutral H$_{n}$-PAHs (pink),
very small H$_{n}$-PAH cations (darkviolet solid), and DPAHs (H=0) (lime
solid). (b)~NIRSpec (3.2--3.6~\micron) at 10$\times$ y-scale compared to (a)
and (c). (c)~MRS portion with the most prominent features. (d)~Breakdown by
hydrogenation state: regular PAHs with 1 H atom per peripheral bonding site
(darkred solid line), H$_{n}$-PAHs with CH$_{2}$ groups (2~H~atoms) on one or
more peripheral bonding sites and no --CH$_{3}$ group (blue dashed, ``Me-PAHs
excluded''), H$_{n}$-PAHs that also have a --CH$_{3}$ group (mediumturquoise
dash-dot-dot-dotted, ``Me-PAHs included''), and DPAHs with zero peripheral H
atoms (lime dash-dot-dotted). (e)~Breakdown by composition, set by the
heteroatoms: unsubstituted PAHs, H$_{n}$-PAHs, DPAHs, along with
C$_{10}$H$_{9}$N$^{+}$ (chocolate solid line), PANHs with skeletal N (darkcyan
dashed), and O-PAHs with peripheral O (blue dash-dot-dot-dotted).
(f)~Breakdown by charge state: PAH neutrals (rebeccapurple solid), PAH cations
(chocolate dashed), and DPAHs (no charge) (lime dash-dot-dot-dotted). (g)
Breakdown by size: large ($N_{\rm C} > 70$, darkred solid), medium ($50 <
N_{\rm C} \leq 70$, chocolate dashed), small ($20 < N_{\rm C} \leq 50$,
darkcyan dash-dot-dot-dotted), and very small ($N_{\rm C} < 20$, blue
dash-dot-dotted). Colorblind-friendly HEX colors are used. The closest
matplotlib names are given to aid the eye.}
\label{fig:figure-4}
\end{center}
\end{figure*}

\begin{figure*}[p]\ContinuedFloat
\figurenum{4}
\begin{center}
\includegraphics[trim=0.07cm 0.07cm 0.07cm 0.15cm, clip, width=0.775\textwidth]{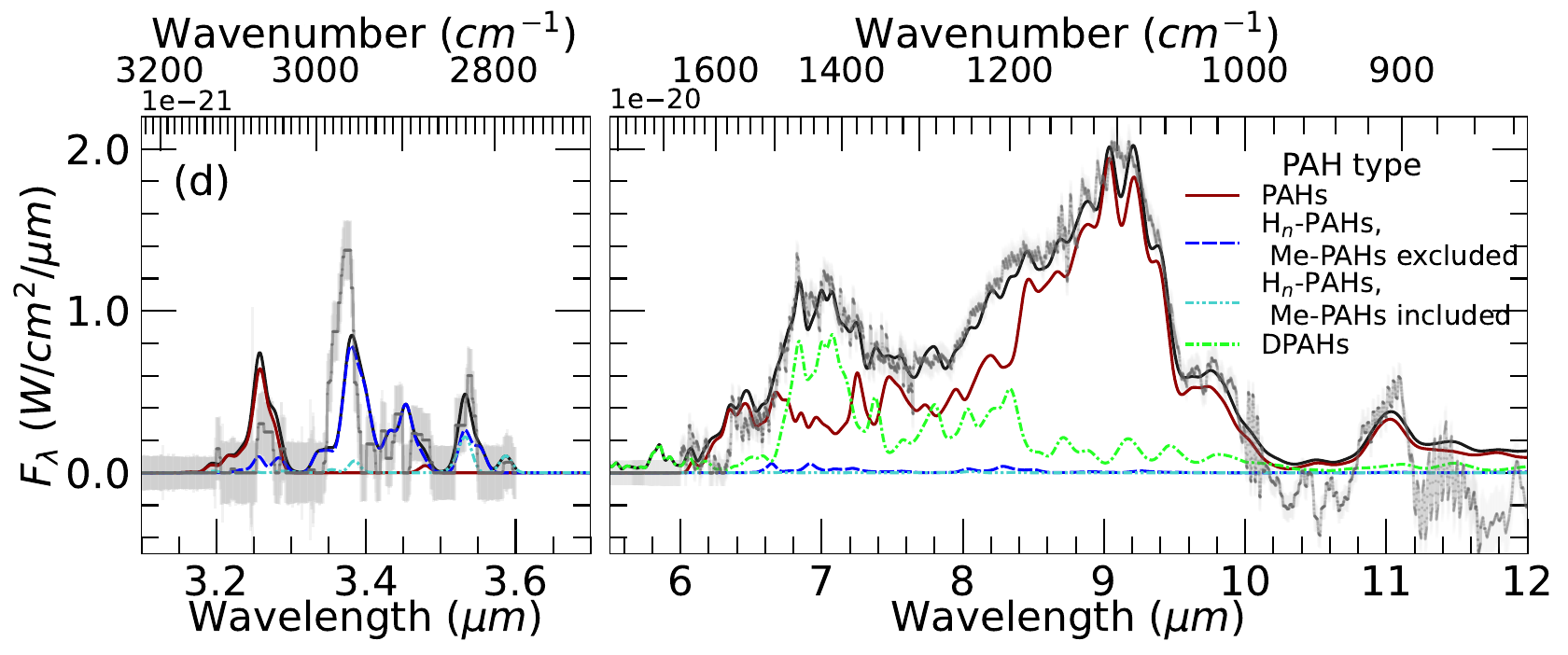}
\includegraphics[trim=0.07cm 0.07cm 0.07cm 0.15cm, clip, width=0.775\textwidth]{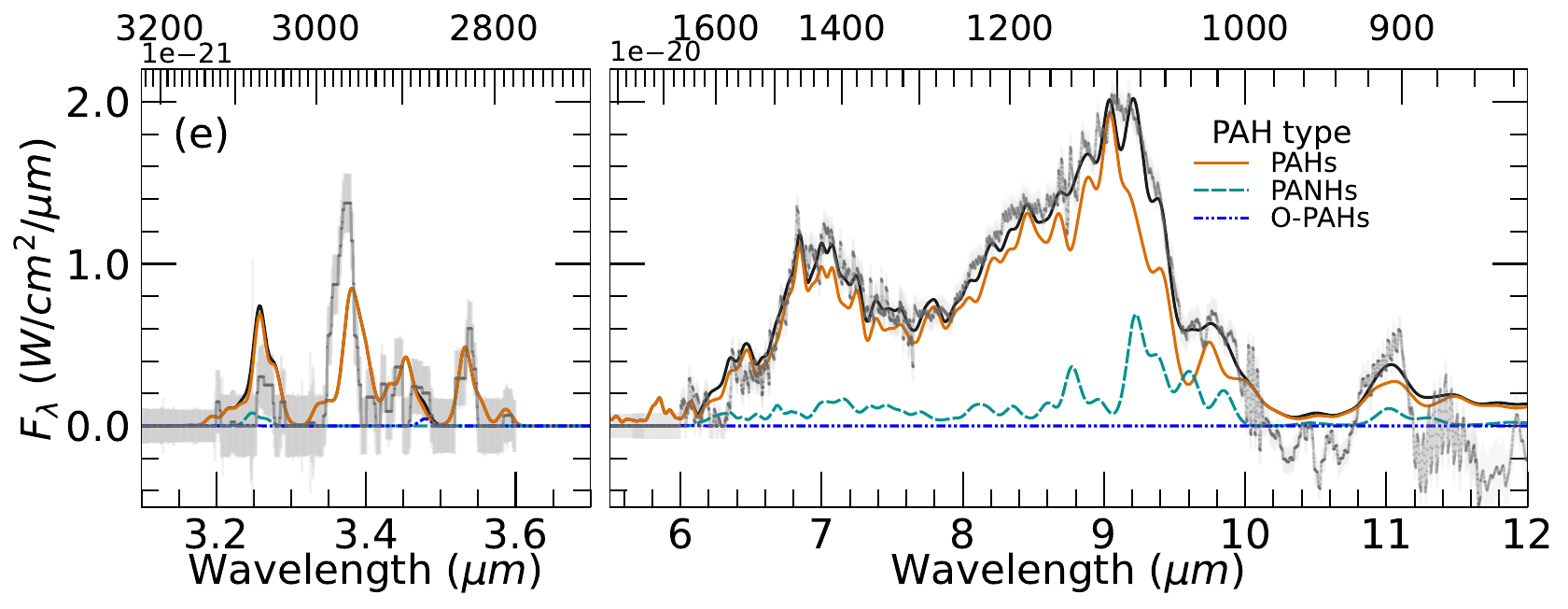}
\includegraphics[trim=0.07cm 0.07cm 0.07cm 0.15cm, clip, width=0.775\textwidth]{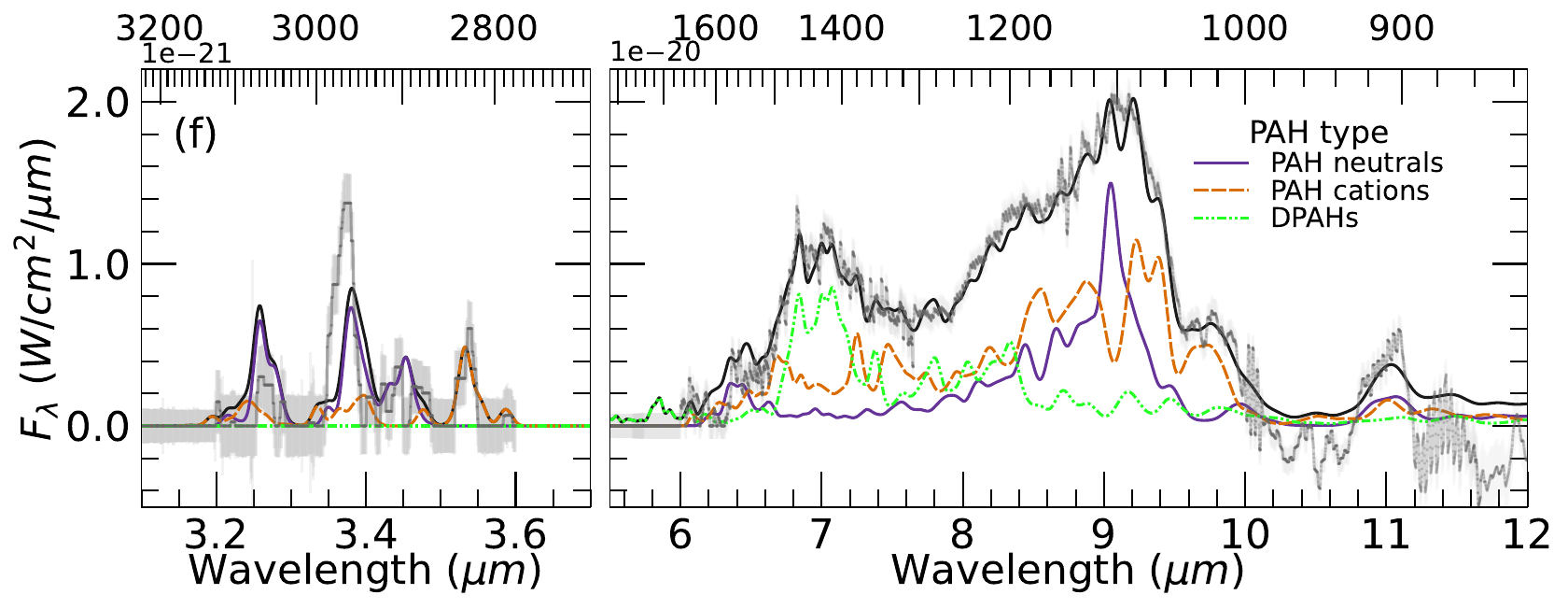}
\includegraphics[trim=0.07cm 0.07cm 0.07cm 0.15cm, clip, width=0.775\textwidth]{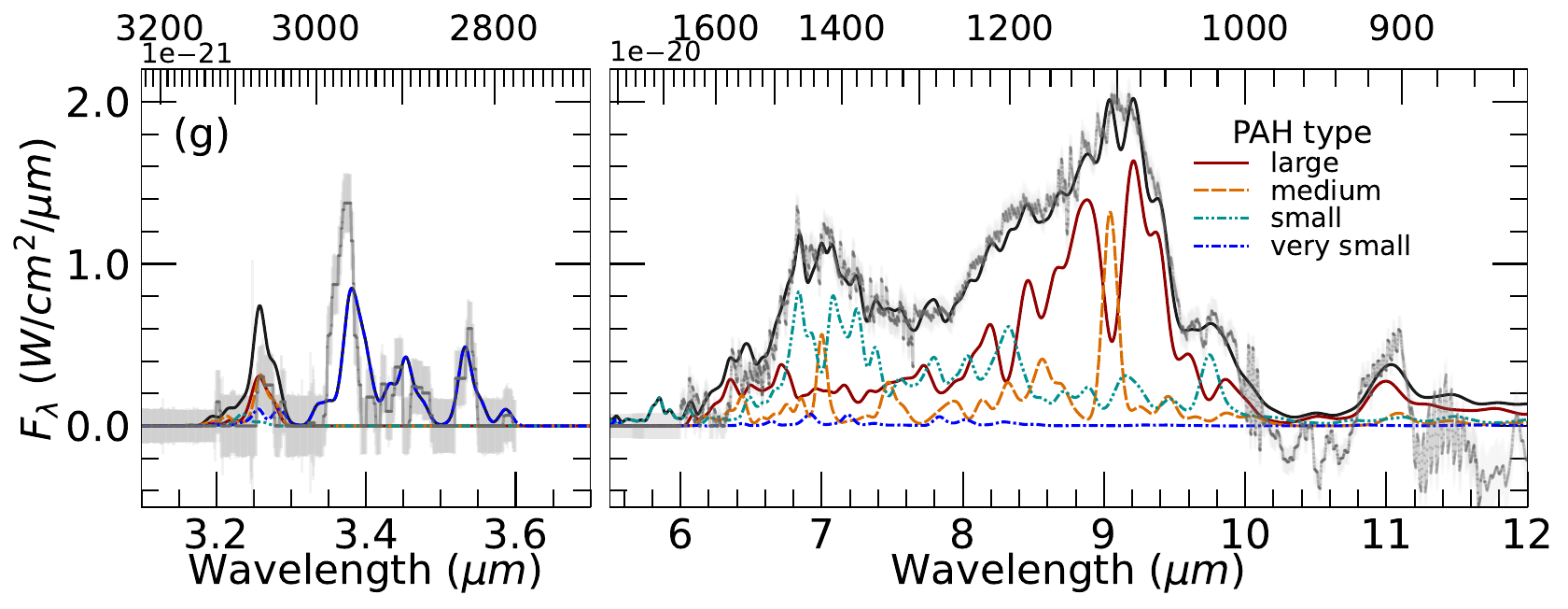}

\caption{\, continued.}
\end{center}
\end{figure*}

The best-fit PAH model builds on the three-subpopulation model of H$_{n}$-PAHs
($N_{\rm{C}} \leq 50$), DPAHs, and large PAHs (Section~\ref{sec:pah:3pop}). It
differs from that model, and from the four-subpopulation model
(Section~\ref{sec:pah:mepah:4pop}), in two ways: it includes very small and
small unsubstituted PAHs (Section~\ref{sec:pah:intro}), and medium PAHs ($50 <
N_{\rm{C}} \leq 70$). Both additions improve the fit to the data. The best-fit
PAH model for comet 81P is shown in Figures~\ref{fig:figure-4}(a)--(c).

In this best-fit model the unsubstituted PAHs appear as very small cations
(contributing minimally) and small cations. There are no unsubstituted PAH
neutrals for $N_{\rm{C}} \leq 50$. The very small cations also include the
O-PAHs C$_{11}$H$_{8}$O$^{+}$ (uid=502) and C$_{11}$H$_{7}$O$^{+}$ (uid=530),
which contribute only 1\% of the NIR flux (Figure~\ref{fig:figure-4}(e)). The
PAHN C$_{10}$H$_{9}$N$^{+}$ (uid=476) is also among the marginal species
(Section~\ref{sec:pah:method}) and contributes 1.1$\times$10$^{-3}$ of the
modeled NIRSpec integrated flux (Section~\ref{sec:pah:panh}).
C$_{10}$H$_{9}$N$^{+}$ is plotted in Figure~\ref{fig:figure-4}(e) with the
PAHs with neither skeletal N nor peripheral O (the unsubstituted PAHs,
H$_{n}$-PAHs, DPAHs, and the one amine PAHN). Medium-sized PAH neutrals
produce strong narrow emission at the central peak of the 9.1~\micron{}
complex, 9.05~\micron{} (1105 cm$^{-1}$). Elsewhere in the MIR the medium PAHs
play a minor role.

The six species with --CH$_{2}$ groups have peripheral CH$_{2}$ fractions
between 0.500 and 0.800 and $N_{\rm{C}} \leq 19$, and six lie inside the
selection box of Figure~\ref{fig:figure-2}(c). The very small neutral
H$_{n}$-PAHs produce the flux near 3.47~\micron{} (2880 cm$^{-1}$), passing
through the observed data points at their low signal-to-noise
(Figures~\ref{fig:figure-2}(a) and \ref{fig:figure-4}(b), pink). The best-fit
model contains no Me-PAH that is not also an H$_{n}$-PAH. The only
methyl-bearing species is C$_{10}$H$_{18}^{+2}$ (uid=4396). It also has
--CH$_{2}$ groups and is counted among the H$_{n}$-PAHs
(Table~\ref{tab:type-size-counts}).

The best-fit model over-predicts the flux at 3.2--3.3~\micron{} relative to
the observations. Within that range, large and medium neutral PAHs generate a
3.27~\micron{} feature that exceeds the observed 1.37$\sigma$ upper limit and
do not contribute to the 3.37~\micron{} feature. The three-subpopulation model
matches the observations at 3.2--3.3~\micron{} to within $\sim$1$\sigma$,
whereas the best-fit model departs by $\sim$2$\sigma$ because it includes the
medium neutral PAHs. The best-fit model improves on both the
three-subpopulation model (Section~\ref{sec:pah:3pop}) and the
four-subpopulation model (Section~\ref{sec:pah:mepah:4pop}). The improvement
comes in the MIR, through the 6.9~\micron{} complex, the 9.1~\micron{}
complex, and the 11.0~\micron{} feature, at the cost of a poorer fit to
NIRSpec. Because the number of PAHs and the number of search parameters differ
between models, we compare them by $\Delta$AIC $=$ AIC$_{\rm min} -$ AIC,
which weighs information content rather than normalizing by the number of data
points (Table~\ref{tab:model-comparison}, Appendix~\ref{sec:appendix1-table}).

The best-fit model does not predict any high-contrast spectral features at
wavelengths greater than $\sim$11.5~\micron{} and therefore presents good
agreement with the data at longer wavelengths. Omitting wavelengths greater
than 12~\micron{} from the fitting process, or underweighting the data points
by amplifying their uncertainties (except for 12.5--16.5~\micron,
Section~\ref{sec:pah:method}), always increases (worsens) the AIC because
there is spectral information at these wavelengths. Such model runs can also
yield predicted features that are in visual disagreement with the observed
comet 81P data at longer wavelengths.

The best-fit PAH model is broken down by hydrogenation, composition, charge
state, and size in Figures~\ref{fig:figure-4}(d)--(g), plotted at the same
scale as Figures~\ref{fig:figure-4}(a)--(c) for comparison.

The breakdown by hydrogenation into H$_{n}$-PAHs, unsubstituted PAHs, and
DPAHs, is shown in Figure~\ref{fig:figure-4}(d) and largely reflects what has
already been described. H$_{n}$-PAHs dominate the NIR, DPAHs dominate the
6.9~\micron{}, and unsubstituted PAHs dominate at wavelengths greater than
8.4~\micron. The DPAHs contribute 54\% of the model flux between 6.7 and
7.6~\micron{}, the 6.9~\micron{} complex, against 26\% integrated over the
6--10~\micron{} range and 32\% integrated over the full 3.2--27~\micron{}
range (explicitly, 3.2--3.6~\micron{} and 5.5--27~\micron{})
(Appendix~\ref{sec:appendix1-nodpah}). The H$_{n}$-PAHs are drawn as two
curves, separating the one species that also carries a --CH$_{3}$ group,
C$_{10}$H$_{18}^{+2}$ (uid=4396), from the five that do not. That species
produces about half the model flux of the 3.53~\micron{} feature and
contributes minimally near 3.59~\micron{} and to the center of the
3.37~\micron{} feature.

The breakdown by composition into PAHs without heteroatoms (the sum of
unsubstituted PAHs, Hn-PAHs, and DPAHs), PANHs, and O-PAHs, set by the
heteroatoms, is shown in Figure~\ref{fig:figure-4}(e). The O-PAHs contribute
at 3.46~\micron{}, with too little contrast to be distinguished in the figure.
The nitrogen is incorporated into PANHs (N in skeletal bonds), and these PANHs
are a subset of the large PAHs in the size range 93 $\leq N_{\rm{C}} \leq 95$,
which constitute a significant fraction ($\sim$70\%) of PAHs of this size in
the PAHdb V4.00 database \citep{2024ApJ...968..128R}. All variations of model
runs that exclude this size range yield higher AIC values (worse fits).

Figure~\ref{fig:figure-4}(f) shows the breakdown by charge state --- neutrals,
cations, and no hydrogens (shown separately for clarity). PAH neutrals and
cations both contribute to the 3.37~\micron{} feature, the 9.1~\micron{}
complex, and the 11.0~\micron{} feature. Excluding the DPAHs, cations
contribute 0.62 of the integrated flux density
(Table~\ref{tab:bestfit-fractions}).

Figure~\ref{fig:figure-4}(g), the breakdown by PAH size, is included for
completeness and for comparison with other PAHdb modeling outputs. Very small
PAHs dominate the 3.37~\micron{} feature. Medium PAHs contribute a narrow
feature in the MIR near the central wavelength of the 9.1~\micron{} complex,
at 9.05~\micron{} (1105 cm$^{-1}$). Large PAHs dominate the 9.1~\micron{}
complex and the 11~\micron{} feature. The large PAHs ($N_{\rm{C}} > 70$)
contribute 69\% of the model flux between 8.6 and 9.6~\micron{}, the
9.1~\micron{} complex, against 52\% over 6--10~\micron{} and 47\% overall
(Table~\ref{tab:bestfit-fractions}).



\begin{deluxetable}{lccccccc}
\tabletypesize{\footnotesize}
\tablecaption{Fractional Breakdown of the Best-fit PAH Model for Comet 81P, by Subpopulation and Ionization\label{tab:bestfit-fractions}}
\tablewidth{0pt}
\tablehead{
\colhead{} & \colhead{large} & \colhead{medium} & \colhead{small} & \colhead{very small} & \colhead{H$_{n}$-PAHs} & \colhead{DPAHs} & \colhead{All but DPAHs}
}
\startdata
\cutinhead{Subpopulation fraction of:}
integrated flux density (W~cm$^{-2}$~$\mu$m$^{-1}$)  & 0.47 & 0.13 & 0.06 & 0.00 & 0.01 & 0.32 & 0.67 \\
number                                               & 0.49 & 0.13 & 0.06 & 0.00 & 0.01 & 0.31 & 0.69 \\
\cutinhead{Ionization fraction $f_{\rm ionized}$ by:}
integrated flux density (W~cm$^{-2}$~$\mu$m$^{-1}$)  & 0.65 & 0.33 & 1 & 1 & 0.75 & 0.01 & 0.62 \\
number                                               & 0.67 & 0.33 & 1 & 1 & 0.76 & 0.01 & 0.63 \\
\enddata
\end{deluxetable}



\begin{deluxetable}{lccccc}
\tabletypesize{\footnotesize}
\tablecaption{Number of PAHs in the Best-fit Model for Comet 81P, by Subpopulation and Size\label{tab:type-size-counts}}
\tablewidth{0pt}
\tablehead{
\colhead{Subpopulation} & \colhead{very small} & \colhead{small} & \colhead{medium} &
\colhead{large} & \colhead{all}
}
\startdata
H$_{n}$-PAHs                      &  5 &  0 & 0 &  0 &  5 \\
H$_{n}$-PAHs, Me-PAHs             &  1 &  0 & 0 &  0 &  1 \\
PAHs                              &  3 &  5 & 4 & 17 & 29 \\
DPAHs                             &  1 & 18 & 2 &  1 & 22 \\
\hline
all                               & 10 & 23 & 6 & 18 & 57 \\
\enddata
\tablecomments{Sizes are very small ($N_{\rm C} < 20$), small ($20 < N_{\rm C} \leq 50$),
medium ($50 < N_{\rm C} \leq 70$) and large ($N_{\rm C} > 70$).
Subpopulation is set by hydrogenation and alkylation. It is a separate axis from composition,
which is set by heteroatoms and is shown in Figure~\ref{fig:figure-4}(e). The two O-PAHs and the
one PAHN are counted here among the PAHs.
C$_{10}$H$_{18}^{+2}$ (uid=4396) carries both --CH$_{2}$ and --CH$_{3}$ groups and so belongs to
the H$_{n}$-PAHs and the Me-PAHs alike. It is given its own row and is not counted twice, so the
H$_{n}$-PAHs number six in total.
Every H$_{n}$-PAH in the best-fit model is very small, and every DPAH is small or medium
except that one is very small and one is large.}
\end{deluxetable}

\begin{figure}[htb]
\figurenum{5}
\begin{center}
\includegraphics[trim=0.07cm 0.05cm 0.07cm 0.20cm, clip, width=\columnwidth]{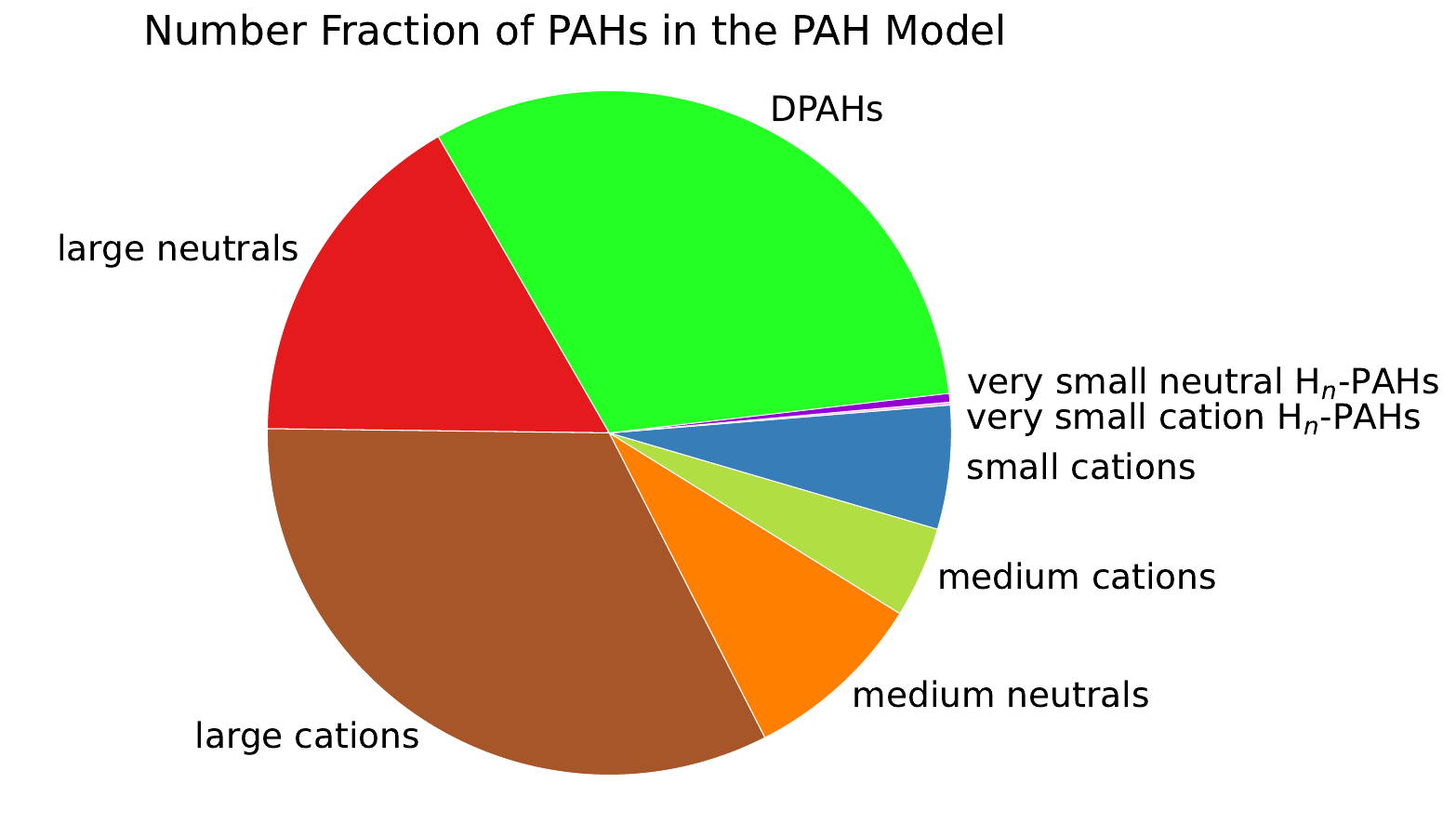}
\caption{Pie diagram of the number fraction of molecules per PAH category in
the best-fit PAH model for comet 81P. Categories and colors are as in
Figure~\ref{fig:figure-4}(a)--(c).}
\label{fig:figure-5}
\end{center}
\end{figure}
\begin{figure*}[t]
\figurenum{6}
\begin{center}
\includegraphics[trim=0.07cm 0.05cm 0.07cm 0.15cm, clip, width=0.980\textwidth]{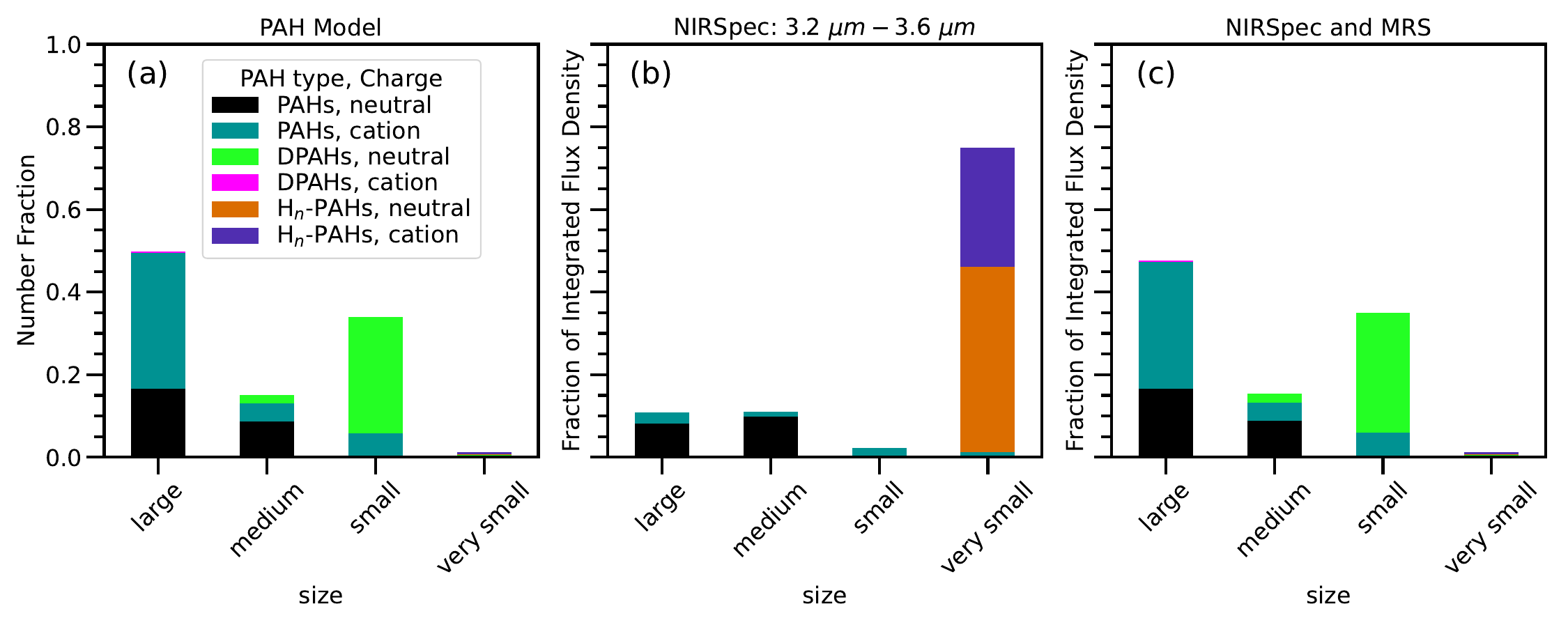}
\caption{Bar plots of fraction versus size bin, with breakouts by PAH type and
charge state: DPAHs (lime), PAH neutrals and cations (black, darkcyan), and
H$_{n}$-PAH neutrals and cations (chocolate, rebeccapurple). (a)~Number
fraction. (b)~Fraction of integrated flux density (F$_\lambda$,
W~cm$^{-2}$~$\mu$m$^{-1}$) for NIRSpec (3.2--3.6~\micron). (c)~Fraction of
integrated flux density for NIRSpec and MRS.}
\label{fig:figure-6}
\end{center}
\end{figure*}

A pie diagram, Figure~\ref{fig:figure-5}, shows the number fraction by PAH
category, using the same categories and colors as in
Figures~\ref{fig:figure-4}(a)--(c). Figure~\ref{fig:figure-6}(a) shows the
number fraction versus PAH size, grouped by charge (neutral or cation) and,
within size, by CH$_{2}>0$ for H$_{n}$-PAHs, H$>$0 for regular PAHs, and H=0
for DPAHs. Table~\ref{tab:bestfit-fractions} gives the corresponding values as
well as the fraction of integrated flux density ($\Sigma
F_{\lambda}\,d\lambda$). The number fractions are derived from the model
weights per PAH, summed over each PAH category and divided by the sum of all
weights. The fraction of integrated flux density is the sum of flux density
over all wavelengths ($\Sigma F_{\lambda}\,d\lambda$) for a given category
divided by the total model flux density.

The DPAHs (H=0) are predominantly small and account for 0.32 of the integrated
flux density and 0.31 by number. The large PAHs account for 0.47 and 0.49,
respectively. They also constitute a greater number fraction than the very
small and small PAHs combined. The medium-sized DPAH C$_{70}$ (fullerene,
D$_{5h}$ symmetry) contributes two percent of the integrated flux density.
C$_{60}$ is not in the best-fit model. Including the DPAHs, the best-fit model
yields number fractions of $\simeq$ 0.35 for the very small and small PAHs,
$\simeq$ 0.15 for the medium PAHs, and $\simeq$ 0.50 for the large PAHs
(Figure~\ref{fig:figure-6}(a)).

The number fraction and the fraction of integrated flux density are nearly
equal for all categories in Table~\ref{tab:bestfit-fractions}, as expected
from the emission model. In the cascade each molecule absorbs one photon and
re-radiates all of it in the infrared, and the NIRSpec and MRS ranges together
capture nearly all of that re-radiated energy. The integrated radiant energy
per molecule is nearly independent of size. The ratio of the fraction of
integrated flux density to the number fraction, measured over the 57 retained
species, varies by only a factor of 1.4, from 1.23 at N$_{\rm C}$ = 10 down to
0.87 at N$_{\rm C}$ = 172. Size determines the wavelengths at which the energy
emerges, not the amount of energy. The two rows are expected to agree. Their
agreement is a check on the model rather than a result. The number fraction
and the flux-density fraction agree wavelength-integrated but not within the
NIRSpec window, where a handful of superhydrogenated very small PAHs produce
most of the flux.

The ratio of the fraction of integrated flux density to the number fraction,
evaluated within the NIRSpec window over the 26 species that sum to 99\% of
the NIRSpec flux, spans more than four orders of magnitude, from 0.014 to 284.
The highest ratios are those of the very small superhydrogenated PAHs. All
Hn-PAHs with N$_{\rm C} \leq$ 20 have ratios above 20, reaching 284 for
C$_{10}$H$_{18}$, and the three very small PAHs that are not H$_{n}$-PAHs run
from 0.59 to 29. The effect of size on this ratio is greater than the effect
of charge, even though charge acts to lower the NIR/MIR flux ratio
(Section~\ref{sec:pah:3pop}).

Figures~\ref{fig:figure-6}(b) and (c) show, with the same grouping as (a) and
at the same vertical scale, the fraction of integrated flux density for
NIRSpec (3.2--3.6~\micron{}) and for NIRSpec and MRS combined. In the NIRSpec
window, Figure~\ref{fig:figure-6}(b), the very small H$_{n}$-PAHs dominate,
and large and medium neutral PAHs appear only through the 3.27~\micron{}
feature as discussed above. The small PAHs are dominated by H$_{n}$-PAH
cations. Over the full range of NIRSpec and MRS, Figure~\ref{fig:figure-6}(c),
the flux-density fractions track the number fractions of (a), as
Table~\ref{tab:bestfit-fractions} shows by category.

\subsection{Search for Very Small PANHs and PAHNs}
\label{sec:pah:panh}

PAHs with N in a peripheral group (PAHNs) were found in Stardust samples
\citep{2010M&PS...45..701C}. The high signal-to-noise JWST remote sensing data
obtained on comet 81P warranted a search for these species, and for PAHs with
N in the aromatic skeleton (PANHs), in the coma emission.

The best-fit model search was not restricted by size or composition. The
absence of very small PANHs from the best-fit model is a result of the fit and
not of the selection. The Stardust samples do not supply an independent test.
Two-step laser mass spectrometry at 266~nm is about 2000 times less sensitive
to N-heterocycles than to their homocyclic isomers
\citep{2010M&PS...45..701C}, and the absence of PANHs from the Stardust mass
spectra is a detection limit rather than a result
(Section~\ref{sec:pah:stardust}). The two non-detections do not confirm each
other.

Of the very small PAHNs searched, only 1-C$_{10}$H$_{9}$N$^{+}$
(1-naphthylamine cation, uid=476) is retained in the best-fit model, and then
only marginally (Section~\ref{sec:pah:method}), at 1.1$\times$10$^{-3}$ of the
modeled NIRSpec integrated flux. The search included PAHs with peripheral
nitrile (--CN) groups as well as amines, and none of the nitrile-bearing
species was retained. At 143~amu, 1-C$_{10}$H$_{9}$N$^{+}$ sits below 185~amu,
the lowest mass at which an amine-compatible peak is reported in the Stardust
sample \citep{2010M&PS...45..701C}. Of the other three,
2-C$_{10}$H$_{9}$N$^{+}$ (uid=474 in the PAHdb V4.00 database) appears in 158
of the 1000 trials and is not retained (Section~\ref{sec:pah:method}), and
9-C$_{14}$H$_{11}$N$^{+}$ (uid=478) and 1-C$_{14}$H$_{11}$N$^{+}$ (uid=480)
are not selected in any trial.

\begin{figure*}[ht!]
\figurenum{7}
\begin{center}
\includegraphics[trim=0.07cm 0.05cm 0.07cm 0.05cm, clip, width=0.850\textwidth]{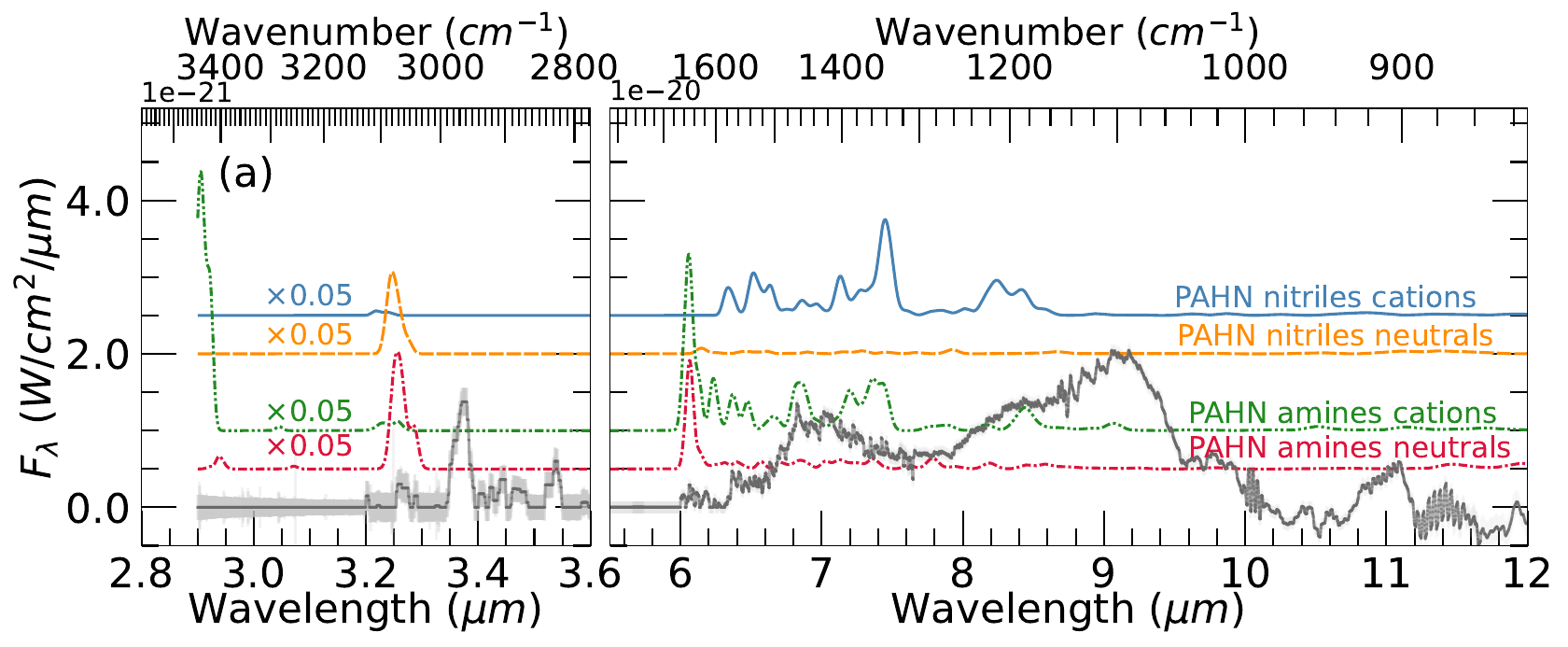}
\includegraphics[trim=0.07cm 0.05cm 0.07cm 0.05cm, clip, width=0.850\textwidth]{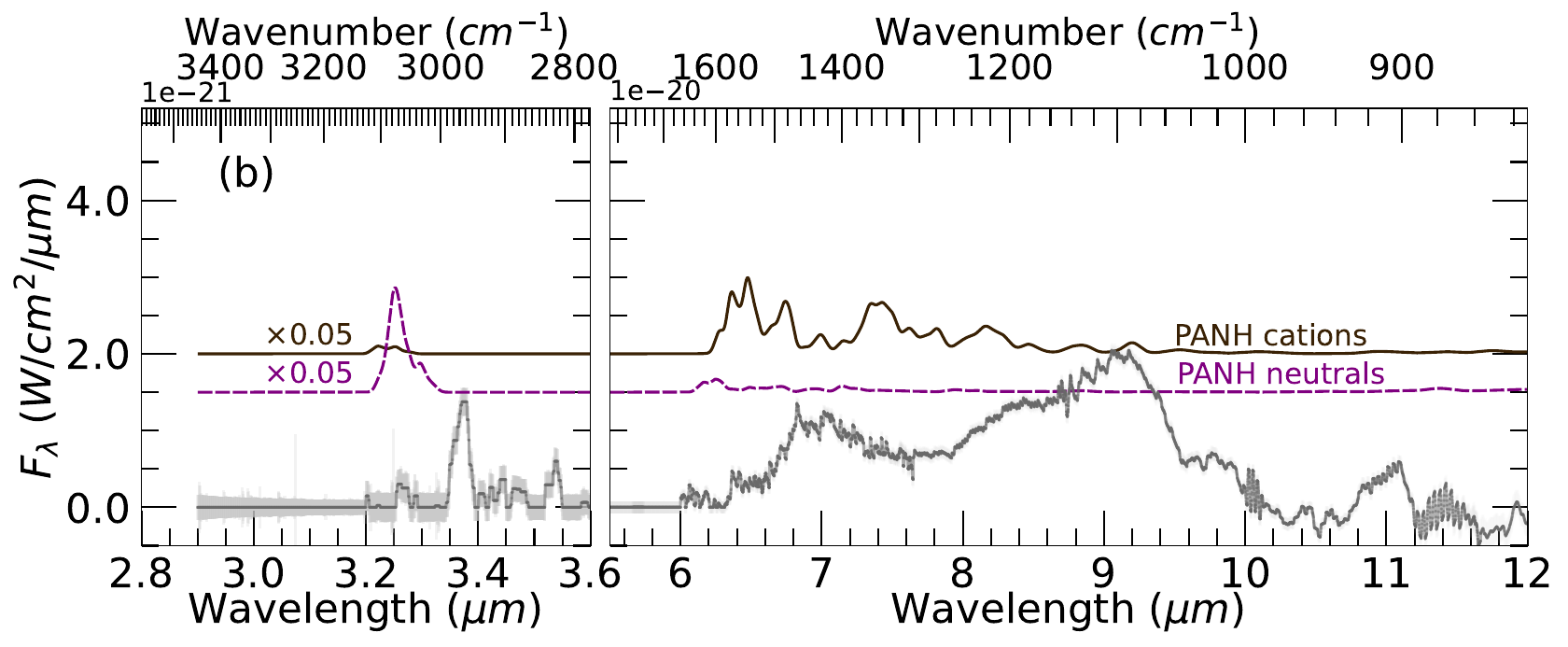}
\caption{Very small PANHs and PAHNs (N$_{\rm C} \leq 17$) in the PAHdb V4.00
database compared to the spectrum of comet 81P. The forward model is computed
over the NIRSpec and MRS wavelength ranges. (a)~PAHNs, which carry the N in a
peripheral group: nitrile (--C$\equiv$N) neutrals and cations (darkorange
dashed and steelblue solid), and amine (--NH$_{2}$) neutrals and cations (red
dash-dotted and forestgreen dash-dot-dotted). (b)~PANHs, which carry the N in
the aromatic skeleton: neutrals and cations (purple dashed and darkbrown
solid). The left (NIRSpec) panel is plotted at 10$\times$ the y-scale of the
right (MRS) panel. The predicted NIR features have high contrast and are
multiplied by 0.05 to fit within the panel.}
\label{fig:figure-7}
\end{center}
\end{figure*}

Figure~\ref{fig:figure-7} shows the averaged spectra of neutral and cationic
PAHNs and PANHs with N$_{\rm C} \leq 17$, the mass range (amu$\leq$244) over
which \citet{2010M&PS...45..701C} give assignments for the Stardust samples
(Section~\ref{sec:pah:stardust}). Figure~\ref{fig:figure-7}(a) shows the
PAHNs, separated into amines (--NH$_{2}$) and nitriles (--C$\equiv$N), and
Figure~\ref{fig:figure-7}(b) shows the PANHs. Forward-predicted flux densities
for fluorescent cascades excited by a 5770~K blackbody (sunlight) are averaged
within each group and plotted stacked with constant offsets. The amine
cations, the class to which the one retained species 1-C$_{10}$H$_{9}$N$^+$
belongs, produce a high-contrast N--H feature at 2.91~\micron{} that peaks at
the short-wavelength edge of the NIRSpec coverage and is clipped in
Figure~\ref{fig:figure-7}(a). The PAH models are fitted over
3.2--3.6~\micron{} in the NIR (Section~\ref{sec:pah:method}), so this
signature of the class lies outside the fitted range and does not constrain
these species. The one retained species, 1-C$_{10}$H$_{9}$N$^+$, is an amine
PAHN cation, and the 2.91~\micron{} N--H feature is a signature of its class.
The amines produce a feature at 6.06~\micron{} that is not observed in the
comet. The amine cations also produce a feature at 7.40~\micron{}, and the
nitrile cations one at 7.45~\micron{}. The PANH cations produce features near
6.4, 6.75 and 7.4~\micron{}. None of these features is in the best-fit model.
In all three groups the neutrals produce features near 3.250--3.255~\micron{},
on the short-wavelength side of the aforementioned 3.27~\micron{} emission,
for which only a 1.37$\sigma$ upper limit is measured. The cations produce
only weak features near 3.2--3.25~\micron{}. This figure shows the challenge
of finding spectral features of PAHNs and PANHs in the very small size range
in the comet's spectrum.

\subsection{Carbon Nanograins and the 14.5~\micron{} Complex}
\label{sec:pah:nanograin}

The high contrast and relatively straight-sided 12.5--16.5~\micron{} emission
feature, which we call the 14.5~\micron{} complex, in comet 81P is in a
spectral region characteristic of the `floppy' bending skeletal modes of large
PAHs, possibly including delocalized non-rigid skeletal bending modes or C-H
out-of-plane bending modes. This region was significantly underweighted by
40$\times$ $\sigma$ in the 12.5--16.5~\micron{} in the PAH model runs
(Section~\ref{sec:pah:method}).

\begin{figure*}
\figurenum{8}
\begin{center}
\includegraphics[trim=0.07cm 0.07cm 0.07cm 0.15cm, clip, width=0.850\textwidth]{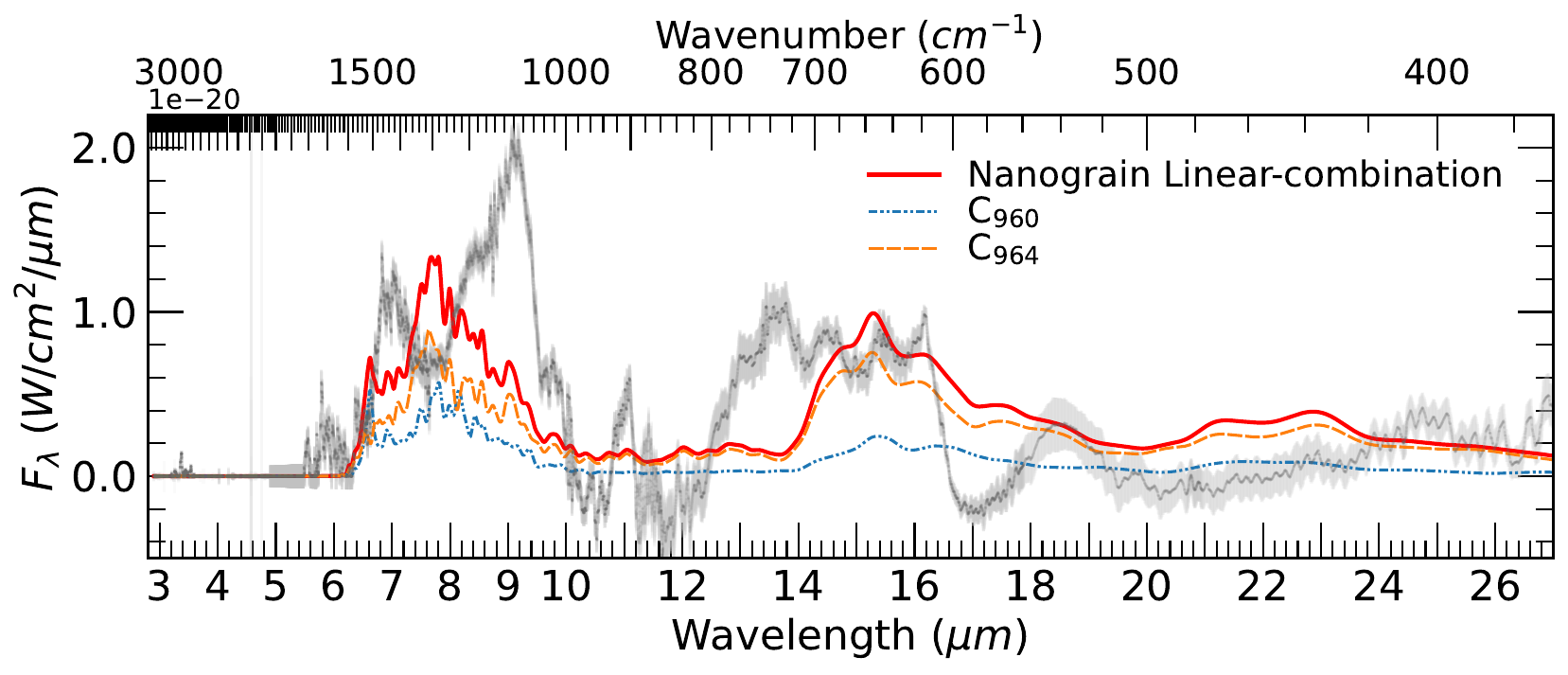}
\caption{Comparison of carbon nano-grains shaped as nested non-spherical buckyballs or fullerenes
\citep{2025ApJ...986...77R} with the full spectrum of comet 81P 
including the 14.5~\micron{} complex with 1$\sigma$ uncertainties. 
Shown is a linear-combination fit (crimson solid line) of 
fluorescent emissions from two nano-grains: C$_{960}$ (steelblue
dash-dot-dotted) and C$_{964}$ (darkorange dashed), excited by 5770~K
blackbody (sunlight) and computed using the Ames NanoGrain database (NGdb), 
with scale factors 0.247 and 0.765, respectively.}
\label{fig:figure-8}
\end{center}
\end{figure*}

\begin{figure*}
\figurenum{9}
\begin{center}
\includegraphics[trim=0.07cm 0.07cm 0.07cm 0.15cm, clip, width=0.850\textwidth]{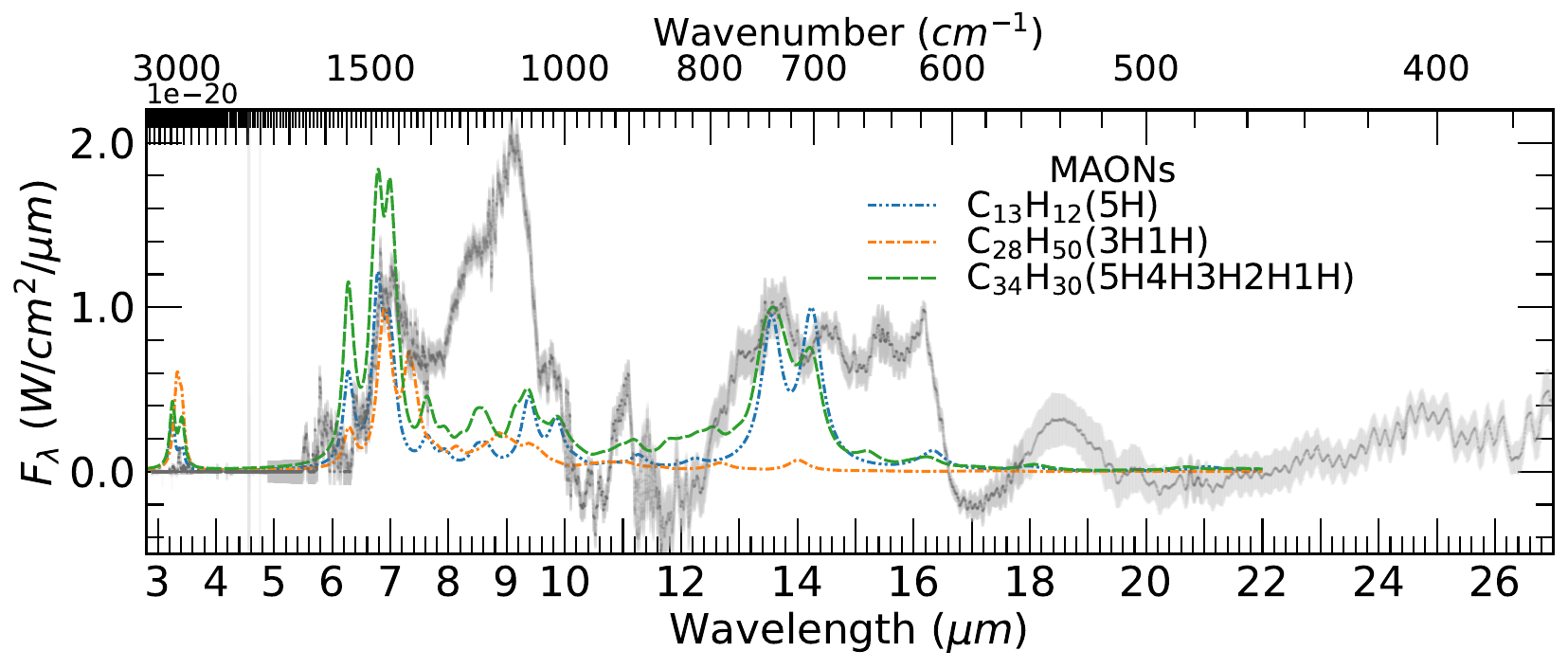}
\caption{Comparison of Mixed 
Aromatic-Aliphatic Organic Nano-grains (MAONs), non-cage-shaped carbonaceous nano-grains, 
with the full spectrum of comet 81P 
including the 14.5~\micron{} complex with 1$\sigma$ uncertainties. Shown are three MAONs: 
C$_{13}$H$_{12}$ (5H) (steelblue dash-dot-dotted), C$_{28}$H$_{50}$ (3H1H)
(darkorange dash-dotted), and C$_{34}$H$_{30}$ (5H4H3H2H1H) (forestgreen
dashed), scaled by 1.45e-2, 1.24e-3, 2.74e-3 so their flux densities near 13.5, 13.5, and 6.9~\micron{} are $10^{-20}~W~cm^{-2}~\micron ^{-1}$, respectively. 
Electronic data \citep[Figure 3]{2016JPhCS.728f2003S}, courtesy of first author S. Sadjadi.
}
\label{fig:figure-9}
\end{center}
\end{figure*}

Carbonaceous nanograins are a natural and predicted extension of PAHs to much
larger sizes. Two candidates are available for comparison, carbon nanograins \citep{2025ApJ...986...77R} (NGdb,
\url{https://nanograin.odr.io/}) and Mixed
Aliphatic--Aromatic Organic Nanograins (MAONs) \citep{2016JPhCS.728f2003S,
2017ApJ...845..123S}. Figure~\ref{fig:figure-8} shows the fluorescent cascade from carbon
nanograins (shaped like concentric buckyballs) excited by sunlight (5770 K).
Figure~\ref{fig:figure-9} shows the flux
spectrum of a linear combination of a subset of MAONs computed at 500 K.
Both are compared to the 2.9--27~\micron{} spectrum of the comet (see
Figure~\ref{fig:figure-4}(a) for best-fit PAH model). On the one hand, carbon
nanograins produce emission in the long-wavelength half of the 14.5~\micron{}
complex, features near 7.7~\micron, and sustained longer-wavelength emission, 
which is a characteristic of this size and shape of nanograins, 
that is stronger than observed in the comet. The nanograins shown here are in
approximately the 3-nm to 5-nm size range. 
On the other hand, MAONs
produce emission in the 12.5--14.5~\micron{} short-wavelength half of the
14.5~\micron{} complex, a 6.9~\micron{} complex similarly shaped to the
comet's, and a weak broad feature in the 3.3--3.4~\micron{} region at the same
flux level as the NIRSpec observations of comet 81P. 
These candidates are shown to spur
further theoretical computations and to support the concept that comet 81P
contains a wealth of aromatic carriers, from small PAHs not expected to
survive in the diffuse interstellar medium (DISM) and ISM,
(Section~\ref{sec:pah:origins}) to nanograins at the upper end of the size
distribution in the ISM.

\section{Discussion}
\label{sec:pah:origins}

\subsection{Overview}
Comet 81P contains a significant variety of PAH types, many as cations, and
this variety represents the many phases of PAH evolution. There are a few
medium and many more large PAHs that survive in the harshest environs of PDRs
in the ISM. In the ISM, large PAHs may break up into dehydrogenated
intermediates that can then form fullerenes \citep{2012PNAS..109..401B}. In
comet 81P, there are dehydrogenated intermediates as small DPAHs that could
not have survived the harsh ISM and likely formed and survived in parts of the
DISM. These species could be precursors to stable PAH cages or fullerenes.

\begin{figure*}[ht!]
\figurenum{10}
\begin{center}
\includegraphics[trim=14pt 177pt 97pt 87pt, clip, width=0.980\textwidth]{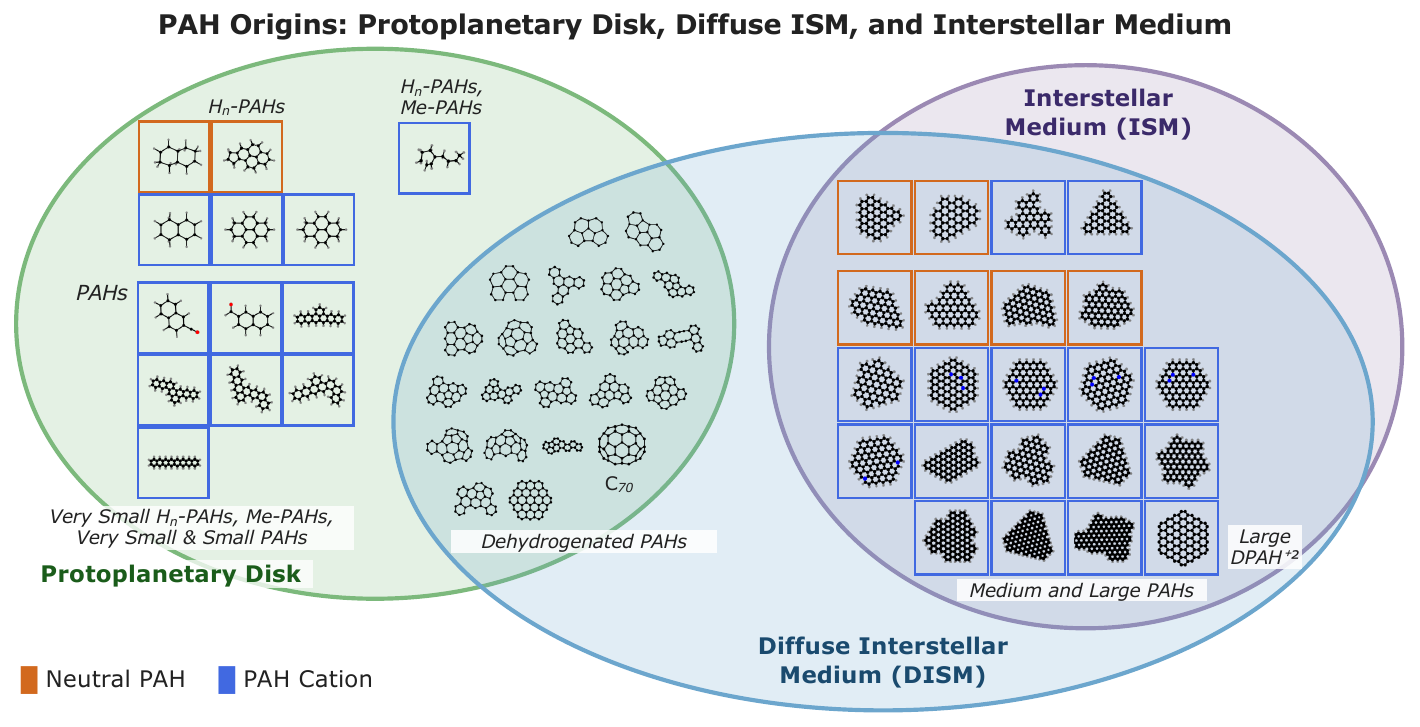}
\caption{PAH Origins. Interpretive Venn diagram of the three astrophysical
environments represented by the best-fit PAH model for comet 81P: the
protoplanetary disk (green ellipse), the diffuse interstellar medium (DISM,
blue ellipse), and the interstellar medium (ISM, purple ellipse). Each PAH
structure is framed in orange (neutral PAH) or blue (PAH cation). The
protoplanetary disk region contains the very small H$_{n}$-PAHs, very small
and small PAHs, and C$_{10}$H$_{18}^{+2}$ (uid=4396) that is in both
subpopulations of H$_{n}$-PAHs and Me-PAHs---species unlikely to survive DISM
or ISM conditions. The DISM region holds the dehydrogenated PAHs (DPAHs),
including C$_{70}$, that form by H-loss from larger PAHs and serve as
intermediates en route to fullerenes and carbon nanograins. The ISM region
holds the medium and large PAHs that survive harsh PDR conditions; PANHs
(skeletal N) appear among them. The DISM ellipse overlaps both flanking
regions, reflecting its role as the bridge between the protoplanetary disk and
the harsher environs of the ISM (see Section~\ref{sec:pah:3pop}).}
\label{fig:figure-10}
\end{center}
\end{figure*}

There are very small H$_{n}$-PAHs (mostly as cations) in comet 81P that are
unlikely to have survived the DISM. These likely arise in the molecular cloud
or cold outer protoplanetary disk. Thus, there are three astrophysical
environments represented by the PAHs in comet 81P (Figure~\ref{fig:figure-10}).
In the subsections below, we briefly discuss the origins of PAHs and their
evolution within each of these environments, starting with the most local
formation environment that is our protoplanetary disk out of which the comet
formed. Between the disk and the DISM we consider the coma environment, how
sunlight impacts PAH lifetimes, and ionization states after their release. The
presence of cations of all sizes is as yet not well understood
(Section~\ref{sec:pah:origins:coma}).

\subsection{PAHs from the protoplanetary disk}
\label{sec:pah:origins:disk}

All H$_{n}$-PAHs in the best-fit model are very small. Too short-lived to have
survived the DISM or the ISM, they likely originated in the natal cloud or the
protoplanetary disk. Among them is C$_{16}$H$_{16}^{+}$. The 14 species in the
protoplanetary disk region contribute 78.9\% of the NIRSpec flux and 5.0\% of
the MRS flux. Very small PAHs (C $<20$) are absent from the DISM and the ISM
of our galaxy. They are likely destroyed via absorption of UV photons
\citep{2013A&A...552A..15M} or cosmic rays \citep{2020JPhCS1412o2034C}. These
same very small PAHs have, however, been found in molecular clouds.
Specifically, 1-ring and 2-ring PAHs with peripheral H atoms substituted by
nitrile (--CN) groups -- benzonitrile (C$_{6}$H$_{5}$CN) and 1- and
2-cyanonaphthalene (C$_{10}$H$_{7}$CN) -- were found in TMC-1 using radio
observations, due to their strong dipole moments \citep{2021Sci...371.1265M,
2024Sci...386..810W}. Pure PAHs were also found in TMC-1, including indene
(C$_{9}$H$_{8}$) with a pentagonal and hexagonal ring
\citep{2021ApJ...913L..18B}, its CN-substituted form 2-C$_{9}$H$_{7}$CN
\citep{2022ApJ...938L..12S}, and the 3-ring acenaphthylene (C$_{12}$H$_{8}$)
\citep{2024A&A...690L..13C}. Because these very small PAHs would not have
survived the DISM stage, their spatial association in TMC-1 points to
bottom-up formation \citep{2024A&A...690L..13C}. The largest CN-PAH found
toward TMC-1 is 7-ring cyanocoronene (C$_{24}$H$_{11}$CN). The CN-PAHs in
TMC-1 have similar column densities, and PAH column densities track CN-PAH
column densities. Although PAHs in molecular clouds are destroyed by ion
reactions or depleted onto grains, the flat distribution of column densities
with increasing PAH size is consistent with bottom-up assembly, potentially
with recycling of smaller PAHs into the larger thermodynamically stable PAH
coronene \citep{2025ApJ...984L..36W}.

Bottom-up formation pathways for very small PAHs include ion-molecule and
neutral-neutral reactions in the gas phase, as well as reactions on dust grain
surfaces \citep{2011PNAS..108..452J}. Top-down formation under protoplanetary
disk conditions may also occur via UV photo processing
\citep{2023A&A...674A.200L} or by plasma processes that produce complex
organic molecules (COMs) and preferentially generate the 4-ring PAH
C$_{16}$H$_{10}$ \citep{2020ApJ...889..101G}. The heavily super-hydrogenated
4-ring H$_{n}$-PAH C$_{16}$H$_{16}^{+}$ (also designated
H$_{6}$-C$_{16}$H$_{10}^{+}$ or H$_{6}$-pyrene$^{+}$, uid=362) is in the
best-fit PAH model for comet 81P. Benzene was found by the JWST MINDS survey
around the M-dwarf star 2MASS-J16053215-1933159 \citep{2023NatAs...7..805T},
providing additional evidence for small aromatic molecules in protoplanetary
disk environments. Whether formed by bottom-up or top-down processes, the very
small and small PAHs in the comet appear associated with a population tied to
the protoplanetary disk.

\subsubsection{H$_{n}$-PAH origins}
Whatever their fate in the coma, the H$_{n}$-PAHs had to survive the interval
between their formation and their assembly into the comet. The H$_{n}$-PAHs
are unlikely to have reached the comet through the gas phase of the DISM or
the ISM. Compared to regular PAHs, H$_{n}$-PAHs are more susceptible to UV
destruction because of their enhanced photoabsorption cross sections, and the
gas-phase lifetime of H$_{6}$-pyrene$^{+}$ (C$_{16}$H$_{16}^{+}$, uid=362), a
species in the best-fit model, is shorter than that of pyrene
\citep{2021A&A...652A..42M}. Their survival to the epoch of comet formation
requires that they were shielded, and ice is the matrix in which they form.
Laboratory irradiation of PAHs in water ice with VUV photons produces
hydrogenated and oxygenated PAH photoproducts \citep{2003ApJ...596L.195G,
2010A&A...511A..33B, 2011EAS....46..251B, 2011A&A...529A..46B,
2011A&A...525A..93B, 2012ApJ...756L..24G, 2015ApJ...799...14C}.

Comet 67P/Churyumov-Gerasimenko (67P), studied in depth over the two-year
Rosetta rendezvous, contains H$_{n}$-PAHs spanning the full range of
hydrogenation states within the mass range of the ROSINA instrument
\citep{2022NatCo..13.3639H}. Comet 81P shares this subpopulation and, as shown
below (Figure~\ref{fig:figure-11}), extends it to larger sizes. These points
together suggest that the H$_{n}$-PAHs formed after PAHs were embedded in ice
and remained within the ice to be assembled into the comet.

\subsubsection{Hydrogenated PAHs: Comets 81P versus 67P} 
\begin{figure*}[ht!]
\figurenum{11}
\begin{center}
\includegraphics[trim=0.07cm 0.05cm 0.07cm 0.05cm, clip, width=0.750\textwidth]{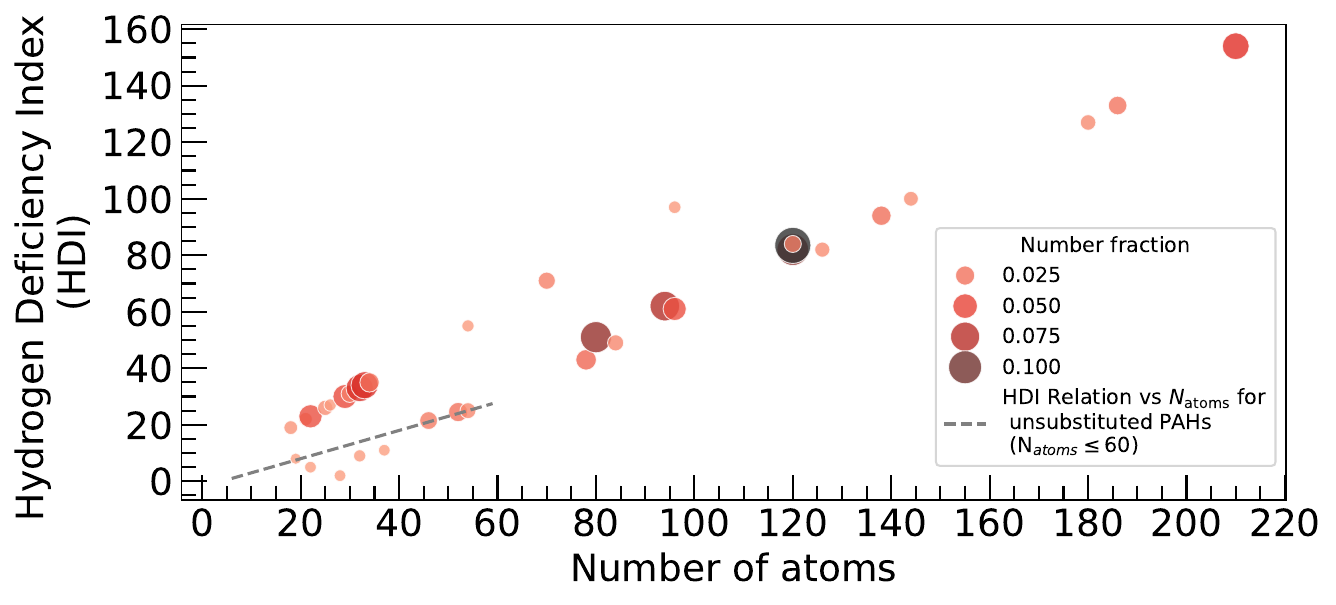}
\includegraphics[trim=0.07cm 0.05cm 0.07cm 0.05cm, clip, width=0.750\textwidth]{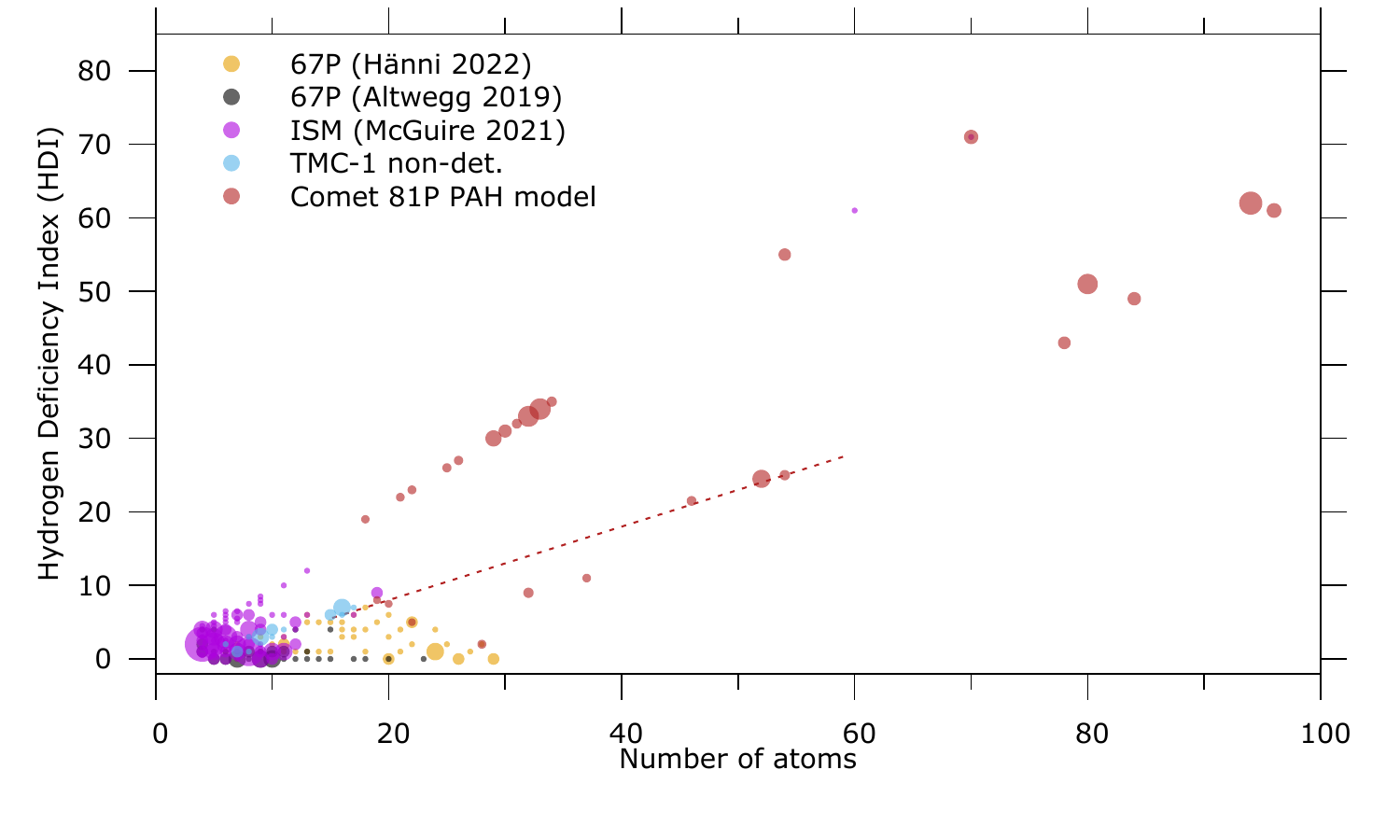}
\caption{Hydrogen Deficiency Index (HDI) versus number of atoms ($N_{\rm
atoms}$). (upper)~HDI vs $N_{\rm atoms}$ for 81P's best-fit PAH model. The
dashed line traces the HDI relation for regular PAHs, $\mathrm{HDI} = 0.75H -
0.5$, plotted for $N_{\rm atoms} \leq 60$; molecules above this line are
dehydrogenated. (lower) Comparison of HDI vs $N_{\rm atoms}$ for comet 81P
with other sources. All symbols are semi-transparent filled circles whose size
traces the number fraction of molecules in the comet 81P PAH model and, for
the other sources, the number of detections. Plotted are: comet 81P's best-fit
PAH model (firebrick); ROSINA detections in comet 67P/Churyumov-Gerasimenko
from \citet{2022NatCo..13.3639H} (orange) and \citet{2019ARA&A..57..113A}
(black); ISM molecules from rotational spectroscopy by
\citet{2021Sci...371.1265M} (darkviolet); and TMC-1 non-detections from
\citet{2022JPCA..126.2716B} (skyblue). Fullerene-C$_{70}$ lies above the
dashed line because it has no peripheral H atoms. Axes are truncated at
$N_{\rm atoms} \leq 100$ ($N_{\rm C} \leq 40$) to make the overlap among the
sources visible (see Section~\ref{sec:pah:origins:disk}). Adapted from
\citet{2022NatCo..13.3639H}, using their data and their data compilations.}
\label{fig:figure-11}
\end{center}
\end{figure*}

Within ROSINA's mass range sensitivity, comet 67P's H$_{n}$-PAHs
spanned the full range of hydrogenation states up to 2-ring PAHs
\citep{2022NatCo..13.3639H}, from regular hydrogenation to complete
super-hydrogenation. Figure~\ref{fig:figure-11} shows the Hydrogen Deficiency
Index (HDI) versus number of atoms for comet 81P's best-fit PAH model. The HDI
is defined as $\mathrm{HDI} = {\rm C} + 1 + {\rm N}/2 - {\rm H}/2 - {\rm
X}/2$, where C, N, H, and X count the carbon, nitrogen, hydrogen, and halogen
atoms; X$=0$ for the species in the PAH model. A lower HDI signals more
peripheral H per heavy atom, since each ring closure and each multiple C--C
bond consumes a peripheral site. PAHs with C$\leq$54 are super-hydrogenated
H$_{n}$-PAHs or are fully dehydrogenated DPAHs. The HDI relation for regular
PAHs, ${\rm HDI} = 0.75{\rm H} - 0.5$, is shown by the dashed line for N$_{\rm
atoms} \leq 60$ as a guide for delineating DPAHs above that line. The
Fullerene-C70 (C$_{70}$) lies above the line because it also is devoid of
peripheral hydrogen atoms.

Plotted alongside the comet 81P model are ROSINA detections in comet 67P
\citep{2022NatCo..13.3639H, 2019ARA&A..57..113A}, ISM detections from
rotational spectroscopy \citep{2021Sci...371.1265M}, and TMC-1 non-detections
\citep{2022JPCA..126.2716B}; $N_{\rm atoms}$ is truncated at $\leq 100$
($N_{\rm C} \leq 40$) so the overlap among the sources is visible. ROSINA's
67P PAH detections do not extend beyond fully super-hydrogenated naphthalene
(C$_{10}$H$_{18}$, uid=353), because PAHs heavier than that produce fragments
without sufficiently distinctive signatures for confident identification in
the ROSINA data. For comet 81P's non-DPAHs, the HDI--$N_{\rm atoms}$ relation
traces the same super-hydrogenated locus that ROSINA identified in comet 67P
but reaches to larger $N_{\rm atoms}$, implying that JWST finds H$_{n}$-PAHs
of greater size than ROSINA's mass range admitted.

\subsubsection{Aromatics in asteroid Bennu versus comet 81P} N-heterocycles
(PANHs) and PAHs with peripheral N bonds (PAHNs) as well as two- through
four-ring PAHs were found in the asteroid Bennu samples. The high nitrogen
content of the PAHs, which are in the soluble organic matter (SOM) that has
large $^{15}$N-enrichments, implies they formed at low temperatures. Origins
in the outer cold protoplanetary disk or potentially through low-temperature
abiotic aqueous reactions in N-rich alkaline fluids on the parent body are
posited \citep{2025NatAs...9..199G}. Nevertheless, the enhanced carbon
isotopic ratios $^{13}$C/$^{12}$C, specifically the clumped enhancements of
$^{13}$C, within the two- through four-ring PAHs suggest that half formed at
extremely cold temperatures ($\sim 10-20$~K) and represent pristine
interstellar cold-cloud materials that were inherited and survived in Bennu
\citep{2023Sci...382.1411Z}.


In the best-fit PAH model for comet 81P, one very small PAHN is retained and
no very small PANH is. Some others appear in fewer than 160 of the 1000 MC
trials in the pre-truncation pool (Section~\ref{sec:pah:panh}). Their absence
from the model is not evidence against their presence. It does indicate that
their number fractions are low compared to the other PAH species contributing
to the modeled flux, too low for their bands to be distinguished in the summed
emission.

Laboratory mass spectrometry of Stardust samples detects N-bearing PAHs with
peripheral nitriles (Section~\ref{sec:pah:stardust}). Very small
nitrile-bearing PAHNs are present in the Stardust samples of comet 81P. More
is known about the origins of the N-bearing PAHs in Bennu, whose PANHs and
PAHNs, measured in the returned samples, carry signatures of their origins in
their C-isotope ratios and N concentrations.



\subsection{PAHs in the coma: lifetimes and charge}
\label{sec:pah:origins:coma}


\subsubsection{Coma lifetimes of very small and small PAHs}
\label{sec:pah:origins:lifetimes}
The detection of very small and small PAHs in the gas phase may bear on how
they are delivered to the coma, from the nucleus or from coma ice or coma dust
grains, because of their photodestruction lifetimes. At a coma
expansion speed of $\sim$0.5~km~s$^{-1}$, a PAH released from the nucleus would need
to survive 1.46$\times$10$^{3}$~s (0.41~hr) to reach the edge of the
0.705\arcsec{} radius extraction aperture, 731~km at the JWST--comet distance
of $\Delta_{\rm JWST}=1.43$~au \citep[Paper~I,][Table~1]{2026arXiv260802190R}.
Photodestruction lifetimes $\tau_{\rm des}$ 
are a steep function of the effective radius $a$, i.e, the radius of an equivalent sphere 
of C atoms. At the 1.85~au distance of comet 81P, for $a=3.5$\AA{} (N$_{\rm C}$=20) 
$\tau_{\rm des}\approx$~50~yr, and for $a=3.71$\AA{} for N$_{\rm C}$=24, e.g., 
C$_{24}$H$_{12}$ (Coronene), $\tau_{\rm des}\approx$~1370~yr, taking T$_\star$=6000~K and 1L$_\odot$ and scaling using $\tau_{\rm des}\propto r^2/L_\odot$ from 
\citep[T$_\star$=6000~K, 10L$_\odot$, r$=10$~au, Figure~16,][]{2017ApJ...835..291S}. If the comet was within 
$\sim$0.4~au of the Sun, which was not the case for 81P, then two photons could have been absorbed 
within the cooling time and then photodestruction lifetimes for these PAH sizes would 
have been shorter than the coma crossing time \citep{1997P&SS...45.1539J}.
Very small and small PAHs 
survive in the coma of 81P.

\subsubsection{Ionization state}
\label{sec:pah:origins:ionization}
The best-fit model is dominated by cations. Excluding the DPAHs, for which the
PAHdb V4.00 database contains only seven charged species, the ionization
fractions by number are 1.0 for the very small and small unsubstituted PAHs,
0.76 for the H$_{n}$-PAHs, 0.67 for the large PAHs, and 0.63 for all PAHs
other than the DPAHs (Table~\ref{tab:bestfit-fractions}). The charge state of
each species is set by the fit to the full spectrum, and the reasons the fit
prefers cations are not captured by a single measure such as the NIR/MIR flux
ratio (Section~\ref{sec:pah:3pop}). Two sources of the charge can be
considered, the ice in which the PAHs were stored and the coma into which they
were released.

The same irradiation that forms H$_{n}$-PAHs ionizes PAHs in water ice
efficiently \citep{2003ApJ...596L.195G, 2010A&A...511A..33B,
2012ApJ...756L..24G, 2014A&A...562A..22C, 2015ApJ...799...14C}. However, the
cations neutralize as the ice warms and crystallizes above about 120~K
\citep{2006ApJ...638..286G, 2007ApJ...664.1264B}, well below the temperature
of subliming ice. The cations in the coma are therefore not the cations made
and stored in the ice.

If not from the ice, the cations would have to be produced in the coma, 
and that is not likely either. Photoionization timescales of $0.7$--$1.7\times10^{5}$~s
are predicted for PAHs of 14 to 32 carbon atoms (Table~\ref{tab:tau-ion}), which are longer than the $1.46\times10^{3}$~s needed to transit the JWST aperture (
v$_{exp}\approx 0.5~km~s^{-1}$). 
Photoionization timescales fall to $3.0\times10^{3}$~s at $N_{\rm C}=172$. 
PAHs released neutral from the nucleus leave the aperture at least 80\% neutral for $N_{\rm C}\leq 120$. 

The gas-phase PAH photoionization timescales are computed from
Equation~(A7) of \citet{2003ApJ...594..987L}, integrated from
$\lambda_{\rm max}=hc/{\rm IP}(a)$ down to 912~\AA{}, where the
ionization potential ${\rm IP}(a)$ depends on the effective radius
$a$. The integrand is the product $\{{\rm Y}_{\rm ion}(E)\,
{\rm C}_{\rm abs}^{\rm PAH}(\lambda)\,F_{\lambda,\odot}\,
(hc/\lambda)^{-1}\}$, where ${\rm C}_{\rm abs}^{\rm PAH}(\lambda)$ is
$N_{\rm C}$ times the absorption cross section per C atom. The cross
sections per C atom are from Equations~(5)--(10) of 
\citet{2001ApJ...554..778L}, adopted unchanged in the ultraviolet by
\citet{2007ApJ...657..810D}, and in this regime ($\lambda <
3030$~\AA{}) they are independent of size and of charge state. PAH
size enters through the factor $N_{\rm C}$ in 
${\rm C}_{\rm abs}^{\rm PAH}(\lambda)$ and through the ionization
potential. We adopt the piecewise linear formulation of
${\rm Y}_{\rm ion}(E)$ from Equation~(A5) of 
\citet{2003ApJ...594..987L}. 

We adopt the Whole Heliosphere Interval (WHI) 
2008\footnote{\url{https://lasp.colorado.edu/lisird/resources/whi_ref_spectra/data/ref_solar_irradiance_whi-2008_ver2.dat}} \citep{2009GeoRL..36.1101W} 
quiet-Sun spectrum, which spans the full integration range 
at 0.1~nm resolution and whose Ly$\alpha$ photon flux integrated over 1205–-1227\AA{} 
($3.673\times10^{11}$~photons~cm$^{-2}$~s$^{-1}$) agrees with 
\citet{2019E&SS....6.2263M} to 0.7\% ($3.7\times10^{11}$~photons~cm$^{-2}$~s$^{-1}$). 
WHI also covers 912--1195~\AA{}
(0.635~mW~m$^{-2}$ at 1~au), which ASTM~E490-00 (AM0) does not \citep{ASTME490_00}.
The ratio of WHI to E490 as tabulated is 0.90 and 0.96 for, respectively,
1195--1700~\AA{} and 1700--2224~\AA{}, and 1.19 within the 10~\AA{} bin
holding Lyman-$\alpha$. The additional short-wavelength coverage of WHI
increases $k_{\rm ion}$ and the lower long-wavelength irradiance lowers it,
and the second dominates, so the WHI-calculated $\tau_{\rm ion}$
are 5--7\% longer than E490-based estimates. 

\begin{deluxetable}{rcccccc}
\tablecaption{Photoionization timescales of neutral gas-phase PAHs in sunlight at 1.85~au\label{tab:tau-ion}}
\tablehead{
\colhead{$N_{\rm C}$} & \colhead{$a$\tablenotemark{a}} & \colhead{IP\tablenotemark{b}} & \colhead{$\lambda_{\rm max}$\tablenotemark{c}} & \colhead{$\tau_{\rm ion}$} & \colhead{$t_{\rm cross}/\tau_{\rm ion}$} & \colhead{$f_{\rm neutral}$\tablenotemark{d}} \\
\colhead{} & \colhead{(\AA)} & \colhead{(eV)} & \colhead{(\AA)} & \colhead{(s)} & \colhead{} & \colhead{}
}
\startdata
14 & 3.10 & 7.44 & 1666 & $1.8\times10^{5}$ & 0.008 & 0.99 \\
24 & 3.71 & 7.29 & 1701 & $1.0\times10^{5}$ & 0.014 & 0.99 \\
32 & 4.08 & 6.71 & 1848 & $7.2\times10^{4}$ & 0.020 & 0.98 \\
54 & 4.86 & 6.25 & 1985 & $3.4\times10^{4}$ & 0.043 & 0.96 \\
72 & 5.35 & 6.05 & 2051 & $2.1\times10^{4}$ & 0.070 & 0.93 \\
94 & 5.85 & 5.88 & 2108 & $1.3\times10^{4}$ & 0.116 & 0.89 \\
172 & 7.15 & 5.57 & 2224 & $3.2\times10^{3}$ & 0.463 & 0.63 \\
ice\tablenotemark{e} & \nodata & \nodata & \nodata & $1.0\times10^{6}$ & 0.0014 & 1.00 \\
\enddata
\tablecomments{Neutral PAHs. For the gas-phase entries, $\tau_{\rm ion}=1/k_{\rm ion}$, 
with $k_{\rm ion}$ from Equation~(A7) of \citet{2003ApJ...594..987L}. 
$t_{\rm cross}=1.46\times10^{3}$~s is the JWST 0\farcs705 (731~km) aperture crossing time at an expansion speed of 0.5~km~s$^{-1}$.}
\tablenotetext{a}{$a$ is the effective radius of a sphere containing $N_{\rm C}$ carbon atoms at the density of ideal graphite, $\rho = 2.24$~g~cm$^{-3}$, $a = 1.286\,N_{\rm C}^{1/3}$~\AA.}
\tablenotetext{b}{Ionization potential: laboratory values for $N_{\rm C}=14$ (anthracene, 7.44~eV), 
24 (coronene, 7.29~eV) and 32 (ovalene, 6.71~eV), from the compilation of  \citet{nist_chemistry_webbook}; Equation~(A6) of \citet{2003ApJ...594..987L} otherwise.}
\tablenotetext{c}{Long-wavelength limit of the integration, $hc/{\rm IP}$.}
\tablenotetext{d}{Fraction still neutral on leaving the aperture, $\exp(-t_{\rm cross}/\tau_{\rm ion})$, for PAHs released neutral at the nucleus.}
\tablenotetext{e}{PAHs in H$_{2}$O ice.
$\tau_{\rm ion, ice}^{-1}\approx \{\sigma_{\rm abs}{\rm Y}\}~(r/1~{\rm au})^{-2}~F({\rm Ly}\alpha,~1~{\rm au})$,
with $r=1.85$~au and
$F({\rm Ly}\alpha,~1~{\rm au})=3.7\times10^{11}$~photons~cm$^{-2}$~s$^{-1}$
(Machol et al. 2019). The rate constant
$\{\sigma_{\rm abs}{\rm Y}\} = 9.0\times10^{-18}$~cm$^{2}$~photon$^{-1}$ was
measured for anthracene through coronene by \citet{2011EAS....46..251B}.}
\end{deluxetable}

If only the Ly$\alpha$ irradiance were considered, the gas-phase PAH
photoionization timescales would be similar for the smallest PAHs but
too long for the larger ones. The fraction of the photoionization rate
carried by Ly$\alpha$ declines steadily with increasing size, from
72\% at $N_{\rm C}=14$ to 16\% at $N_{\rm C}=172$, as
$\lambda_{\rm max}$ moves to longer wavelengths (Table~\ref{tab:tau-ion}).

The ionization rate measured for PAHs in 
water ice is slower than the gas-phase estimate. The average PAH photoionization rate
constant of anthracene through coronene in water ice, is $\sigma_{\rm ion, ice}{\rm Y}=
9\times10^{-18}~{\rm cm}^{2}~{\rm photon}^{-1}$ \citep{2011EAS....46..251B}.
When combined with the Ly$\alpha$ flux at 1.85~au of $1.081\times10^{11}$~photons~cm$^{-2}$~s$^{-1}$ 
\citep{2019E&SS....6.2263M}, the photoionization timescale
for these PAHs in water ice gives $10^{6}$~s, six to ten times longer
than the gas-phase photoioniztion estimate. 

For comet 81P, photoionization is too slow to produce the cation fractions of
the PAH model, and the laboratory experiments show that the cations made in
ice neutralize before the ice sublimates. Neither route accounts for the
ionization fraction of the coma PAHs that the observations reveal and the PAH
model quantifies, and it is not yet understood. Modeling how coma PAHs obtain
their charge may resolve this question but is beyond the scope of this paper.

\subsection{PAHs in the DISM}
\label{sec:pah:origins:dism}

Both H$_{n}$-PAHs and Me-PAHs are aliphatic PAHs that contribute to the
3.4~\micron{} feature found in some astronomical sources. This association has
stimulated laboratory and theoretical study of PAH evolution in the DISM and
ISM. In PDRs, however, neither H$_{n}$-PAHs nor Me-PAHs are likely to survive,
as erosion of C--H bonds follows absorption of the first VUV photon almost
immediately \citep{2021A&A...652A..42M}. H$_{n}$-PAHs are more prone to
fragmentation than Me-PAHs, and fragmentation of Me-PAHs may itself produce
H$_{n}$-PAHs. The gas-phase lifetime of H$_{6}$-Pyr$^{+}$
(C$_{16}$H$_{16}^{+}$) is shorter than that of pyrene (C$_{16}$H$_{10}$)
because VUV photoabsorption leads to some destruction of the carbon backbone
(transforming hexagonal rings to pentagonal rings) and because of its higher
photoabsorption cross section \citep{2021A&A...652A..42M}. PAHs containing
pentagonal ring structures may be abundant in PDRs \citep{2021A&A...652A..42M,
2016ApJ...826...33P, 2017PCCP...19.2974D}.

Smaller PAHs with 30--50 carbon atoms likely cannot survive in the ISM and
DISM as fully hydrogenated PAHs. They become strongly or completely
dehydrogenated (DPAHs) \citep{2000A&A...363L...5V, 2013A&A...552A..15M,
2016A&A...595A..23A, 2025A&A...696A.180O}. H-loss is the fastest process by
which small PAHs evolve in the DISM \citep{2025A&A...696A.180O}. Continued UV
irradiation under DISM conditions may further evolve DPAHs into carbon
`necklaces,' carbon clusters \citep{2015ApJ...799..131M} that are structurally
distinct from DPAHs with hexagonal or pentagonal rings. In NGC~7023's cavity
toward the central source, where density is lower and radiation is harsher
(\textit{A}$_{V}$ $\lesssim$ 2), all PAHs are predicted to be completely
dehydrogenated \citep{2013A&A...552A..15M, 2015ApJ...799..131M}. The formation
of carbon clusters, including C$_{24}^{0/+}$, is strongly suggested by
theoretical simulations that include multi-photon events
\citep{2013A&A...552A..15M}. In NGC7023, the high-radiation cavity
(\textit{A}$_{V}$ $\lesssim$ 2) hosts large PAHs and relatively small carbon
clusters, conditions that potentially promote C$_{60}$ formation
\citep{2012PNAS..109..401B}.

Carbon clusters take various forms including non-planar bowls and cages. The
most stable forms of C$_{20}$ and C$_{24}$ are the bowl and cage, respectively
\citep{doi:10.1021/acs.jpca.5b10266}. Full dehydrogenation of small PAHs in
the ISM and DISM can yield carbon clusters or graphene-like flakes (carbon
nano-flakes) containing some pentagonal rings; those pentagonal rings result
in more energetically stable, non-planar bowl shapes. Dehydrogenation renders
the molecules highly reactive \citep{2012PNAS..109..401B}. Laboratory
experiments starting from a graphene sheet produce a bowl-shaped DPAH that
subsequently becomes fullerene \citep{2010NatCh...2..450C}, though the DPAHs
in those experiments are larger than those assessed for comet 81P. Theoretical
calculations of PAH H-photolysis and isomerization show that C$_{n}$ with
\textit{n} $\simeq 35$--50, when heated to 1000~K, forms convex or conical
structures as well as discrete short chains. Such large convex or conical
DPAHs can then incorporate gas-phase C$^{+}$ to form the stable cage C$_{60}$
\citep{2021A&A...650A.193O}.

Spatially resolved observations of NGC7023 identified a weak 10.6~\micron{}
feature as the best tracer of DPAHs because it shows systematic variation with
distance from the central star, decreasing toward the ionizing source while
fullerenes are concentrated in the central high-ionization zone. This spatial
pattern implies that the presence of DPAHs leads to fullerene formation
\citep{2015ApJ...799..131M}. Comet 81P has a high relative number of DPAHs
with a wide variety of shapes. The average number of pentagons per DPAH is
1.3, and most pentagons are not on the periphery.

An alternative to inheritance from the DISM is dehydrogenation in the coma
itself, where the modeled lifetimes (Section~\ref{sec:pah:origins:coma}, Coma
lifetimes) imply H loss within seconds to minutes. Whether sequential H loss
could strip the dozen or more hydrogens off a small PAH before its carbon
skeleton fragments, while the very small H$_{n}$-PAHs in the same aperture
retain theirs, has not been calculated, so the coma cannot be excluded as the
site where some of the small DPAHs formed.

This DPAH population may represent the first high-contrast IR spectroscopic
evidence of DPAHs in an astronomical source, molecules hypothesized to be a
dominant component of the DISM that, until now, had not been revealed by a
strong and dominant IR spectroscopic signature.

\subsection{Large PAHs in the ISM, in PDRs}
\label{sec:pah:origins:pdr}

DPAHs are the probable intermediate species between large PAHs and fullerenes.
For large PAHs, H$_{2}$ formation via loss of peripheral H atoms becomes the
dominant fragmentation channel, making these large PAHs precursors to both
fullerenes and nano-grains \citep{2018A&A...616A.167C}. Partial
dehydrogenation creates dangling bonds that promote accretion of C atoms and
small radicals from the gas phase, driving the continued chemical evolution of
these molecules. Compact structures without bays are more stable for large
PAHs.

Comet 81P's largest medium-size PAHs and its large PAHs have structures
similar to those of comparable size found in fitting the JWST
spectrum of the PDR in the Orion Bar with large PAHs
\citep{2024ApJ...968..128R}. Those structures are classified as zigzag,
armchair, and edge defect, where edge-defect structures have bays. For comet
81P, the fractions of integrated flux contributed by armchair, zigzag, and
edge-defect (C $\gtsimeq 50$) PAHs are 28\%, 19\%, and 52\%, respectively,
computed in Jy for direct comparison with \citet{2024ApJ...968..128R}. For the
Orion Bar, the corresponding fractions are 15\%, 21\%, and 64\%. The
integrated flux fractions are broadly similar, with 81P showing a higher
armchair fraction. Large PAHs with fewer bays are more stable, and if they
survive in the Orion Bar, they can be transported to our protoplanetary disk.

Lastly, beyond the sizes represented in the PAHdb, the aromatic-bond carriers
continue as carbon nanograins. The 14.5~\micron{} complex in comet 81P, which
was deliberately underweighted in the PAH model runs
(Section~\ref{sec:pah:method}), finds plausible carriers in MAONs and in
carbon nanograins shaped like concentric buckyballs
(Section~\ref{sec:pah:nanograin}). These candidates extend the size
distribution of aromatic carriers upward from the large PAHs discussed here,
consistent with survival of the largest species in ISM and PDR conditions and
with their subsequent transport to the protoplanetary disk.

\subsection{Comparison with the Stardust samples}
\label{sec:pah:stardust}

PAHs in Stardust samples from comet 81P were measured by 2-step laser
ionization mass spectrometry with a $\sim$5~\micron{} IR laser spot size
\citep{2010M&PS...45..701C}. The samples measured include two entire Stardust
tracks, impact craters on foils, and terminal particles. This technique
measures intact molecules with minimal fragmentation, is orders of magnitude
more sensitive than other mass spectrometry or spectroscopy techniques, and
identifies specific molecules rather than the functional groups or molecular
classes sensed by IR, Raman, XANES, and EELS.


Aromatic molecules were detected over a mass envelope of approximately
100--350~amu, with possible assignments given for parent species up to 244~amu
\citep[Table~2,][]{2010M&PS...45..701C}. A weaker secondary envelope extends
to approximately 700~amu in one track and is attributed to free-radical
polymerization induced by impact heating rather than to indigenous cometary
material. The simple PAHs observed include naphthalene, phenanthrene, pyrene,
chrysene, and benzopyrene, along with their alkylation series
\citep[Table~2,][]{2010M&PS...45..701C}. Alkylation series appear as peaks
separated by 14~AMU, where each step corresponds to replacement of a
peripheral H atom by a --CH$_{3}$ (methyl) functional group (Me-PAHs); Me-PAHs
appear to dominate. A separate group of odd-mass PAHs that are not simple
hydrocarbons is also identified. These are interpreted as N-bearing PAHs
(PAHNs), where the N is not in the aromatic ring but is bound as aromatic
nitriles at peripheral sites \citep{2010M&PS...45..701C}. The PAHs measured in
the Stardust samples had similarities with meteoritic PAHs but were more
similar to those in interplanetary dust particles (IDPs). The peripheral-N
PAHNs are one of these observations that is more similar to IDPs than to
meteorites \citep{2010M&PS...45..701C}.

The detection of pyrene and its Me-PAH series as major species in Stardust
samples strongly supports the inclusion of small PAHs in the best-fit PAH
model for comet 81P (Section~\ref{sec:pah:bestfit}) and the importance of
small PAHs for tracing a PAH subpopulation that cannot survive in the DISM
(Section~\ref{sec:pah:origins:dism}). However, the best-fit PAH model for
comet 81P lacks the methylated aromatic PAHs that dominate the Stardust
alkylation series. Its one species returned by both the CH$_{2}$ and CH$_{3}$
searches of the database, C$_{10}$H$_{18}^{+2}$ (uid=4396,
Section~\ref{sec:pah:mepah:models}), is a saturated ring with an alkyl side
chain rather than a methylated aromatic. The model instead has H$_{n}$-PAHs of
similar small sizes. Experimental data show that di-Methyl-Anthracene (DMA)
has features in the region of comet 81P's 3.37~\micron{} feature and
highlights the lack of anharmonicity in the theoretical computations currently
available in the PAHdb V4.00 database (Section~\ref{sec:pah:mepah},
Figure~\ref{fig:figure-3}). This current computational deficiency may
contribute to a modeling bias that disfavors Me-PAHs. Future access to
anharmonicity calculations may resolve the apparent discrepancy between the
comet 81P PAH model and the mass spectrometry results for the Stardust
samples. The peripheral-N PAHNs are the PAHN class that was searched for in
the best-fit PAH model. One very small peripheral-N PAHN is retained, at the
level of a tenth of a percent of the modeled NIRSpec flux
(Section~\ref{sec:pah:panh}). That species, 1-naphthylamine (uid=476), carries
an amino group rather than a nitrile. Nitrile-bearing PAHs, the class that
dominates the Stardust odd-mass peaks, were included in the search and none
was retained (Section~\ref{sec:pah:panh}). These PAHNs are invoked for
prenatal-cloud and outer-disk chemistry (Section~\ref{sec:pah:origins:disk}).


\section{Conclusions}
\label{sec:pah:summary}

The infrared emission of comet 81P's coma over 3.2--3.6~\micron{} and
5.5--27~\micron{} is reproduced by a single population of PAHs, modeled with
the PAHdb PythonSuite and the PAHdb V4.00 database under solar excitation, in
which the same set of molecules produces both the NIR and the MIR flux. The
best-fit model retains 57 species (Appendix~\ref{sec:appendix2}) and is built
from three subpopulations that each dominate a region of the spectrum,
together with unsubstituted PAHs of very small to medium size. 
The photodissociation lifetimes of gas-phase PAHs in the coma exceed the
aperture crossing time by at least six orders of magnitude for
$N_{\rm C}\geq 20$. The photoionization lifetimes are long enough that PAHs
released neutral leave the aperture at least 80\% neutral for
$N_{\rm C}\leq 120$ (Table~\ref{tab:tau-ion}). The spectrum therefore
represents the PAH reservoir of the nucleus rather than a population ionized
in the coma. Its main findings are these.

\begin{itemize}

\item The 3.37~\micron{} feature is carried by very small ($N_{\rm{C}} \leq
19$) H$_{n}$-PAHs, in which at least 35\% of the peripheral bonding sites
carry a CH$_{2}$ group. These species produce the 3.37 and 3.53~\micron{}
features and little or no 3.3~\micron{} emission, which is why the comet shows
a weak 3.37~\micron{} feature and no 3.3~\micron{} feature. Their identity is
robust. When the MIR model is rebuilt without DPAHs, the same very small
H$_{n}$-PAHs still carry the 3.37~\micron{} feature
(Appendix~\ref{sec:appendix1-nodpah}). Too fragile to have crossed the DISM or
ISM in the gas phase, they were shielded in the ice in which they form, and
they share this subpopulation with comet 67P, where ROSINA found the full
hydrogenation sequence (Section~\ref{sec:pah:origins:disk}). Methyl-PAHs
cannot account for the feature with the harmonic spectra now available,
although they cannot be ruled out until anharmonic calculations cover a wider
range of Me-PAH sizes and structures (Section~\ref{sec:pah:mepah}).

\item The 6.9~\micron{} complex (6.7--7.6~\micron{}) is dominated by
dehydrogenated PAHs (DPAHs, H=0), mostly small ($20 < N_{\rm{C}} \leq 50$),
which contribute 54\% of the modeled flux density in that range and 0.32 of
the integrated flux density overall (Table~\ref{tab:bestfit-fractions}).
Excluding DPAHs worsens the fit by $\Delta$AIC $= -9504.7$ 
(Appendix~\ref{sec:appendix1-nodpah}, Table~\ref{tab:model-comparison}), four
thousand times the run-to-run fluctuation in AIC, i.e., the Monte Carlo uncertainty that arises because each run uses a different draw of random numbers  (Appendix~\ref{sec:appendix1-aic}). No other
component of the model is as strongly required. Formation of the DPAHs in the
coma has not been excluded but would require loss of a dozen or more
peripheral C--H bonds within the aperture
(Section~\ref{sec:pah:origins:dism}). This DPAH population may be the first
high-contrast infrared spectroscopic evidence of DPAHs in an astronomical
source.

\item The 8.4--11~\micron{} emission, including the 9.1~\micron{} complex, the
9.76~\micron{} shoulder, and the 11.0~\micron{} feature, is carried by large
PAHs ($N_{\rm{C}} > 70$), with medium PAHs ($50 < N_{\rm{C}} \leq 70$)
supplying the narrow 9.05~\micron{} peak. Nitrogen enters the model only as
skeletal N in large PANHs with $93 \leq N_{\rm{C}} \leq 95$.

\item Very small N-bearing PAHs are effectively absent. Of the PANHs and PAHNs
searched, only the amine cation 1-C$_{10}$H$_{9}$N$^{+}$ is retained, and none
of the nitrile-bearing species is retained, even though nitrile PAHNs dominate
the odd-mass peaks of the Stardust samples (Sections~\ref{sec:pah:panh} and
\ref{sec:pah:stardust}). Their absence from the model is not evidence against
their presence, but their number fractions must be low relative to the species
that carry the modeled flux.

\item The PAH population records three astrophysical environments
(Figure~\ref{fig:figure-9}). The very small H$_{n}$-PAHs, the very small and
small unsubstituted PAHs, and C$_{10}$H$_{18}^{+2}$ would not survive the DISM
or the ISM and point to the prenatal cloud or the protoplanetary disk, with
the H$_{n}$-PAHs formed by UV irradiation of PAHs in ice. Their hydrogen
deficiency index follows the super-hydrogenated locus that ROSINA measured in
comet 67P and extends it to larger sizes (Figure~\ref{fig:figure-10}). The
DPAHs, including C$_{70}$, are the dehydrogenated intermediates expected in
the DISM en route to fullerenes and carbon clusters. The medium and large
PAHs, with their PANHs, are the survivors of PDR conditions in the ISM. MAONs
and carbon nanograins, which are candidates for the unexplained 14.5~\micron{}
complex (12.5--16.5~\micron{}) in the comet, extend that population upward in
size (Section~\ref{sec:pah:nanograin}).

\end{itemize}

Two questions remain open.

\begin{itemize}

\item Compared with the Stardust mass spectrometry of the same comet, the
model agrees on the presence of very small and small PAHs of pyrene size but
lacks the methylated aromatics that dominate the Stardust alkylation series;
its one methyl-bearing species is an alkylated saturated ring. Whether this
reflects the comet or the absence of anharmonic spectra for Me-PAHs in the
current database is not settled.

\item PAH charge state in the coma is the other open question. Setting aside
the DPAHs, the model is dominated by cations, 0.63 by number. The laboratory
experiments imply that PAH cations made in ice neutralize before the ice
sublimates, and the estimated coma photoionization timescales imply there is
insufficient time for ionization for $N_{\rm C}\leq 120$ in the model
(Section~\ref{sec:pah:origins:coma}, Ionization state). Modeling how coma PAHs
obtain their charge may resolve this mystery but is beyond the scope of this
work.

\end{itemize}


\begin{acknowledgments}
This work is based on observations made with the NASA/ESA/CSA James Webb Space Telescope. The data were obtained from the Mikulski Archive for Space Telescopes at the Space Telescope Science Institute, which is operated by the Association of Universities for Research in Astronomy, Inc., under NASA contract NAS 5-03127 for JWST. These observations are associated with program \#1897. The specific observations analyzed can be accessed via doi:10.17909/py2d-2z26. Support for program \#1897 was provided by NASA through a grant from the Space Telescope Science Institute, which is operated by the Association of Universities for Research in Astronomy, Inc., under NASA contract NAS 5-03127. N.X.R, C.E.W, D.H.W. and M.S.P.K. acknowledge support from JWST-GO-01897. N.X.R., M.A.C., S.B.C., S.N.M., and P.A.G. were also supported by the NASA Planetary Science Division Internal Scientist Funding Model program through the Fundamental Laboratory Research work package (FLaRe). D.H.W. was also supported by the NASA Planetary Science Division Internal Scientist Funding Model through the Cold Solar System Objects (CSSO) ISFM. We gratefully acknowledge insightful discussions with Keiko Nakamura-Messenger, Don E. Brownlee, John P. Bradley, and Hope A. Ishii regarding the Stardust returned samples.
\end{acknowledgments}

\software{Astropy \citep{astropy:2013, astropy:2018, astropy:2022},
Astroquery \citep{Ginsburg2019}, jwstComet \citep{Roth2026b},
sbpy \citep{Mommert2019},
SciPy \citep{Virtanen2020}}

\appendix
\restartappendixnumbering
\twocolumngrid


\section{Model Selection and Model Comparison}
\label{sec:appendix1}

\subsection{The Akaike Information Criterion}
\label{sec:appendix1-aic}
The Akaike Information Criterion \citep[AIC,][]{doi:10.1177/0049124104268644,
2023PSJ.....4..242H} was used in the analysis of the best-fit PAH populations
models. For these analyses we formulate the AIC as:

\begin{equation}
AIC = -2 ln(\mathcal{L}) + 2k
\end{equation}

\noindent where
\begin{equation}
k = k_{\rm transitions} + k_{\rm search} + k_{\rm PAHs},
\end{equation}
\noindent with $k_{\rm transitions}=3$ (excitation temperature 5770~K,
wavelength shift, and FWHM), $k_{\rm search}$ the number of conditions in the
database search string, and $k_{\rm PAHs}$ the number of species retained in
the model.





%

For the best-fit model $k_{\rm transitions} = 3$, $k_{\rm search} = 14$, and
$k_{\rm PAHs} = 57$ with N$_{points,~\rm NIRSpec} =602$, N$_{points,~\rm MRS}
=9749$, and N$_{points} =10351.$
Hence:

\begin{equation}
\begin{split}
-2 ln(\mathcal{L}) &= \chi^{2} + 2\times \Sigma_{i}\text{ln}(\sigma_{i}) \\
 &\quad + N ln(2 \pi)
\end{split}
\label{eq:lnL}
\end{equation}

\noindent Four best-fit model
runs, each of 1000 trials, yield AIC values of $-946518.838$, $-946517.710$,
$-946517.048$ and $-946516.555$, a range of 2.28. We take this range as the run-to-run fluctuation in AIC, that is, the Monte Carlo uncertainty that arises because each run uses a different draw of random numbers. The best-fit model also yields a $\chi^2$(NIRSpec)=1257.8, $\chi^2$(MRS)=23398.9, $\chi^2 = 24656.7$, which translates to a reduced $\chi^{2}_{\nu} = \chi^{2}/(N_{\rm points}-k) = 2.399$. 

The reduced $\chi^{2}$ is approximately the average $\chi^{2}$ per point and barely separates models fitted to $10^{4}$ points. The AIC uses $\chi^{2}$, calculated over all points, and charges two units per free parameter, so the evidence ratio (relative likelihood) for one model over another accumulates across the spectrum. 
The three- and four-subpopulation models have the same combined $\chi^{2}_{\nu}$ to three
decimals, 2.661, and differ by 7.95 in AIC 
(Appendix~\ref{sec:appendix1-table}), 3.5 times the 2.28 range of AIC across four best-fit runs. For this reason the models are fitted to the data at
native spectral resolution rather than binned to raise the signal-to-noise
ratio per point, which would remove spectral information and add none. A model
fitted to fewer data points, or to binned data, may reach a lower
$\chi^{2}_{\nu}$, but the term $2\Sigma_{i}\ln(\sigma_{i})$ in
Equations~(\ref{eq:lnL}) and (\ref{eq:aic}), a sum of negative terms in the
flux units used here, loses a term for every point removed, and the AIC
becomes significantly less negative (worse). AIC values are therefore compared
only among models fitted to the same points, as all models in
Table~\ref{tab:model-comparison} are. Models that may succeed these need to
demonstrate lower AICs.


\begin{equation}
\begin{split}
AIC &= \chi^{2} + 2 \times \Sigma_{i}\text{ln}(\sigma_{i}) \\
 &\quad + N ln(2 \pi) + 2k
\end{split}
\label{eq:aic}
\end{equation}

\noindent With $2 \times \Sigma_{i}$ ln($\sigma_{i}$)$_{\rm NIRSpec}$ =
$-60267.2$, $2 \times \Sigma_{i}$ ln($\sigma_{i}$)$_{\rm MRS}$ = $-930078.0$,
and $2 \times \Sigma_{i}$ ln($\sigma_{i}$) = $-990345.1$,
Equation~(\ref{eq:aic}) gives AIC$_{\rm NIRSpec}$=$-57754.982$, AIC$_{\rm
MRS}$=$-888613.573$, and AIC$=-946516.555$. Note that the per-instrument AIC
values each carry the same penalty $2k$, so they do not sum to the combined
AIC.

\begin{deluxetable*}{lccccccccc}[t]
\tabletypesize{\footnotesize}
\tablecaption{Comparison of the PAH Models for Comet 81P\label{tab:model-comparison}}
\tablewidth{0pt}
\tablehead{
\colhead{} & \multicolumn{3}{c}{Free parameters} & \multicolumn{3}{c}{$\chi^{2}_{\nu}$} &
\multicolumn{3}{c}{$\Delta$AIC} \\
\colhead{Model} & \colhead{$k_{\rm search}$} & \colhead{$k_{\rm PAHs}$} & \colhead{$k$} &
\colhead{Total} & \colhead{NIRSpec} & \colhead{MRS} &
\colhead{Total} & \colhead{NIRSpec} & \colhead{MRS}
}
\startdata
\shortstack[l]{\rule{0pt}{3.0ex}Three-subpopulation:\tablenotemark{a,e}\\ H$_{n}$-PAHs, DPAHs, large} & 16 & 54 & 73 & 2.661 & 2.005 & 2.717 & $-2689.4$ & $+201.2$ & $-2888.6$ \\ \hline
\shortstack[l]{\rule{0pt}{3.0ex}Four-subpopulation:\tablenotemark{b,e}\\ H$_{n}$-PAHs, Me-PAHs, DPAHs, large} & 19 & 55 & 77 & 2.661 & 2.020 & 2.717 & $-2681.5$ & $+193.5$ & $-2881.0$ \\ \hline
No DPAHs\tablenotemark{c} & 14 & 47 & 64 & 3.323 & 3.376 & 3.342 & $-9504.7$ & $-538.7$ & $-8945.9$ \\ \hline
Best-fit (unrestricted search)\tablenotemark{d} & 14 & 57 & 74 & 2.399 & 2.382 & 2.418 & 0 & 0 & 0 \\
\enddata
\tablecomments{$\Delta$AIC $=$ AIC$_{\rm min}-$AIC, where AIC$_{\rm min}$ is the AIC of the
best-fit model, combined NIRSpec and MRS. The total number of free parameters is
$k = k_{\rm transitions} + k_{\rm search} + k_{\rm PAHs}$, where $k_{\rm transitions}=3$
(excitation temperature 5770~K, wavelength shift, and FWHM) for every model.
$\chi^{2}_{\nu} = \chi^{2}/(N_{\rm points}-k)$, with $N_{\rm points} = 602$ for NIRSpec,
9749 for MRS, and 10351 in total. The NIRSpec and MRS values each carry the full penalty
$2k$ and therefore do not sum to the combined value. AIC$_{\rm min}$ is $-946516.6$,
$-57755.0$, and $-888613.6$ for the instrument-combined, NIRSpec, and MRS, respectively.
Search strings are given in the notes as counted for $k_{\rm search}$, one condition per
comparison. Each string was preceded by \texttt{mg=0 o=0 fe=0 si=0} on its large-PAH
branch, excluding metal-bearing and oxygenated large PAHs; these four conditions are common
to every model, are not counted in $k_{\rm search}$, and cancel in every $\Delta$AIC.
Placeholders \texttt{uids\_...} stand for lists of uids defined in note (e).}
\tablenotetext{a}{\texttt{charge>=0 o=0 n=0 ch3=0 ch2=0 h=0 \&
uids\_ch2GT0\_ch2FracPeriphBonds\_GE\_0pt35\_cLE50 \& uids\_cGE70}}
\tablenotetext{b}{As (a) with \texttt{\& uids\_cLE50\_ch3GT0} inserted before
\texttt{uids\_cGE70}.}
\tablenotetext{c}{\texttt{c>=70 \& charge>=0 \& h>0 \& n>=0 | c<70 mg=0 fe=0 si=0
charge>=0 \& h>0 \& (ch2>0 | ch3>0 | n>=0 | o>=0)}}
\tablenotetext{d}{\texttt{c>=70 \& charge>=0 \& h>=0 \& n>=0 | c<70 mg=0 fe=0 si=0
charge>=0 \& (ch2>0 | ch3>0 | n>=0 | h>=0 | o>=0)}}
\tablenotetext{e}{\texttt{uids\_ch2GT0\_ch2FracPeriphBonds\_GE\_0pt35\_cLE50}, the
H$_{n}$-PAHs with \texttt{ch2>0}, CH$_{2}$ fraction of peripheral bonds $\geq 0.35$, and
N$_{\rm C} \leq 50$, counted as three conditions: uids 15, 16, 17, 332, 333, 336, 339, 350,
351, 352, 353, 355, 357, 362, 363, 584, 4376, 4377, 4380, 4383, 4394, 4395, 4396, 4397,
4399, 4401. \texttt{uids\_cLE50\_ch3GT0}, the Me-PAHs: \texttt{c<=50 charge>=0 ch3>0}.
\texttt{uids\_cGE70}, the large PAHs: \texttt{charge>=0 c>=70 o>=0 n>=0 h>=0 ch2>=0
ch3>=0}.}
\end{deluxetable*}

\subsection{Comparison of the Models}
\label{sec:appendix1-table}

Table~\ref{tab:model-comparison} shows for four models their free parameters,
reduced $\chi^{2}$, and $\Delta$AIC. $\chi^{2}_{\nu}$ is computed with $N_{\rm
points}=602$ for NIRSpec, 9749 for MRS, and 10351 in total. $\Delta$AIC $=$
AIC$_{\rm min}-$AIC, where AIC$_{\rm min}$ is the combined AIC of the best-fit
model. All models are fitted to the same data at native spectral resolution,
with identical $N_{\rm points}$ and identical $2\Sigma_{i}\ln(\sigma_{i}) =
-990345.1$, so their AIC values are directly comparable. 
The smallest $\Delta$AIC in
Table~\ref{tab:model-comparison} is 85 times that fluctuation (i.e., Monte Carlo uncertainty), so none of the
comparisons with the best-fit model rest on it. The four-subpopulation model
has an AIC lower than that of the three-subpopulation model by 7.95, 3.5 times
the fluctuation (Section~\ref{sec:pah:mepah:4pop}).

\subsection{The No-DPAHs Model}
\label{sec:appendix1-nodpah}

\begin{figure*}[h]
\figurenum{A1}
\begin{center}
\includegraphics[trim=0.07cm 0.05cm 0.07cm 0.05cm, clip, width=0.850\textwidth]{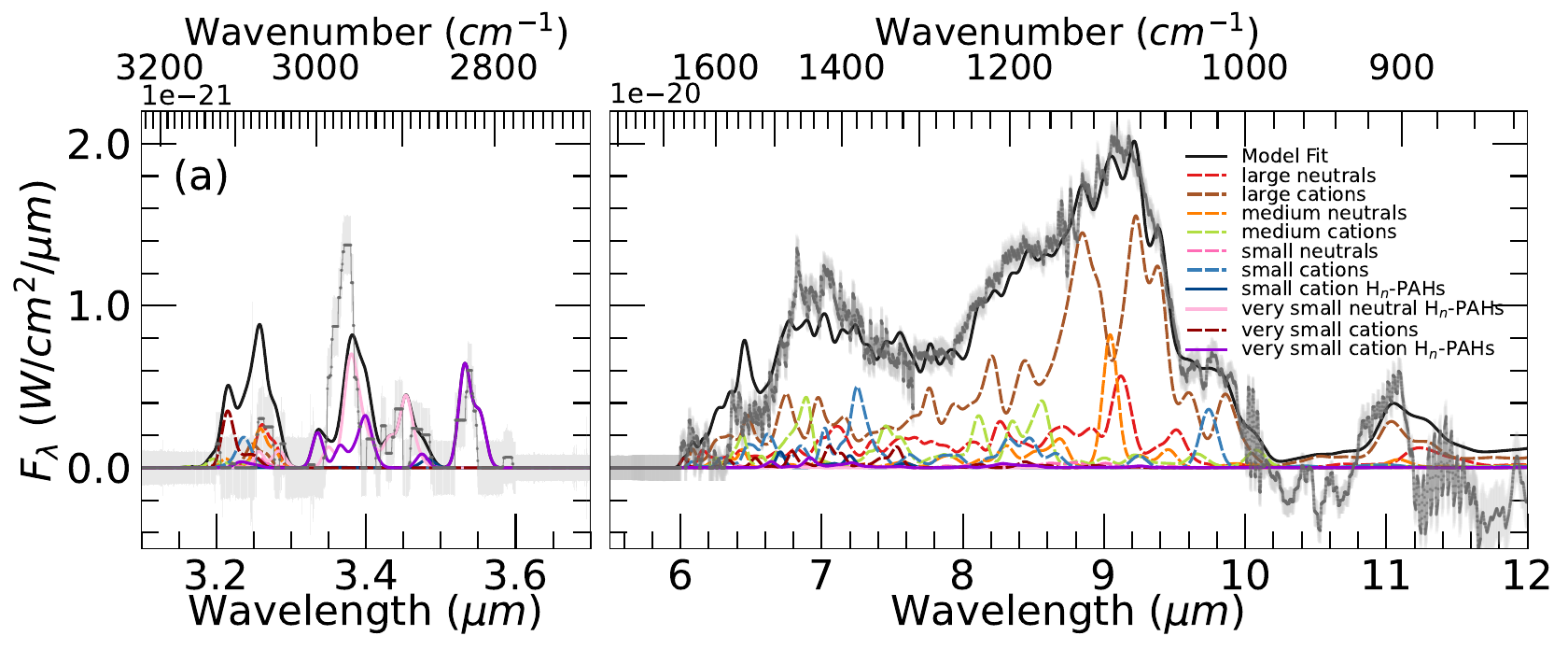}
\includegraphics[trim=0.07cm 0.05cm 0.07cm 0.05cm, clip, width=0.850\textwidth]{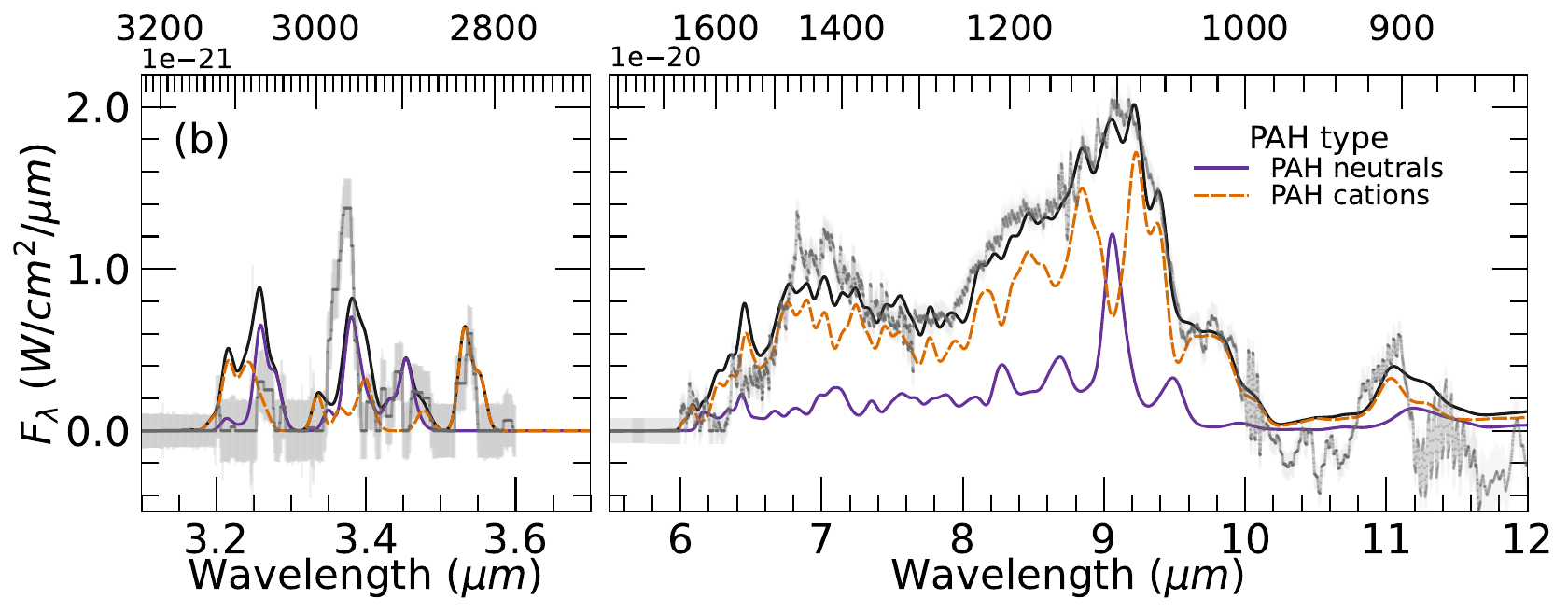}
\includegraphics[width=0.850\textwidth]{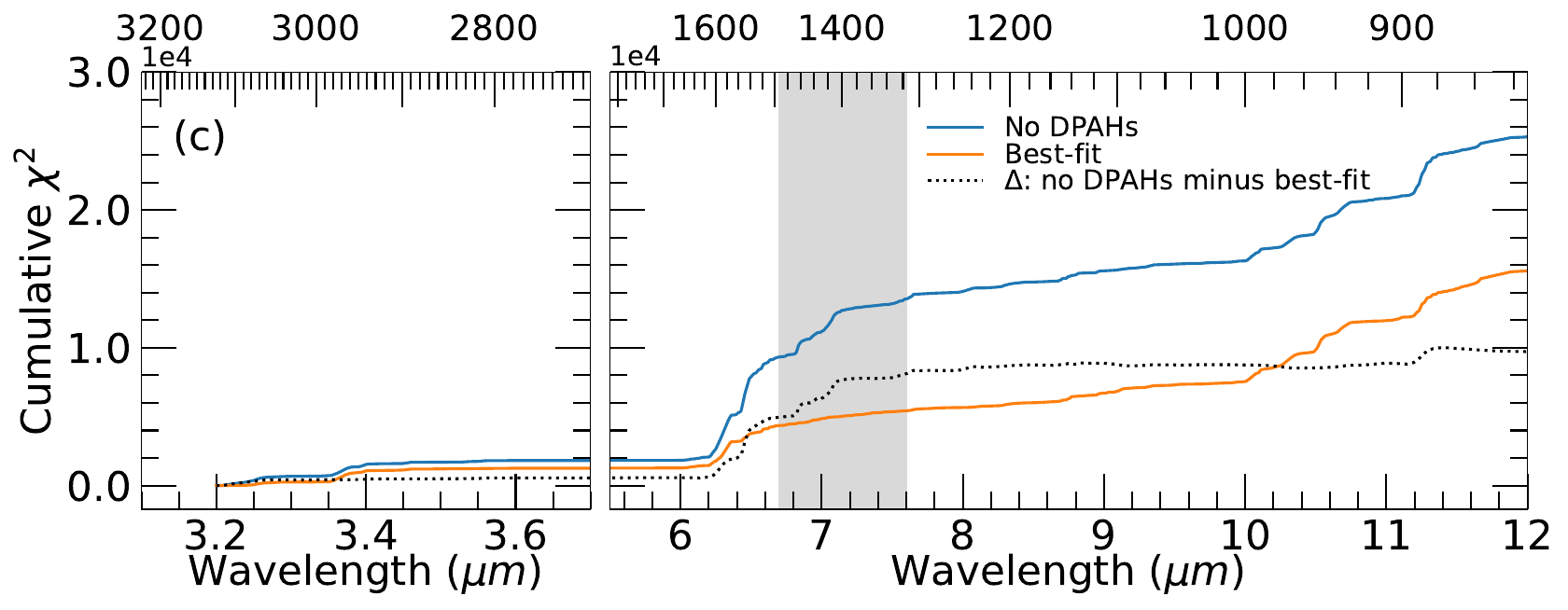}
\caption{The no-DPAHs model for comet 81P, in which PAHs with zero peripheral
hydrogen are excluded. In (a)--(c) the left panel is NIRSpec over
3.2--3.6~\micron{} and the right panel is MRS over 5.5--12~\micron{}, the
model fit is the black solid line, and the comet 81P data are gray with their
uncertainties shaded. (a)~Components, with line colors as in
Figure~\ref{fig:figure-4}(a--c) and two additions: small neutrals (hotpink
dashed) and small cation H$_{n}$-PAHs (midnightblue solid), neither of which
is retained in the best-fit model. The DPAHs (lime solid) of
Figure~\ref{fig:figure-4} are absent by construction. (b)~Breakdown by charge
state, as in Figure~\ref{fig:figure-4}(f): PAH neutrals (rebeccapurple solid)
and PAH cations (chocolate dashed). (c)~Cumulative $\chi^{2}$ versus
wavelength over the fitted points shown, for the no-DPAHs model (blue solid)
and the best-fit model (orange solid), and their difference, no-DPAHs minus
best-fit (black dotted). The gray band is the 6.9~\micron{} complex,
6.7--7.6~\micron{}. The dotted curve rises in a single step between 6.0 and
7.6~\micron{}, the interval in which the no-DPAHs model departs from the data
and which contains 0.79 of the total $\Delta\chi^{2}$. Colorblind-friendly HEX
colors are used; the closest matplotlib names are given to aid the eye.}
\label{fig:appendix-1}
\end{center}
\end{figure*}

The three-subpopulation and best-fit models have the same $k$, so their
$\Delta$AIC is $\Delta\chi^{2}$ alone. The three- and four-subpopulation
models fit the NIRSpec better than the best-fit model does but they fit the
MRS significantly worse. The no-DPAHs model (H$>$0) fits significantly worse
than the best-fit model and is presented here as a demonstration of the
importance of DPAHs in the best-fit model.

Figure~\ref{fig:appendix-1}(a) shows the no-DPAHs model. Removing the DPAHs
does not subtract a component from the best-fit model, it reorganizes the fit.
The no-DPAHs model retains 47 species against the best-fit model's 57, and
only 20 are common to both. Figure~\ref{fig:appendix-1}(b) shows the breakdown
into charge, and there is no DPAH curve compared to the same figure for the
best-fit model (Figure~\ref{fig:figure-4}(f)). The no-DPAHs model fits worse
over the whole spectrum, $\Delta$AIC $=-9504.7$ combined and $-8945.9$ on the
MRS alone (Table~\ref{tab:model-comparison}).

Figure~\ref{fig:appendix-1}(c) shows where the fit is lost. The cumulative
$\chi^{2}$ of the two models, and their difference, rise together in a single
step between 6.0 and 7.6~\micron{}, which holds 0.79 of the total
$\Delta\chi^{2}$ of 9525; the 6.9~\micron{} complex as defined
(6.7--7.6~\micron{}) holds 0.33 and 6.0--6.7~\micron{} holds 0.46. The large
and medium cations that replace the DPAHs in the complex emit more strongly at
6.0--6.7~\micron{}, relative to 6.7--7.3~\micron{}, than the DPAHs do. Raised
to fill the complex, they overshoot the data by 23--30\% over
6.0--6.7~\micron{} and still fall 10--12\% short over 6.7--7.3~\micron{},
where the best-fit model is within a few percent. Outside 6.0--7.6~\micron{}
the two models differ little, and longward of 12~\micron{} the no-DPAHs model
fits marginally better.

Over 6.7--7.6~\micron{}, the fractions of the modeled flux density change as
follows between the best-fit and the no-DPAHs model: DPAHs 0.54 $\rightarrow$
0, large cations 0.19 $\rightarrow$ 0.33, medium cations 0.07 $\rightarrow$
0.18, large neutrals 0.06 $\rightarrow$ 0.17, and small cations 0.10
$\rightarrow$ 0.18. The very small cation H$_{n}$-PAHs barely move between the
best-fit and no-DPAHs models, 0.01 $\rightarrow$ 0.02. What changes among the
H$_{n}$-PAHs is which of them contributes to the complex. Every H$_{n}$-PAH
retained in the best-fit model is very small. In the no-DPAHs model a small
H$_{n}$-PAH cation that the best-fit model does not retain,
C$_{24}$H$_{24}^{+2}$, is retained in 1000 of 1000 trials and contributes 0.02
to the 6.9~\micron{} complex. The 6.9~\micron{} complex switches from
DPAH-dominated to cation-dominated. Over 6.7--7.6~\micron{} the PAH cations
rise from 0.37 to 0.80 of the modeled flux density while the PAH neutrals rise
only from 0.09 to 0.20 (Figure~\ref{fig:appendix-1}(b)).

Two features do not move. The 3.37~\micron{} feature is carried entirely by
very small H$_{n}$-PAHs in both models, and the regular PAHs contribute
nothing to it. In the best-fit model the carriers are C$_{10}$H$_{18}$ (0.39),
C$_{19}$H$_{18}$ (0.38), C$_{16}$H$_{16}^{+}$ (0.11) and C$_{10}$H$_{18}^{+2}$
(0.06), and in the no-DPAHs model the same molecules switch places,
C$_{19}$H$_{18}$ (0.43), C$_{10}$H$_{18}$ (0.29), C$_{16}$H$_{16}^{+}$ (0.15)
and C$_{10}$H$_{12}^{+}$ (0.09). The 3.53~\micron{} feature is also carried by
H$_{n}$-PAHs in both models, but C$_{10}$H$_{18}^{+2}$ is not among the 47
species of the no-DPAHs model. There the feature is filled in by
C$_{10}$H$_{12}^{+}$ (0.69) and C$_{16}$H$_{16}^{+2}$ (0.31). The carriers are
H$_{n}$-PAHs in both models, the same molecules for the 3.37~\micron{} feature
but different molecules for the weaker 3.53~\micron{} feature, where only
cations contribute.

The presence of these H$_{n}$-PAHs does not depend on which PAHs contribute to
the MIR. The two models share only 20 of their 57 and 47 retained species, and
over 6.7--7.6~\micron{} the DPAHs are replaced by cations at every size. Yet
both models place the 3.37 and 3.53~\micron{} features entirely on very small
H$_{n}$-PAHs, and each of those carriers is retained in at least 999 of 1000
trials.

\clearpage

\section{The Best-fit PAH Model, Species by Species}
\setcounter{table}{0}
\setcounter{figure}{0}
\label{sec:appendix2}

Table~\ref{tab:bestfit-species} lists the 57 PAHs retained by the best-fit
model, in the order in which the selection admitted them: first by NIRSpec
flux to 99\% of the NIRSpec total, then by MRS flux to 95\% of the MRS total.
The \textit{counts} column gives the number of Monte Carlo trials out of 1000
in which the species was retained.

All species with n$_{\rm N} > 0$ are PANHs, with the nitrogen in skeletal
bonds, except C$_{10}$H$_{9}$N$^{+}$ (uid=476, a PAHN with the nitrogen in a
peripheral amino group).

\startlongtable
\begin{deluxetable*}{rlllrrrrr}
\tabletypesize{\footnotesize}
\tablecaption{The Best-fit PAH Model for Comet 81P, Species by Species\label{tab:bestfit-species}}
\tablewidth{0pt}
\tablecolumns{9}
\tablehead{
\colhead{uid} & \colhead{formula} & \colhead{size} & \colhead{subpopulation} & \colhead{counts} &
\colhead{$f_{\rm NIR}$} & \colhead{$\Sigma f_{\rm NIR}$} &
\colhead{$f_{\rm MRS}$} & \colhead{$\Sigma f_{\rm MRS}$} \\
\colhead{} & \colhead{} & \colhead{} & \colhead{} & \colhead{of 1000} &
\colhead{(\%)} & \colhead{(\%)} & \colhead{(\%)} & \colhead{(\%)}
}
\startdata
\cutinhead{Block 1 --- ordered by NIRSpec flux, to 99\% of the NIRSpec total}
351    & C$_{19}$H$_{18}$               & very small & H$_n$-PAHs           & 1000 &  24.35 &  24.35 &   0.05 &   0.05 \\
352    & C$_{10}$H$_{18}$               & very small & H$_n$-PAHs           & 1000 &  20.81 &  45.16 &   0.01 &   0.05 \\
4396   & C$_{10}$H$_{18}^{+2}$          & very small & H$_n$-PAHs, Me-PAHs  & 1000 &  10.92 &  56.08 &   0.04 &   0.09 \\
17     & C$_{10}$H$_{12}^{+}$           & very small & H$_n$-PAHs           &  999 &   8.89 &  64.97 &   0.06 &   0.15 \\
362    & C$_{16}$H$_{16}^{+}$           & very small & H$_n$-PAHs           & 1000 &   6.33 &  71.30 &   0.09 &   0.25 \\
10206  & C$_{60}$H$_{20}$               & medium     & PAHs                 & 1000 &   5.86 &  77.15 &   4.83 &   5.08 \\
4380   & C$_{16}$H$_{16}^{+2}$          & very small & H$_n$-PAHs           & 1000 &   4.13 &  81.28 &   0.04 &   5.12 \\
9911   & C$_{72}$H$_{22}$               & large      & PAHs                 & 1000 &   3.57 &  84.85 &   4.63 &   9.74 \\
10194  & C$_{60}$H$_{20}$               & medium     & PAHs                 & 1000 &   3.33 &  88.18 &   1.86 &  11.60 \\
821    & C$_{34}$H$_{20}^{+}$           & small      & PAHs                 & 1000 &   1.63 &  89.81 &   0.77 &  12.37 \\
9669   & C$_{94}$H$_{26}$               & large      & PAHs                 & 1000 &   1.51 &  91.32 &   3.77 &  16.14 \\
9947   & C$_{72}$H$_{22}$               & large      & PAHs                 &  999 &   1.31 &  92.63 &   1.77 &  17.90 \\
9743   & C$_{94}$H$_{26}$               & large      & PAHs                 & 1000 &   1.16 &  93.80 &   3.62 &  21.52 \\
8517   & C$_{93}$H$_{24}$N$_{3}^{+}$    & large      & PAHs                 & 1000 &   1.02 &  94.81 &   5.43 &  26.95 \\
502    & C$_{11}$H$_{8}$O$^{+}$         & very small & PAHs                 &  861 &   1.01 &  95.82 &   0.01 &  26.97 \\
8527   & C$_{93}$H$_{24}$N$_{3}^{+}$    & large      & PAHs                 & 1000 &   0.69 &  96.51 &   2.14 &  29.10 \\
6290   & C$_{54}$H$_{24}^{+2}$          & medium     & PAHs                 & 1000 &   0.59 &  97.10 &   2.47 &  31.57 \\
6258   & C$_{60}$H$_{24}^{+2}$          & medium     & PAHs                 & 1000 &   0.38 &  97.48 &   1.27 &  32.84 \\
6184   & C$_{72}$H$_{24}^{+2}$          & large      & PAHs                 & 1000 &   0.26 &  97.74 &   4.28 &  37.13 \\
8524   & C$_{93}$H$_{24}$N$_{3}^{+}$    & large      & PAHs                 & 1000 &   0.26 &  98.00 &   1.10 &  38.23 \\
4890   & C$_{33}$H$_{19}^{+2}$          & small      & PAHs                 & 1000 &   0.24 &  98.23 &   1.55 &  39.78 \\
530    & C$_{11}$H$_{7}$O$^{+}$         & very small & PAHs                 &  185 &   0.20 &  98.44 &   0.00 &  39.78 \\
5777   & C$_{29}$H$_{17}^{+2}$          & small      & PAHs                 & 1000 &   0.17 &  98.61 &   1.85 &  41.64 \\
8466   & C$_{93}$H$_{24}$N$_{3}^{+}$    & large      & PAHs                 &  947 &   0.15 &  98.75 &   0.60 &  42.24 \\
7725   & C$_{94}$H$_{24}$N$_{2}^{+}$    & large      & PAHs                 & 1000 &   0.12 &  98.88 &   1.37 &  43.61 \\
476    & C$_{10}$H$_{9}$N$^{+}$         & very small & PAHs                 &  379 &   0.12 &  99.00 &   0.00 &  43.61 \\
\cutinhead{Block 2 --- ordered by MRS flux, to 95\% of the MRS total}
2376   & C$_{22}$                       & small      & DPAHs                & 1000 &   0.00 &  99.00 &   5.61 &  49.22 \\
6424   & C$_{172}$H$_{38}^{+}$          & large      & PAHs                 & 1000 &   0.07 &  99.06 &   5.35 &  54.57 \\
3000   & C$_{33}$                       & small      & DPAHs                & 1000 &   0.00 &  99.06 &   5.12 &  59.69 \\
2789   & C$_{32}$                       & small      & DPAHs                & 1000 &   0.00 &  99.06 &   3.85 &  63.54 \\
3084   & C$_{34}$                       & small      & DPAHs                & 1000 &   0.00 &  99.06 &   3.63 &  67.17 \\
2923   & C$_{33}$                       & small      & DPAHs                & 1000 &   0.00 &  99.06 &   2.87 &  70.04 \\
2541   & C$_{29}$                       & small      & DPAHs                & 1000 &   0.00 &  99.06 &   2.66 &  72.70 \\
6401   & C$_{108}$H$_{30}^{+2}$         & large      & PAHs                 & 1000 &   0.08 &  99.14 &   2.47 &  75.17 \\
2807   & C$_{32}$                       & small      & DPAHs                & 1000 &   0.00 &  99.14 &   2.32 &  77.49 \\
735    & C$_{70}$                       & medium     & DPAHs                & 1000 &   0.00 &  99.14 &   2.31 &  79.80 \\
2615   & C$_{30}$                       & small      & DPAHs                & 1000 &   0.00 &  99.14 &   2.19 &  82.00 \\
2523   & C$_{29}$                       & small      & DPAHs                & 1000 &   0.00 &  99.14 &   2.18 &  84.18 \\
2525   & C$_{29}$                       & small      & DPAHs                & 1000 &   0.00 &  99.14 &   1.82 &  85.99 \\
2762   & C$_{32}$                       & small      & DPAHs                & 1000 &   0.00 &  99.14 &   1.71 &  87.71 \\
3061   & C$_{33}$                       & small      & DPAHs                & 1000 &   0.00 &  99.14 &   1.69 &  89.40 \\
6328   & C$_{150}$H$_{36}^{+2}$         & large      & PAHs                 & 1000 &   0.03 &  99.17 &   1.65 &  91.05 \\
6104   & C$_{96}$H$_{30}^{+2}$          & large      & PAHs                 & 1000 &   0.10 &  99.27 &   0.96 &  92.01 \\
2397   & C$_{25}$                       & small      & DPAHs                & 1000 &   0.00 &  99.27 &   0.94 &  92.95 \\
2765   & C$_{32}$                       & small      & DPAHs                &  998 &   0.00 &  99.27 &   0.86 &  93.81 \\
6340   & C$_{144}$H$_{36}^{+2}$         & large      & PAHs                 & 1000 &   0.02 &  99.29 &   0.82 &  94.63 \\
7435   & C$_{114}$H$_{30}^{+}$          & large      & PAHs                 & 1000 &   0.06 &  99.36 &   0.79 &  95.42 \\
2371   & C$_{21}$                       & small      & DPAHs                & 1000 &   0.00 &  99.36 &   0.78 &  96.20 \\
2363   & C$_{18}$                       & very small & DPAHs                & 1000 &   0.00 &  99.36 &   0.75 &  96.95 \\
704    & C$_{54}$                       & medium     & DPAHs                & 1000 &   0.00 &  99.36 &   0.54 &  97.50 \\
2628   & C$_{30}$                       & small      & DPAHs                &  988 &   0.00 &  99.36 &   0.45 &  97.94 \\
5102   & C$_{33}$H$_{19}^{+2}$          & small      & PAHs                 &  821 &   0.05 &  99.40 &   0.32 &  98.26 \\
2433   & C$_{26}$                       & small      & DPAHs                &  856 &   0.00 &  99.40 &   0.29 &  98.55 \\
710    & C$_{96}^{+2}$                  & large      & DPAHs                & 1000 &   0.00 &  99.40 &   0.22 &  98.78 \\
4864   & C$_{33}$H$_{19}^{+2}$          & small      & PAHs                 &  904 &   0.03 &  99.44 &   0.20 &  98.97 \\
9470   & C$_{94}$H$_{26}^{+}$           & large      & PAHs                 & 1000 &   0.03 &  99.47 &   0.15 &  99.13 \\
2949   & C$_{33}$                       & small      & DPAHs                &  337 &   0.00 &  99.47 &   0.15 &  99.28 \\
\enddata
\tablecomments{\textbf{The row order is the selection method and should not be re-sorted.}
Species were ranked by their mean fractional contribution to the integrated NIRSpec flux
(3.2--3.6~\micron{}) and accepted until the cumulative NIRSpec flux reached 99\%; that is Block~1,
26 species. The remaining species were then ranked by their contribution to the integrated MRS
flux and accepted until the cumulative MRS flux reached 95\%; that is Block~2, 31 species.
Block~1 already supplies 43.61\% of the MRS flux before Block~2 begins.
The 57 species together account for 99.47\% of the NIRSpec flux and 99.28\% of the MRS flux.
$\Sigma f$ columns are running totals down the table in this order.
Fractions were computed in Jy and converted to W~cm$^{-2}$~\micron{}$^{-1}$.
Sizes are very small ($N_{\rm C} < 20$), small ($20 < N_{\rm C} \leq 50$),
medium ($50 < N_{\rm C} \leq 70$) and large ($N_{\rm C} > 70$). Subpopulation is set by
hydrogenation and alkylation, a separate axis from composition, which is set by heteroatoms
(Table~\ref{tab:type-size-counts}).
C$_{10}$H$_{18}^{+2}$ (uid=4396) carries both --CH$_{2}$ and --CH$_{3}$ groups and so belongs to
the H$_{n}$-PAHs and the Me-PAHs alike; it is not counted twice.}
\end{deluxetable*}

\clearpage
\bibliographystyle{aasjournalv7}

\section{Methanol Flourescence Modeling with the NASA Planetary Spectrum Generator}\label{sec:methanol}

The methanol (CH$_3$OH) fluorescence models implemented in the NASA Planetary Spectrum Generator (PSG) are primarily based on the spectroscopic and fluorescence framework developed by \cite{Villanueva2012a} for the $\nu_3$ C--H stretching band. That work provided a detailed quantum-mechanical description of the rotation--torsion--vibration structure of CH$_3$OH, targeting one specific ro-vibrational band, and demonstrated that the resulting fluorescence model could accurately reproduce high-resolution cometary spectra. Within PSG, this framework has been retained for the $\nu_3$ band and adapted to represent the neighboring $\nu_2$ and $\nu_9$ C--H stretching systems. Because these bands differ substantially in the completeness of their laboratory characterization and quantum assignments, their implementation necessarily employs different levels of spectroscopic approximation. The $\nu_3$ system is represented by the detailed model of \cite{Villanueva2012a}, whereas the $\nu_2$ and particularly the $\nu_9$ systems are modeled using extensions of this framework constrained by available laboratory and astronomical spectra. Together, these models provide a unified description of the principal CH$_3$OH fluorescence emission across the 3~$\mu$m C--H stretching region and enable forward modeling of this complex spectral interval within PSG.

Modeling the infrared spectroscopic properties and fluorescence emission of methanol (CH$_3$OH) is particularly challenging because of the molecule's asymmetric structure and hindered internal rotation. The coupling between overall molecular rotation, torsional motion, and small-amplitude vibrations produces a dense and strongly perturbed rovibrational spectrum, particularly in the C--H stretching region near 3~$\mu$m. Until relatively recently, quantum assignments for individual transitions within the three principal C--H stretching fundamentals, $\nu_2$, $\nu_3$, and $\nu_9$, were restricted to a comparatively small subset of the transitions contributing to this spectral region \citep[e.g.,][]{Xu1997}. This limitation is particularly important for astronomical applications because the 2--5~$\mu$m interval is accessible to many ground- and space-based infrared spectrometers and contains some of the strongest fluorescent signatures of methanol.

The difficulty in constructing a complete spectroscopic model becomes especially apparent between approximately 3.35 and 3.6~$\mu$m, where thousands of CH$_3$OH transitions contribute to the observed spectrum. Unambiguous quantum assignments of individual lines in this region have historically proven difficult \citep{Hunt1991,Bignall1994,Xu1997}. Accurate calculation of the rotation--torsion--vibration energy structure requires a Hamiltonian capable of accounting for strong interactions among the C--H stretching vibrations, torsional motion, and nearby low-lying vibrational states. These interactions produce substantial shifts and mixing of the nominal zeroth-order energy levels. Detailed analysis of the $\nu_2$ system by \cite{Xu1997}, for example, demonstrated strong mixing with nominally ``dark'' background or bath states. The resulting perturbations redistribute intensity among transitions and disrupt otherwise recognizable rovibrational progressions, making a conventional isolated-band description inadequate.

Early fluorescence models developed for the interpretation of cometary spectra therefore necessarily adopted substantial simplifications. The first attempts to represent the C--H stretching bands at relatively low resolving powers ($R \sim 1000$) used known ground-state rotational constants to approximate the rotational structure of the excited vibrational states while neglecting, or only approximately representing, molecular asymmetry and internal rotation \citep{Hoban1991,Reuter1992,BockeleeMorvan1995}. Despite these limitations, these pioneering calculations established the basic interpretation of the 3~$\mu$m methanol emission observed in comets and provided fluorescence efficiencies ($g$-factors) that have subsequently been used to retrieve CH$_3$OH production rates and abundances. Their limitations become increasingly apparent at higher spectral resolving powers ($R \gtrsim 10^4$), however, where the approximate models no longer reproduce the detailed morphology, relative intensities, and positions of the observed emission features \citep[e.g.,][their Fig.~1b.]{DiSanti2009}.

A major improvement was achieved with the development of a full quantum model for the $\nu_3$ band by \cite{Villanueva2012a}. In contrast to the earlier approximate treatments, this model explicitly incorporated the torsional structure of CH$_3$OH and accounted for perturbations to the rovibrational levels arising from interactions with nearby levels and dark background states. The resulting spectroscopic model was benchmarked against high-resolution laboratory and astronomical observations and demonstrated excellent agreement with the measured spectral structure \citep{Villanueva2012a}. This model provides the most mature description of the three C--H stretching fundamentals considered here and forms an important component of the CH$_3$OH fluorescence treatment implemented within the NASA Planetary Spectrum Generator (PSG).

The remaining C--H stretching bands present greater challenges. The $\nu_9$ system is spectroscopically distinct from the $\nu_2$ and $\nu_3$ systems, but its high line density and extensive perturbations have historically prevented the construction of a comparably complete quantum model. Only a limited number of upper-state rovibrational levels and transitions have been cataloged \citep{Xu1997}. Using high-resolution cometary spectra, Villanueva et al.\ (in preparation) have also developed an approximate representation of the $\nu_2$ system that reproduces a substantial fraction of the observed emission features. Together with the detailed $\nu_3$ model, this provides a framework for representing much of the methanol emission across the C--H stretching region.

The brightest and most structurally complex component of this region nevertheless remains difficult to reproduce in detail. Strong perturbations involving torsional and vibrational states result in highly irregular spectral structure, such that the observed transitions do not follow the simple rovibrational progressions expected for an approximately asymmetric-top molecule. Consequently, a direct extrapolation from conventional rotational constants or from a small set of assigned transitions can lead to substantial errors in both line positions and intensities.

To obtain a first-order representation of the $\nu_9$ contribution suitable for fluorescence calculations in PSG, we developed an empirical spectroscopic model constrained by both laboratory and astronomical observations. The laboratory constraints were provided by room-temperature spectra from the Pacific Northwest National Laboratory (PNNL) spectral database, while astronomical constraints were obtained from JWST observations of comet C/2022 E3 \citep{Milam2026}. The combination is particularly useful because the two datasets probe substantially different rotational populations: the PNNL spectrum represents a rotational temperature of approximately $T_{\rm rot}=296$~K, whereas the cometary spectrum is characterized by a considerably lower rotational temperature of approximately $T_{\rm rot}=80$~K.

Starting from an approximate upper-state description of the $\nu_9$ system, we introduced perturbations to the lower-$K$ ladders and explored candidate rovibrational progressions that could simultaneously reproduce features present in the laboratory and cometary datasets. The objective was not to derive a definitive effective Hamiltonian for the $\nu_9$ state, but rather to construct a physically motivated empirical representation that captures the dominant spectral structure and integrated fluorescence of the band over the range of temperatures relevant to astronomical observations. This distinction is important: the present treatment should be regarded as a practical spectroscopic approximation for forward modeling rather than a complete quantum-mechanical solution of the highly perturbed $\nu_9$ manifold.

An important result of this exercise is that the previously reported assignments of \cite{Xu1997} could not be independently validated against the combined PNNL and astronomical datasets. Adopting those assignments directly as the basis for the present model resulted in significant discrepancies in the predicted spectral structure. We therefore did not enforce these assignments when constructing the empirical $\nu_9$ representation. Instead, the model was optimized to reproduce consistent progressions identifiable across the laboratory and astronomical spectra. This approach provides improved agreement with the available observations, although definitive quantum assignments will require additional high-resolution laboratory spectroscopy and a more complete theoretical treatment of the interacting torsion--rotation--vibration states.

The resulting spectroscopic description is incorporated into the methanol fluorescence framework in PSG. For a specified rotational temperature, PSG calculates the populations of the relevant lower states and propagates these populations through the adopted rovibrational transition structure to predict the fluorescent spectrum. The simultaneous treatment of the $\nu_2$, $\nu_3$, and $\nu_9$ systems is particularly important at moderate spectral resolving powers, where transitions from the three bands overlap and collectively determine the morphology of the observed 3~$\mu$m emission complex. 

The present $\nu_9$ treatment should nevertheless be regarded as provisional. Its successful reproduction of the overall PNNL and JWST spectral morphology demonstrates that it captures much of the relevant spectroscopic structure, but agreement at moderate resolving power does not uniquely validate the underlying quantum assignments. Further benchmarking against laboratory spectra spanning a broader range of temperatures, together with high-resolution astronomical observations and improved theoretical calculations of the coupled torsion--rotation--vibration system, will be necessary to establish a fully predictive line-by-line model. Such work will be particularly important for applications at high resolving power, where small errors in upper-state energies and perturbations translate directly into measurable offsets in individual transition frequencies.

\begin{figure*}[ht!]
\figurenum{C1}
\begin{center}
\includegraphics[trim=0.07cm 0.05cm 0.07cm 0.07cm, clip, width=0.800\textwidth]{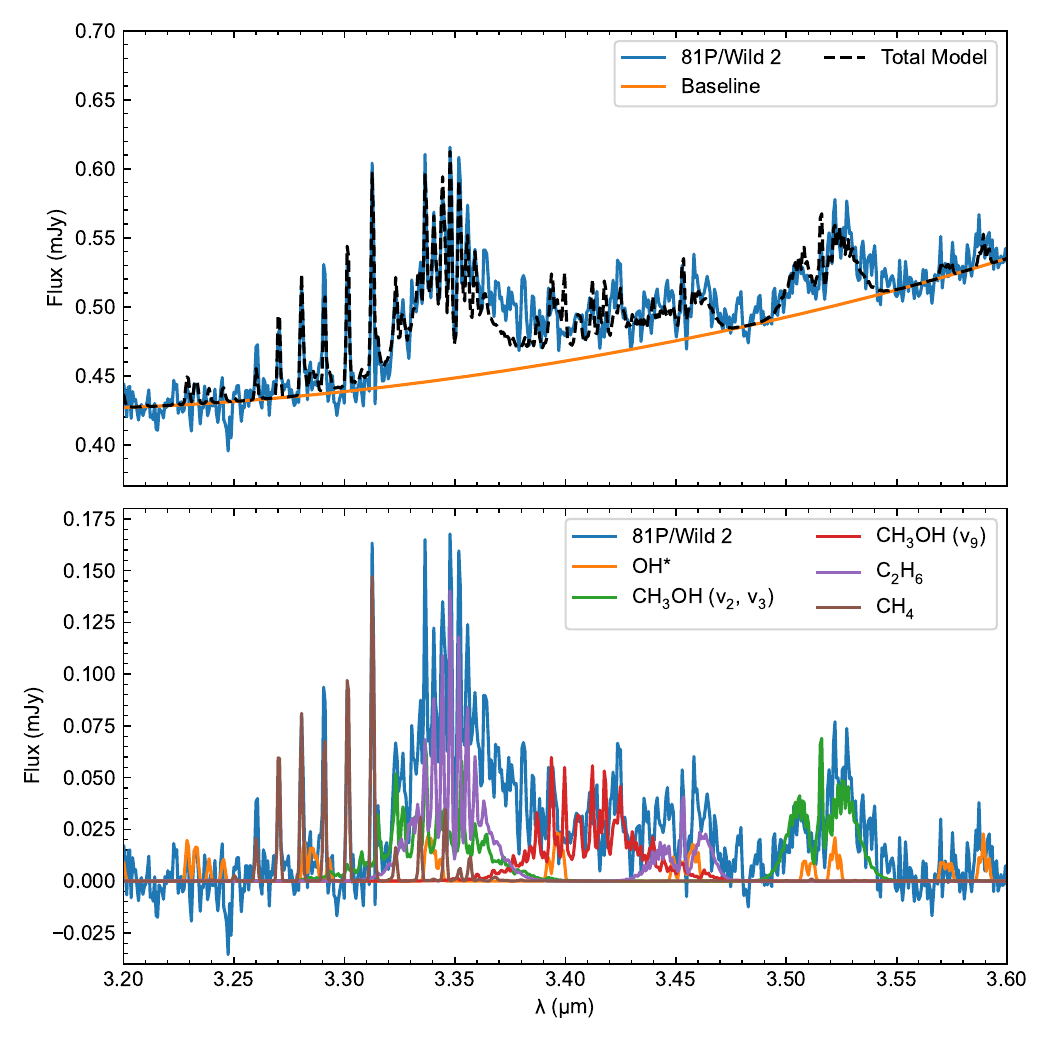}
\caption{Spectrum of comet 81P extracted in a 1\farcs41 diameter aperture and showing detections of \ce{CH3OH}, \ce{C2H6}, \ce{CH4}, and OH* on March 24, demonstrating the application of the PSG \ce{CH3OH} model to generate a residual spectrum. The upper panel shows the total fluorescence model and the spectral baseline. The lower panel shows the baseline-subtracted 81P spectrum, with individual molecular fluorescence models overlaid. The \ce{CH3OH} models are broken out by band assignment. Recreated from Paper I \citep{2026arXiv260802190R}.}
\label{fig:figure-c1}
\end{center}
\end{figure*}

Figure~\ref{fig:figure-c1} shows our application of the PSG \ce{CH3OH} model to the 81P JWST spectrum, with the $\nu_9$ model separated from the $\nu_2$ and $\nu_3$ models. The combined fluorescence models are able to account well for all major features except for the clear residual near 3.37 \um{}, which is associated with H$_n$-PAHs in this work.

\clearpage

\bibliography{Citations-v2}{}



\end{document}